\documentclass[12pt,a4paper]{report}
\makeatletter
\renewcommand\listoftables{%
    \section*{\listtablename}%
    \@mkboth{\MakeUppercase\listtablename}%
        {\MakeUppercase\listtablename}%
    \@starttoc{lot}%
}
\renewcommand\listoffigures{%
    \section*{\listfigurename}%
    \@mkboth{\MakeUppercase\listfigurename}%
        {\MakeUppercase\listfigurename}%
    \@starttoc{lof}%
}
\makeatother

\usepackage{verbatim}
\usepackage{fancyvrb}
\usepackage{graphicx}
\usepackage[numbers]{natbib}
\usepackage{latexsym,amssymb,amsmath,amsfonts,mathtools}

\usepackage{lmodern}	
\usepackage[usenames,dvipsnames,pdftex]{xcolor}
\usepackage{multirow}
\usepackage{siunitx}
\usepackage{float}
\usepackage[autostyle]{csquotes}
\usepackage[paper=a4paper,top=1in,bottom=1in,margin=1in]{geometry}
\usepackage{ragged2e}
\usepackage{textcomp}

\usepackage{hyperref}
\usepackage[acronyms,symbols,nonumberlist,section,toc,nopostdot]{glossaries}
\usepackage[automake]{glossaries-extra}

\usepackage{xargs}

\usepackage{subcaption}
\usepackage{titlesec}
\titleformat
{\chapter} 
[hang] 
{\bfseries\LARGE} 
{\thechapter } 
{.5em} 
{} 
[] 
\titlespacing*{\chapter}{0pt}{-50pt}{30pt}

\usepackage{fancyhdr}
\usepackage{pdflscape}

\usepackage{setspace}

\usepackage{enumitem}
\setlist[itemize]{noitemsep,topsep=0pt}

\usepackage{lastpage}

\definecolor{javared}{rgb}{0.6,0,0} 
\definecolor{javagreen}{rgb}{0.25,0.5,0.35} 
\definecolor{javapurple}{rgb}{0.5,0,0.35} 
\definecolor{javadocblue}{rgb}{0.25,0.35,0.75} 
\usepackage{listings}
\makeglossaries

\newcommand{\mathgloss}[3]{%
\newglossaryentry{#1}{
  type=symbols,
  name={\ensuremath{#2}},
  description={#3},
  symbol={\ensuremath{#2}}
}}
\mathgloss{efficiency}{\mathcal{E}}{Efficiency of the serverless platform}
\mathgloss{product}{\mathcal{P}}{Workload completed by a serverless platform}
\mathgloss{cost}{\mathcal{C}}{Resource cost of a serverless platform}
\mathgloss{sojourn}{S(e)}{Sojourn time of an event e}
\mathgloss{queue}{W(e)}{Queue delay of an event e}
\mathgloss{setup}{I(e)}{Setup time (initialisation) to an event e}
\mathgloss{exec}{B(e)}{Execution time of an event e}
\mathgloss{e:set}{e^k_i \in E^k}{The $i$th event in the set of events of class $k$}
\mathgloss{e:time}{t^k_i \in T^k}{Arrival time of the $i$th event of class $k$}
\mathgloss{e:decision}{x^k_{i,c} = \{0,1\}, x^k_{i,c} \in X}{Decision variable to schedule the $i$th event of class $k$ events on instance $c$ in a scheduling solution $X$}
\mathgloss{c:set}{c \in C^k}{Instance in the set of instances of class $k$}
\mathgloss{c:decision}{y^k_{c,w} = \{0,1\}, y^k_{c,w} \in Y}{Decision variable to allocate instance $c$ of class $k$ on worker $w$ in an allocation solution $Y$}
\mathgloss{w:set}{w \in W}{Worker $w$ in the set of workers}
\mathgloss{w:active}{N_w(t)}{Set of active instances on worker $w$ at time $t$}
\mathgloss{w:maxactive}{z_N}{Maximum number of active instances at a worker}
\mathgloss{w:total}{C_w(t)}{Set of instances (idle or active) on worker $w$ at time $t$}
\mathgloss{w:maxtotal}{z_C}{Maximum number of instances at a worker}
\mathgloss{w:setup}{t_w}{Time to obtain a worker from infrastructure resource management}
\mathgloss{maxarrival}{\Lambda}{Maximum arrival rate of a function in an experiment}

\newacronym{NIST}{NIST}
{US National Institute of Science and Technolgy}

\newacronym{BoT}{BoT}
{Bag of Tasks}

\newacronym{CNCF}{CNCF}
{Cloud-Native Computing Foundation}

\newacronym{MQTT}{MQTT}
{Message Queuing Telemetry Transport (ISO/IEC PRF 20922)}

\newacronym{AMQP}{AMQP}
{OASIS Advanced Message Queuing Protocol}

\newacronym{RDMA}{RDMA}
{Remote Direct Memory Access}

\newacronym{JVM}{JVM}
{Java\texttrademark\,Virtual Machine}

\newacronym{SR-IOV}{SR-IOV}
{Single Root Input/Output Virtualisation}

\newacronym{NVMe}{NVMe}
{Non-Volatile Memory Host Controller Interface Specification}

\newacronym{SSD}{SSD}
{Solid State Disk}

\newacronym{HDD}{HDD}
{Hard Disk Drive}

\newacronym{CPU}{CPU}
{Core Processing Unit}

\newacronym{DDR3}{DDR3 1600}
{Double Data Rate Type 3 Synchronous Dynamic Random-Access Memory with 1600 million transfers per second}

\newacronym{FCFS}{FCFS}
{First-Come First-Served}

\newacronym{PS}{PS}
{Processor-Sharing}

\newacronym{LRU}{LRU}
{Least Recently Used}

\newacronym{TPU}{TPU}
{Tensor Processing Unit}

\newacronym{GPU}{GPU}
{Graphical Processing Unit}

\newacronym{AWS}{AWS}
{Amazon Web Services, Inc.}

\newacronym{API}{API}
{Application Programming Interface}

\newacronym{URL}{URL}
{Uniform Resource Locator}

\newacronym{VM}{VM}
{Virtual Machine}

\newacronym{PaaS}{PaaS}
{Platform-as-a-Service}

\newacronym{IaaS}{IaaS}
{Infrastructure-as-a-Service}

\newacronym{NodeJS}{NodeJS}
{Node.js\textregistered\,JavaScript runtime}

\newacronym{BF}{BF}
{Best-Fit Heuristic}

\newacronym{FF}{FF}
{First-Fit Heuristic}

\newacronym{NF}{NF}
{Next-Fit Heuristic}

\newacronym{10GbE}{10GbE}
{10 Gigabit Ethernet}

\newacronym{PTAS}{PTAS}
{Polynomial-Time Approximation Scheme}

\newacronym{SLA}{SLA}
{Service-Level Agreement}

\newglossaryentry{cloud-native}{name={cloud-native},%
  text={cloud-native},%
  description={is a synonym of the \gls{CNCF} model for container packaged, dynamically managed and microservice-oriented applications on container platforms}%
}%

\newglossaryentry{Serverless}{name={Serverless},%
  text={Serverless},%
  description={The fictional term serverless stems from the platform offering complete transparency (location, access, etc.) of event execution that relieves the customer from compute resource planning and operation, hence the name. {\citet{adzic:2017}} explain the term with the lack of a server process (event loop) in the deployed application that only consists of function logic.}%
}%

\newglossaryentry{greenthreads}{name={green threads},%
  text={green threads},%
  description={(or green threading) emulates concurrency in a user process, e.g. the runtime provides threads that are not natively manages by the kernel but only in user-space.}%
}%

\newcommand*{\newdualentry}[4]{%
  \newglossaryentry{main-#1}{name={#2},%
  text={#2\glsadd{#1}},%
  description={{#4}}%
  }%
  \newglossaryentry{#1}{
  type=\acronymtype,
  first={#2 (#1)},
  name={#1\glsadd{main-#1}},
  description={\glslink{main-#1}{#2}}
  }%
}

\newdualentry{DPDK}{Data Plane Development Kit}{Data Plane Development Kit}{is a set of libraries for fast packet processing. The libraries run on dedicated core processing units and have exclusive access to dedicated network interfaces.}

\newdualentry{REST}{Representational state transfer}{Representational state transfer}{this best practise has been developed for interoperability between web APIs and organises the service interface as web resources (identified by URLs) that can be manipulated using HTTP methods (GET, POST, PUT, DELETE).}

\newdualentry{ETL}{Extract-Transform-Load}{Extract-Transform-Load}{is a data warehousing term that describes data processing which extracts information from a source, transforms it to a common representation and loads it into a structured result set. Big Data Analytics jobs can be considered ETL jobs although the pipeline is typically segmented to store intermediate results, e.g. MapReduce.}

\newdualentry{IoT}{Internet of Things}{Internet of Things}{the term is generally applied to describe distributed sensing and acting with the use of many devices (things) that communicate in a common spectrum (Internet).}

\newdualentry{SaaS}{Software-as-a-Service}{Software-as-a-Service}{provides use of applications hosted in a Cloud ecosystem but without the need to manage or control any of the resource aspects and no means to change its capabilities except to control access modalities.}

\newdualentry{LBaaS}{Load-Balancing-as-a-Service}{Load-Balancer-as-a-Service}{provides state-of-the-art configurable load balancing for virtualised application deployments.}

\newdualentry{CaaS}{Container-as-a-Service}{Container-as-a-Service}{is a service model that provisions OS-sharing, isolated execution contexts using kernel namespacing and often comprises autoscaling and load balancing services.}

\newdualentry{FaaS}{Function-as-a-Service}{Function-as-a-Service}{refers to the service model to provision code execution on demand for a resource-transparent deployment. The service aspect separates the provider role (e.g. corporation, institution, IT department) from the customer role (e.g. developer, application designer). }

\begin{document}
\pagenumbering{gobble}

\setlength{\topmargin}{-0.5cm} 
\addtolength{\textheight}{1.5cm} 
\begin{titlepage}
\newgeometry{top=1in,bottom=1in,right=1in,left=1in}



\begin{flushright}
\includegraphics[width=0.25\textwidth]{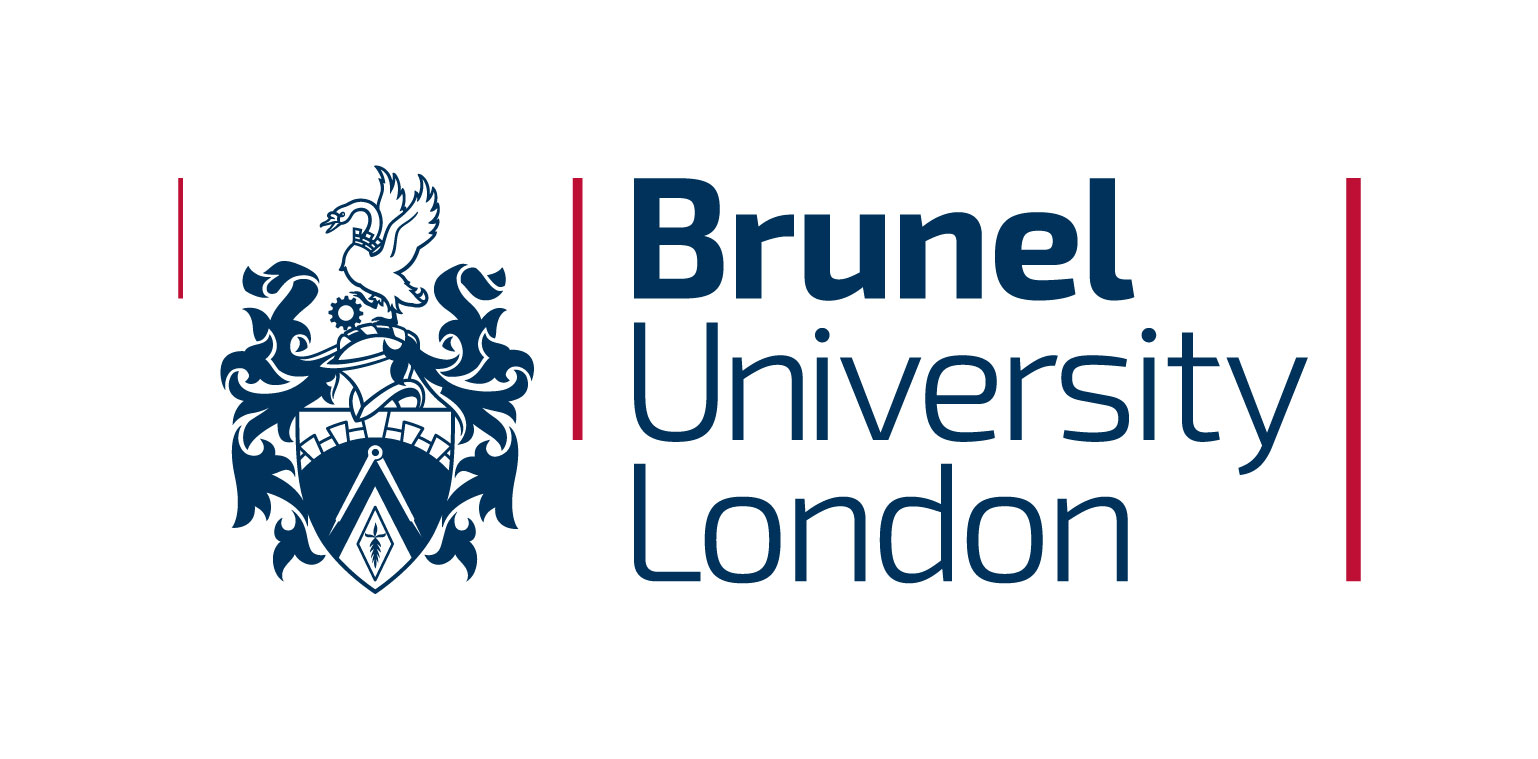}
\end{flushright}
\center 
\Large{\textbf{College of Engineering, Design and Physical Sciences}}

\vspace{0.2cm}

\large{\textbf{Department of Electronic \& Computer Engineering}}

\vspace{0.2cm}

\large{MSc Distributed Computing Systems Engineering}

\vspace{1.4cm}

\LARGE{\textbf{Brunel University London}}

\vspace{2cm}

{ \huge \bfseries Adaptive Event Dispatching in Serverless Computing
Infrastructures}

\vspace{2cm}

\LARGE{\textbf{Manuel Stein}}

\vspace{1.4cm}

\LARGE{\textbf{Prof. Maozhen Li}}

\vspace{1.4cm}

\large{March 2018}

\vspace{1.4cm}

A Dissertation submitted in partial fulfillment of the \\
requirements for the degree of Masters of Science

\end{titlepage}

\newgeometry{top=1in,bottom=1in,right=20mm,left=40mm}
\fancypagestyle{plain}{%
  \renewcommand{\headrulewidth}{0pt}%
  \fancyhf{}%
  \fancyfoot[C]{\footnotesize page \thepage}%
}
\pagestyle{fancy}
\fancyhf{}
\fancyhead[R]{\leftmark}
\fancyfoot[C]{\footnotesize page \thepage}


\chapter*{Acknowledgements}
I like to thank my loving wife, the best wife of all, for her relentless support, 
the countless times she took care of things, her constant caring for my wellbeing 
and her endless understanding and appreciation for my work.

Special thanks go to my supervisor, Prof. Maozhen Li, for taking up with the project idea 
and for consulting and guiding me on this endeavour.

I'd also like to thank Dr. Volker Hilt, who has welcomed and encouraged my pursuit of
the masters degree and I thank my employer, Nokia Bell Labs, for co-sponsoring
the part-time masters course.

\vspace{6cm}
\centerline{\large\emph{to Julia, Raphael and Noah}}

\newpage

\pagenumbering{roman}

\tableofcontents
\addcontentsline{toc}{chapter}{Contents}
\newpage

\listoffigures
\addcontentsline{toc}{section}{List of Figures}

\listoftables
\addcontentsline{toc}{section}{List of Tables}
\newpage

\printsymbols[title={List of Symbols}]
\newpage

\printacronyms[title={List of Acronyms}]
\newpage

\printglossary
\newpage


\pagenumbering{arabic}
\doublespacing

\chapter{Introduction}

Serverless computing is an emerging Cloud service model.
It is currently gaining momentum as the next step in the evolution of hosted
computing from capacitated machine virtualisation and microservices towards
utility computing.
The term \emph{serverless} has become a synonym for the entirely
resource-transparent deployment model of cloud-based event-driven distributed
applications.
This work investigates how adaptive event dispatching can improve serverless
platform resource efficiency and contributes a novel approach that allows for
better scaling and fitting of the platform's resource consumption to actual
demand.

Following the investigation of economical aspects and the state-of-the-art
overview on serverless given in the interim report \cite{interim}, the thesis report opens with the definition of a
common terminology to discuss design and integration aspects in order to synthesise a
platform design reference and an event dispatching reference architecture
(section~\ref{systems}) used in the remainder of the thesis.

A thorough literature review has been conducted
to analyse the problem space in chapter \ref{design}.
Distributed system scheduling taxonomies are applied to classify the serverless scheduling problem (section
\ref{design:taxonomy}). Stakeholders and survey papers have been consulted
where necessary to identify the design goal and applicable methods.
The classification is followed by formalisation of a multi-objective optimisation
problem (section \ref{design:moop}) in accordance with the serverless platform design reference.
The literature review then extends to a compilation 
of heuristics including the existing OpenWhisk serverless load balancer and the 
analysis of two game theoretic approaches (sections \ref{design:heuristics}f).

The core contribution of this thesis is the adaptive event
dispatching solution presented in chapter \ref{noah}, that is based on the insights of the
thorough problem analysis. To evaluate the design, it has been found necessary to
also design, implement and verify discrete event simulation of a serverless system (chapter \ref{simulation}).
The simulator is applied to compare the new solution with the OpenWhisk heuristic load balancer and a 
game-theoretic noncooperative load balancer. Experiment design and simulation results are laid out and discussed in chapter \ref{evaluation}.
Chapter \ref{discussion} concludes with a critical assessment of the achievements, future directions and a summary of the contributions.


\section{Aims \& Objectives}
\label{intro:aims}

The aim of this thesis is to improve efficiency of serverless platforms 
by designing an adaptive event dispatching. Efficiency is 
the ratio of produced output over incurred cost and as such comprises 
several starting points for improvement.

The first objective is to exploit data locality as a means to reduce data
transfer, because data replication and context synchronisation can incur
costly overheads.
The second objective is to adapt the scheduling policy to varying demand. 
Controlling the amount of employed resources is required to curb cost.
The third objective is to find a suitable trade-off between response times and
resource allocation. With less resources allocated, the utilisation can be increased.
However, a higher utilisation may also cause degradation of the system response times.

\section{Motivation}
\label{intro:motivation}

Serverless event processing on demand is credited potential to supersede the traditional Cloud deployment model.
In place of traditional resource rental, the Cloud provider services the execution of 
user-provided code, which comprises demand dispatching and automated resource scaling.

A typical Cloud application deployment would be accompanied by event loops, load
balancers and orchestration tools that automate planning, allocation and
configuration of the deployment. Serverless radically reduces the deployment
model to the registration of user-provided function code. A serverless provider
only charges for the number of dispatched events and the resource time
covered by the event execution.

Data center efficency has been criticised for comatose and idling
servers~\cite{koomey:2015,koomey:2017}. Virtual infrastructure
providers can not repurpose resources blocked by allegedly comatose servers 
without jeopardising the \gls{SLA} that binds them to provide virtual
machine capacity that the customer is paying for.

The shift from a rental model to pay per use allows providers to increase
utilisation while it allows customers to fit cost of a service deployment to
actual demand.


\section{Problem Statement}

The customer entrusts the serverless platform with demand scheduling and
resource scaling, so the provider needs to solve the workload scheduling 
problem in lieu of cloud-native application orchestration, where application-specific load
balancers would usually have been engineered to find a suitable trade-off
between response time guarantees and proactive resource allocation. Serverless
is supposed to ease the orchestration and operation automation challenge as it
finds the application decomposed into a functional (i.e. event-, or data-driven)
set of non-blocking executions that are structurally similar tasks which can be
scaled independently. The challenge is to find a demand scheduling that fits all
applications' performance requirements.

The majority of existing serverless platforms scales reactively. Upon a request, 
they either find an idling instance of the function to dispatch the event to or they 
launch a new instance. Upon bursts, this scaling behaviour leads to large numbers of instances. 
Few platforms decouple the scaling from event dispatching completely and scale by secondary metrics, such as CPU load or queue lengths.
To scale in, idling user function instances are either evicted after an idle
timeout or displaced by instances of other functions.
Controversely, customers implement function pings to keep idling instances from
being timed out because initialising a new instance causes long setup times. A
better event dispatching is required that considers platform efficiency and
manages instantiation according to the function's demand.



\begin{figure}[H]
\centering
\includegraphics[width=\textwidth]{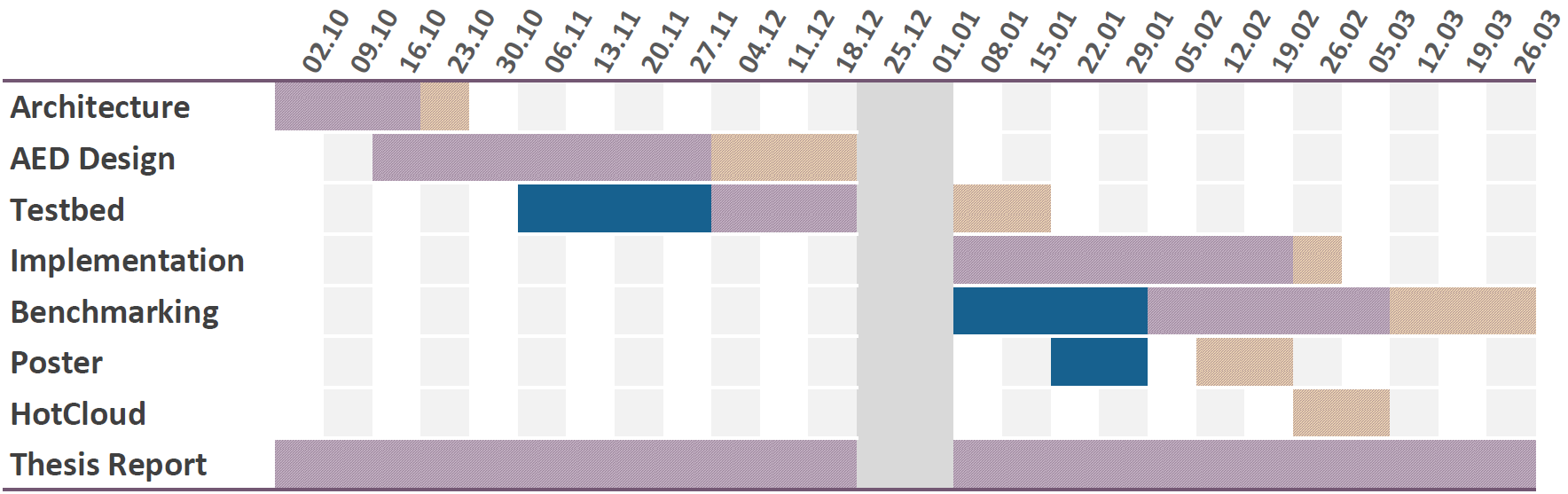}
\caption{Project timeline}\label{fig:timeline}
\end{figure}

\section{Project Management}

Figure~\ref{fig:timeline} shows the updated timeline of the project.
Changes in the architecture task required to synthesise a generic platform design and a reference scheduling model from publicly available material. 
The design, including literature review and analysis of methods, has been given more time to raise proficiency on the topic.
In turn, testbed work has been reduced to hands on experience and measurements with OpenWhisk as well as simulation tool evaluation with the result to design and verify a new serverless simulator.
The implementation task covered the design of the discrete event simlator and integration of compared approaches for evaluation.
The benchmarking task specifies the experiment, simulation runs and evaluation of the results, though implementation and benchmarking were tightly interwoven.
Quick progress on the practical work has provided initial results for the poster delivery.
The results had motivated the submission of a short paper to the Usenix HotCloud'18 workshop\cite{hotcloud18}, which wasn't planned initially.
During the entire project, the thesis report was updated as a background task, but required most attention towards the end of the project which concludes with the submission of this report.

In retrospective, the ample literature review has helped to build up a good understanding of the problem and it has raised confidence in the design of the solution.
The quick progress of the implementation can be attributed to the preparation.
A more comprehensive analysis would be desirable to evaluate the gap between the developed heuristic and optimal solutions.
Initially, the work should have accompanied an industrial project, but the setting was changed.
External dependencies pose the highest risks and need to be managed continuously. 
The choice of an emerging research topic requires constant monitoring of the field.
\chapter{Serverless Systems}
\label{systems}

This chapter synthesises a reference architecture for the design of event dispatching, i.e. a serverless scheduling framework. A serverless terminology is introduced and existing serverless platforms are reviewed for their design motivations and architectural decisions to eventually derive a common event dispatching reference design.

The report from the first International Workshop on Serverless Computing\cite{wosc} lists many open questions on the understanding of serverless. The novelty of the trend requires to outline a common terminology (section~\ref{systems:terminology}) used with established systems, as no common standard exists that defines a domain specific language for serverless to be used in this thesis\footnote{The CNCF has released a whitepaper \cite{cncf:serverlesswhitepaper} on 14/02/18 that may become a milestone for standardisation though biased to microservices}.
In the meantime, both commercial and academic platforms have tried to qualify as serverless solutions.
The realisations vary, as emerging \gls{FaaS} architectures are being designed for the ecosystem they integrate with and for the use cases they need to support.
Section~\ref{systems:operation} discusses the critical architecture design criteria of a serverless platform to provide the expected operational behaviour.
Section~\ref{systems:comparison} compares architectures of existing systems with respect to the important design criteria.
Section~\ref{systems:reference} synthesises a design reference and an event dispatching reference architecture.
The design reference delineates the boundaries and integration points of a serverless system within a Cloud ecosystem.
The event dispatching reference architecture is used in the remainder of this thesis to design and evaluate a serverless event scheduling approach.

\section{Serverless Terminology}
\label{systems:terminology}

In essence, a serverless platform provides execution of custom \textbf{functions}.
The user-provided code is registered with an event type.
The occurrence of an \textbf{event} and its context \textbf{parameters} trigger the execution of the function, also referred to as \textbf{invocation}.
The parameters may contain \textbf{data} (i.e. by value) or \textbf{references} to named data that are read and/or written during execution.
The function execution requires a \textbf{runtime}, e.g. an interpreter in case of a script language, a JVM for Java byte code or an OS architecture for executables of different formats.
Upon execution, the container-, library-, and runtime-\textbf{dependencies} of a function need to be presently loaded.
Each such loaded environment ready to process events is called a function \textbf{instance}.
Serverless \textbf{event dispatching} assigns compute resources to events, a decision that comprises the full lifecycle of an event from its occurence until the successful completion of the function execution.

Let aside the business roles, serverless is a technology that manages user-provided functions and their execution in a multi-tenant resource-sharing infrastructure alongside existing services. The environment is called an \textbf{ecosystem} and can contain various technologies. A serverless architecture design needs to adapt and consider reuse of existing infrastructure services, e.g. available resource provisioning technology to host the function instances, means of communication to exchange events and data stores to access shared execution contexts, among others.

In a Cloud ecosystem, infrastructure management provides capacitated resource slices using varying isolation technologies. Isolation ranges horizontally across compute (\gls{CPU}, \gls{GPU}, \gls{TPU}), storage (\gls{HDD}, \gls{SSD}, \gls{NVMe}) and networking (Ethernet, Infiniband) technologies and vertically from dedicated resource access (\gls{DPDK}, bare-metal) to temporal isolation (Hypervisor, \gls{SR-IOV}) to operating-system isolation (containers, processes, control groups, OpenVSwitch) to \gls{greenthreads} (\gls{JVM}, \gls{NodeJS}).
Capacitated compute resource shares may be commonly referred to as \textbf{workers} regardless the isolation technology, as they are pooled and scaled by the serverless platform. The worker terminology is common with many application platforms (MapReduce, Spark) and reasons well with the \emph{Twelve-Factor App}\cite{12factorapp} methodology suggested for web apps and \gls{SaaS}, which suggests pooling of worker processes. The mapping of instances to workers can have different realisations, e.g. runtime \emph{process instances} sharing a \emph{worker container} or runtime \emph{unikernel instances} mapping to a \emph{\gls{VM} worker} each.

The term \textbf{serverless infrastructure} is chosen to refer to serverless as a core Cloud provider technology. 
Serverless is sometimes interpreted solely as an operations-free application development cycle and used to pitch \gls{SaaS} and \gls{PaaS} services. 
Cloud ecosystems offer various platform services, such as object storage (Amazon S3, Google Cloud Storage, OpenStack Switft), key-value stores (Amazon Dynamo DB), messaging services (Amazon SQS, Google Firebase, OpenStack Zaqar/Monasca) and stream processing (Amazon Kinesis, Google Cloud Dataflow) among others. 
The technology of many is being developed as open source under the Apache Software Foundation (e.g. Beam, Flink, Kafka, Spark, Storm) and many platform technologies have made efforts to position themselves as serverless solutions by allowing custom user-provided code\footnote{e.g. Databricks offers serverless on Spark \url{go.databricks.com/announcing-databricks-serverless}, Google positions several products as serverless \url{cloud.google.com/serverless}}. In these cases, the design of serverless workload scheduling is predetermined by the technology's \emph{workload management}, e.g. to stream routing (Beam, Storm), batch job scheduling (Spark) or hash-based partition load balancing (Kafka). Although data processing services have pioneered custom user function deployment, it is difficult for them to be used outside their data context as generic event processing platforms.

When serverless is designed as a core infrastructure service, the event \textbf{source} can be any service in the ecosystem. To make function execution available externally, it is very common to provide a web technology front-end that terminates web sessions, such as a web \gls{API} \textbf{gateway}. In that case, event types (or function names) are embedded in the \gls{URL} target of a web request.
While web technolgy uses requests in a client/server fashion, \gls{IoT} applications use messaging protocols for communication.
The front-end technology may then comprise a message broker (e.g. Amazon Greengrass has functions subscribed to \gls{MQTT} topics).
Furthermore, worker pooling would commonly use the terms \emph{job} and \emph{task}, but serverless has established the term \textbf{event} to indicate the distributed and seemingly arbitrary occurrence of triggers to execute a user-provided function.

The architecture design may consider geographical distribution as data center installations become increasingly distributed. Several research strains, both industrial and academic, have proposed the wide-area distribution of cloud computing nodes.
From standardisation of multi-access edge computing\cite{etsi:mec} and fog computing\cite{nist:fog} to various new consortia (Edge Computing Consortium, Automotive Edge Computing) and public funding programmes (Cloud Computing in the recent Horizon 2020 ICT work programme\cite{H2020}), the term \textbf{edge computing} is being established to refer to compute nodes in metropolitan regions, residential areas or on customer premises that provide computing in access and local networks and requires distributed application platforms. Serverless resource transparency provides a desirable infrastructure abstraction level.

\section{Resource Allocation and Event Isolation}
\label{systems:operation}

Using the terminology, operational design questions of a serverless platform in resource-sharing environments can be discussed. To meet execution demand, a serverless platform requires a technology that provisions infrastructure resource allocations. To accomodate multiple users' functions, the platform needs to isolate event execution.

\noindent\textbf{Worker allocation.}
Resource allocations have a capacity (e.g. memory size, \gls{CPU} cores, network interface bandwidth) and service-level objectives concerning their availability. Such resource allocation is regardless of how the serverless platform is going to isolate events but rather how it integrates with the ecosystem. For example, the serverless platform could make an allocation for every event and release it afterwards or it may acquire a set of bare-metal servers and manage event processing across these. The ecosystem predetermines the available isolation technologies, which are typically \glspl{VM} or containers.

Recently introduced technologies such as Unikernel and Hypercontainers try to lower initialisation time of hypervisor resource isolation. For example, Unikernels link the application with kernel libraries to bundle the minimal required functionality (e.g. memory management, multithreading, networking or file system access) with an application for execution in a virtual machine. Modern hypervisors require little to no initialisation of the virtual hardware, so simple Unikernels can boot in \SI{4}{ms} \cite{unikernel,lightvm}. Recently introduced runtimes such as Kata Containers\cite{katacontainers} and hypercontainers\cite{hypercontainers} provide hypervisor isolation for containers. Although most current serverless platforms target container virtualisation, a recent publication tries to make a case to consider Unikernels for stronger security context isolation\cite{enddominance}.

\begin{figure}
\centering
\includegraphics[width=\textwidth]{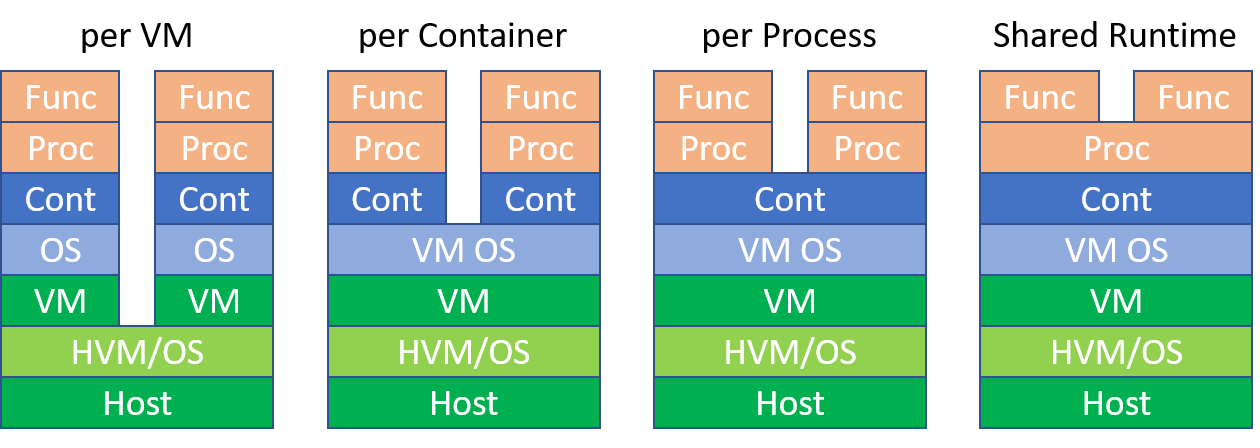}
\caption{Worker isolation levels}
\label{fig:isolation}
\end{figure}

Figure~\ref{fig:isolation} shows common mappings of function isolation. Colours indicate different technology layers, i.e. host virtualisation, OS context isolation (container) and POSIX process/memory isolation. 
To have a function \textbf{per \gls{VM}} would be the Unikernel or Kata Container approach, whereby only the latter stack uses a container layer to reuse container images. A popular model is to have a function \textbf{per container}, i.e. to isolate process and user IDs, file system, inter-process communication, networking and control groups.
Traditional complex event processing systems use a \textbf{shared runtime} (e.g. Java\texttrademark). 
Of course, higher layer mappings can do without hypervisor or container context isolation which is only shown for completion. Traditional \gls{IaaS} provides \glspl{VM}, \gls{CaaS} provides containers and \gls{PaaS} provide management of processes, threads, jobs or tasks at runtime level.

\noindent\textbf{Event isolation.}
Besides selecting a stack that provides the worker pool, event isolation needs to be architected.
Each serverless event execution is a stateless, independent action and is expected not to jeopardise any other event execution.
Hence, the worker is expected to isolate invocations, e.g. using OS process isolation or runtime means of context isolation.
Public serverless offerings also require accountable resource consumption metrics, so they often use a container worker per function and OS performance isolation (e.g. \gls{CPU} pinning without preemption or real-time scheduling) and isolate events temporally, i.e. only a single event runs in a container at a given time.
To save overhead, the container context would be reused for the next event (warm start).
For instance, OpenWhisk uses container isolation for concurrently running instances and also isolates event execution within an instance \cite{behindopenwhisk}. 
To isolate memory contents of sequential executions from one another, it starts a fresh runtime for every event. 
However, the default configuration only assigns \gls{CPU} weights for performance isolation.

\noindent\textbf{Resource management.}
The mode of isolation affects serverless resource management.
If the platform allocated a new worker for every event, the serverless platform would itself scale by the demand that it processes, but worker the initialisation overhead could render the operations inefficient.
When the platform allocates resources in bulk, it needs to tune event dispatching for utilisation of resources that it acquires and releases in a separate control loop.
The scaling of workers can also be delegated completely. For example, some infrastructures offer to size a pool based on their \gls{CPU} utilisation. Or, autoscaling services may collect serverless metrics (e.g. event queue lengths, demand prediction) as in traditional Cloud application orchestration to scale the number of workers.
Adaptive event dispatching developed in this thesis has to consider elastic scaling of the worker pool.

\section{Architecture Comparison}
\label{systems:comparison}

This section compares existing architectural design approaches. Several open-source serverless platforms are based on \gls{cloud-native} architectures. Their design is described and OpenFaaS\cite{openfaas} is used as one instance of a \gls{cloud-native} serverless platform.
Other platforms are designed for VMs or bare-metal hosts. OpenWhisk is used here as a popular representative of traditional \gls{VM} or bare-metal design. Amazon's Greengrass platform is mentioned for its use in IoT, which runs on capacitated appliances as an extension to the Amazon Lambda service.

\noindent\textbf{Cloud-native serverless platforms}.
The serverless platform can itself be designed as a \gls{cloud-native} container deployment (\gls{CNCF}).
A \gls{cloud-native} platform such as Kubernetes, Rancher or Apache Mesos provisions operating-system containers and offers rigid or flexible resource isolation.
\Gls{cloud-native} platforms also provide services to scale the number of container instances based on resource utilisation, either by monitoring the resource consumption or the service quality using best practises of Cloud application orchestration.
The \gls{CNCF} has recently released its own serverless whitepaper~\cite{cncf:serverlesswhitepaper}.
The integration with \gls{CaaS} platforms provides monitoring, resource allocation, automated scaling and \gls{LBaaS}.
The open source serverless projects OpenFaaS, Kubeless, Fission and Funktion are designed for \gls{cloud-native} computing platforms (\gls{CNCF}) and scale workers by the container.
\begin{figure}
\centering
\includegraphics[width=.5\textwidth]{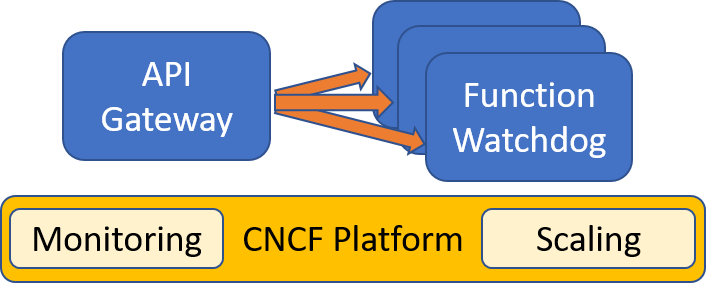}
\caption{OpenFaaS architecture}
\label{fig:openfaas}
\end{figure}
For instance, OpenFaaS adopts the microservice model and embeds each user-provided function in a service container image (see Figure~\ref{fig:openfaas}).
Each function is packaged with a \emph{function watchdog} - a tiny HTTP server implementation that offers execution of a particular function through a web service \gls{REST} \gls{API}. The watchdog launches a process instance for every request.
The OpenFaaS \emph{\gls{API} Gateway} reverse-proxy load balances events and uses resource monitoring to decide on scaling of the number of containers.
In general, integration with a \gls{cloud-native} platform can leverage existing container monitoring (Prometheus) or \gls{CPU}-utilisation based scaling (Kubernetes) but leaves the serverless platform with less control over the (co-)location of instances.

\begin{figure}
\centering
\includegraphics[width=.88\textwidth]{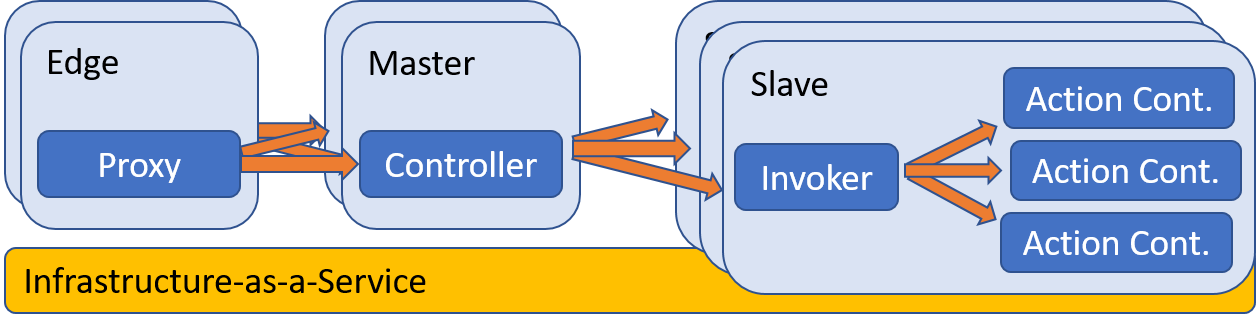}
\caption{OpenWhisk architecture}
\label{fig:openwhisk}
\end{figure}

\noindent\textbf{OpenWhisk} also runs containerised functions, but its design requires control over the local container runtime on each host. 
Unlike \gls{cloud-native} serverless designs, OpenWhisk balances events on a fixed (virtual) host resource allocation. 
The architecture (see Figure~\ref{fig:openwhisk}) uses a so called \emph{invoker} that manages a container pool on a local container runtime (e.g. Docker, containerd) of each slave \gls{VM} or bare-metal server. 
The invoker is configured to maintain a maximum number of containers, but does not actively manage the local memory and compute resources. 
When the invoker launches instances, it passes resource allocation parameters to the local runtime for performance isolation of events (\gls{CPU} share weights and memory limits). 
To deploy OpenWhisk, one would typically plan resource requirements and allocate hosts (\glspl{VM}) that are managed by the platform.
OpenWhisk uses an Nginx proxy that terminates client sessions and dispatches web requests to controllers. 
A controller stores the request in a shared data layer, looks up the references functions (called actions) for access control and dispatches the invocation event to an invoker. 
The commercial IBM Bluemix Cloud ecosystem is based on Apache OpenWhisk.


Other architectures exist that assume static resource configurations. 
For example, \textbf{Greengrass} is designed to run on appliances, such as an embedded platform (Raspberry Pi) or a small \gls{VM}. Similar to OpenWhisk, it is preconfigured to utilise a maximum amount of resources per appliance (32 containers, including management functions) and even uses its own container management (based on Linux kernel namespaces). 
Greengrass would scale by the appliance at the edge and is backed by the Amazon Lambda (serverless) service. 
Like OpenWhisk and its counterpart Amazon Lambda, it uses temporal isolation and reuse of containers. 
In case of congestion, idle containers are evicted to launch instances of requested functions.
\textbf{\citet*{McGrath}} design a serverless architecture on Microsoft Azure services, which uses worker \glspl{VM} and Azure Storage queues to implement workload polling (work stealing). 
The client-facing \gls{REST}ful front-end is implemented as a web service, but they do not discuss scaling of the worker pool.

\section{Serverless Design Reference}
\label{systems:reference}

\begin{figure}
\centering
\includegraphics[width=\textwidth]{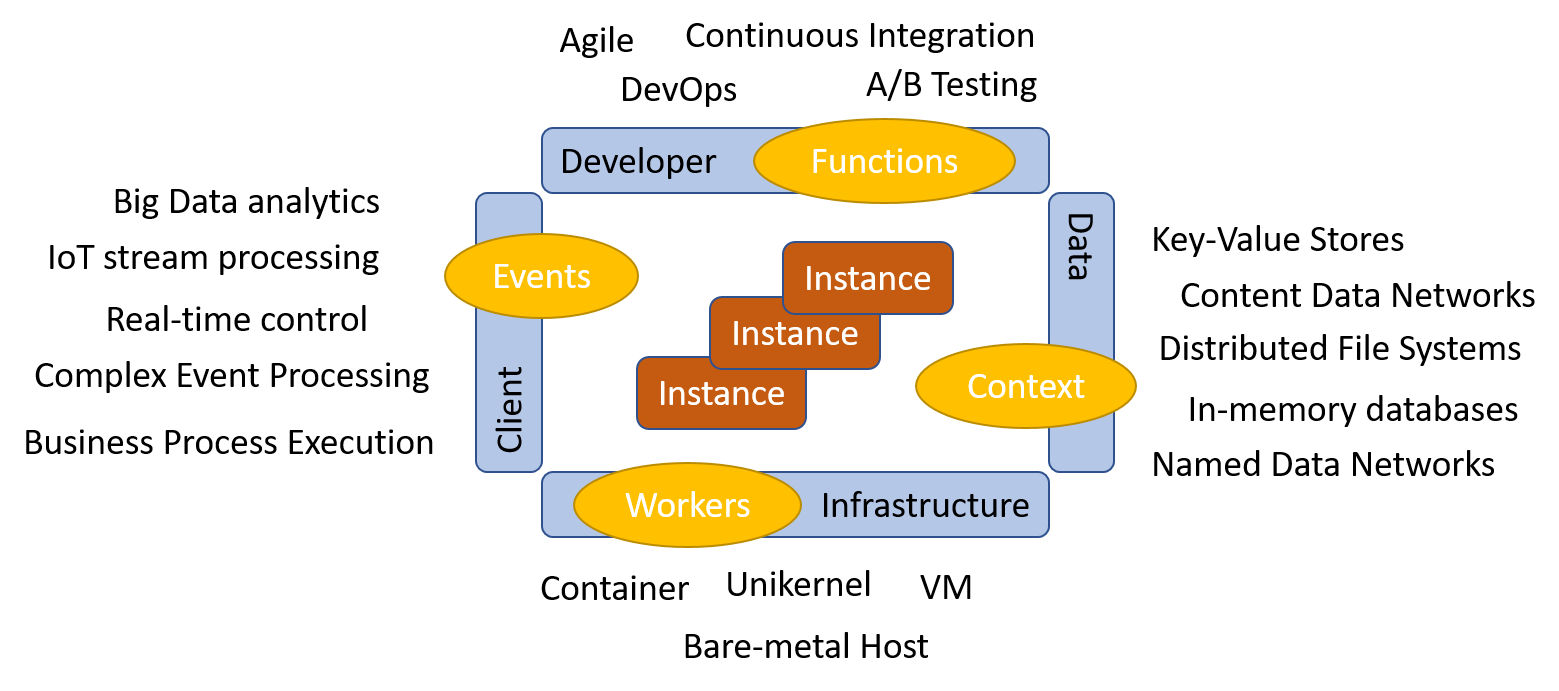}
\caption{Serverless design reference}
\label{fig:serverless_design}
\end{figure}

Given the many domains that serverless is discussed in, we can extract four boundaries to consider for the design of a serverless architecture as sketched in Figure~\ref{fig:serverless_design}.
Serverless is prominent in data analytics, IoT data processing, complex event processing and workflow execution to allow for customised, user-provided functions in these platforms. The \textbf{client access} technologies may vary across the set and comprise message-based technology (Kafka, \gls{AMQP}, \gls{MQTT}), request-based web technology (HTTP/\gls{REST}, webhooks) or stream protocols.
The selling point to speedup the release cycle (continuous integration) and to support agile software engineering methods with integrated testing (e.g. A/B testing) and automated operations (DevOps) is repetitive across domains. The \textbf{developer} merely provides application logic as packaged \emph{functions} and their configuration (event subscription, access control, etc.).
To externalise program state, event processing relies on \textbf{data} storage services to hold execution contexts. Both function input and output may be exchanged referring to named data items. For example, big data analytics use cases require the customised functions to access, process and store data (cmp. \gls{ETL}). This implies that functions can not be considered idempotent as the user-provided function may modify external data sources in the process.
Serverless architectures run in resource sharing environments and need to integrate with \textbf{infrastructure} provisioning services, which may be container-based, hypervisor-based (\gls{VM}/Unikernel) or bare-metal servers without virtualisation. The utilisation of acquired resources and the efficiency of the platform is important to consider.
Reviewed architectures are different implementations within this design reference.

\begin{figure}[H]
\centering
\includegraphics[width=.8\textwidth]{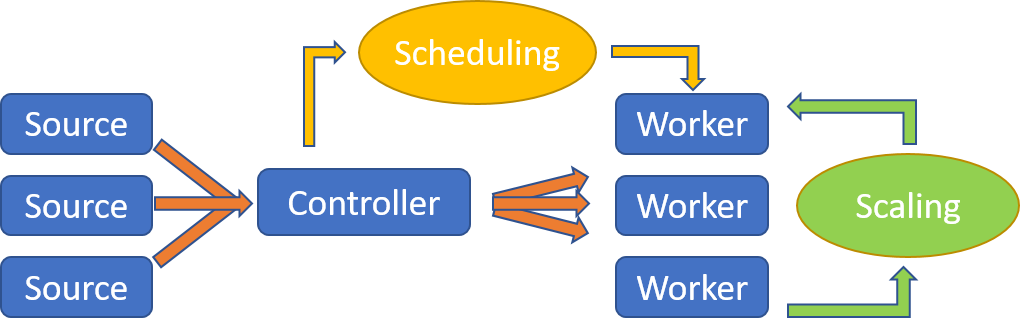}
\caption{Event dispatching reference architecture}
\label{fig:reference_architecture}
\end{figure}

\textbf{Event dispatching} designs across architectures share few common elements as depicted in Figure~\ref{fig:reference_architecture}. An important notion is that the scheduler appears as a single (distributed) function that dispatches events across workers and thereby affects the scaling behaviour of the system. OpenWhisk uses a distributed controller, OpenFaaS uses a non-distributed API gateway (microservice), and each Amazon Greengrass appliance implements an \gls{MQTT} broker that \gls{IoT} devices need to connect to.
Discovery of the controller is outside the scope of this thesis, as it depends largely on the client protocol and whether the controller is implemented in a distributed fashion.
Worker scaling in the ecosystem is often considered to be independent from event scheduling. Although the OpenFaaS API gateway controls the container instance count, its utilisation monitoring control loop is independent from event dispatching (\emph{load-scaling}). OpenWhisk does not actively scale the worker pool, and Greengrass is limited to the appliance. This work argues that event dispatching should consider adaptive scheduling for elastic resource scaling.
The worker is a capacitated resource (container, \gls{VM} or host) that controls the launch of function instances and processing of events. The worker pool designs vary from using worker queues for backpressure scheduling (e.g. OpenWhisk), per-function queues for work stealing (\cite{McGrath}) to processor sharing (OpenFaaS).

\chapter{Scheduling Literature Review}
\label{design}

As concluded in chapter~\ref{systems:terminology}, serverless event dispatching covers scheduling and resource scaling aspects. 
This chapter classifies and describes the serverless scheduling problem using two popular scheduling taxonomies for distributed scheduling and derives design goals and applicable methods.
Based on the design goals to improve efficiency, section~\ref{design:moop} formalises the multi-objective optimisation problem and discusses abstractions and optimality.
Section~\ref{design:heuristics} discusses heuristics, i.e. approaches known from web application load balancing and the OpenWhisk serverless scheduling heuristic.
Section~\ref{design:game} explores game theoretic approaches to load balancing (\cite{grosu:2005}) and job scheduling (\cite{duan:2014}).

\section{Scheduling Taxonomy}
\label{design:taxonomy}
The generic taxonomy by \citet*{taxonomy:scheduling} is applied with respect to the reference architecture and yields potential approaches to the serverless scheduling problem. 
To specify the problem in more detail, a recent taxonomy on job scheduling \citep*{taxonomy:jobscheduling} is applied that positions serverless scheduling quite uniquely among popular scheduling problems researched in from 2005-2015.


\citet*{taxonomy:scheduling} provide a taxonomy to classify scheduling in general-purpose distributed computing systems.
Their bibliography covers \emph{process} scheduling but the taxonomy abstracts from the operation and regards scheduling merely a mechanism \enquote{used to efficiently and effectively manage the access to and use of a resource by its various consumers}.
This makes the taxonomy a remarkably generic classification system.
As opposed to user processes, serverless function execution typically has a limited total execution time and rather resembles job/task scheduling. The taxonomy can be applied without loss to serverless event dispatching, because it doesn't detail the characterisation of workload. 


\citet*{taxonomy:jobscheduling} provide a novel taxonomy for job scheduling on distributed computing systems.
The work classifies the scheduling problem and scheduling solutions separately. They have applied the taxonomy on the top cited approaches between 2005 and 2015.
Their distinct classification system uses 17 features to distinguish the scheduling problem, but falls short on classifying the solutions as compared to \cite{taxonomy:scheduling}.
The following applies both taxonomies to classify the serverless scheduling problem and identifies the scheduling goals and solution space.

\subsection{The Serverless Scheduling Problem}
\label{design:taxonomy:problem}

\begin{figure}
\centering
\includegraphics[width=.8\textwidth]{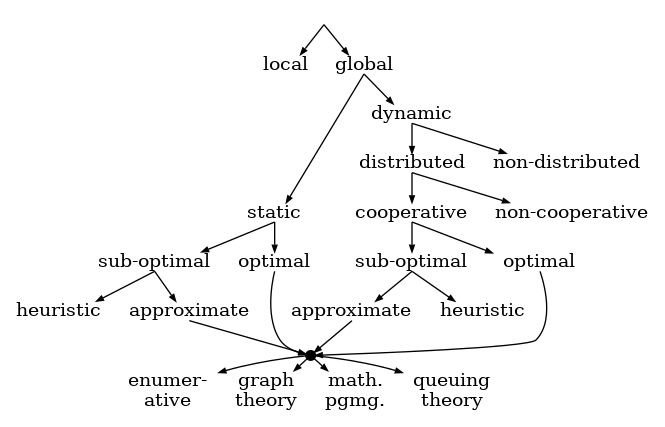}
\caption{Taxonomy of scheduling in distributed computing systems\cite{taxonomy:scheduling}}
\label{fig:taxonomy}
\end{figure}

Figure~\ref{fig:taxonomy} shows the taxonomy of \citet*{taxonomy:scheduling} which serves to broadly classify the field.
Regarding the distributed and sporadic occurence of events that need to be dispatched to suitable execution locations in a single domain, serverless event dispatching can be considered a \emph{global dynamic} scheduling system.

Event scheduling could be solved \emph{non-distributed}, \emph{cooperative distributed} or \emph{non-cooperatively distributed scheduling}. Current serverless platforms use centralised or distributed solutions (\ref{systems:comparison}), e.g. the OpenFaaS API gateway makes a central scheduling decision, the OpenWhisk controller is a distributed implementation.
Cloud platforms are economies of scale and varying demands, which requires demand scheduling to scale for multiple tenants. Hence, the controller logic should be realized as a replicable, \emph{distributed} implementation.

Contemporary work still uses differing terms to classify distribution of the scheduling task, e.g. the latest attempt to a grid taxonomy~\cite{taxonomy:autonomicmanagement} categorises application scheduling into \emph{centralised}, \emph{hierarchical} and \emph{decentralised} to classify scheduling responsibility, whereby hierarchical means a cooperative master-worker star topology.
Lopes and Menasc\'e \cite{taxonomy:jobscheduling} simply distinguish from \emph{centralised} and \emph{decentralised}, but many patterns exist to organise schedulers, such as peer-to-peer distributed hash tables (cmp. Chord) or consensus protocols, which are subsumed by \emph{decentralised}.
Casavant and Kuhl have a slightly more distinctive notion of decentralisation and discuss the difference of decentralised vs. distributed.
They argue, that any shared responsibility of scheduling decisions should primarily be called \emph{distributed} and only a system that also distributes authority should be called \emph{decentralised}.
Their differentiation of decentralisation is not based on communication topology but on \enquote{the authority to change past or make future decisions}.

The growing interest in edge computing with massively distributed data centers may require a \emph{decentralised} solution to manage workload of geographically distributed deployments. The service discovery and distribution of load across edge data centers is out of scope of this thesis, i.e. when an event enters the domain of the event dispatcher, its task is to schedule it to its known worker pool (cmp. reference architecture section~\ref{systems:reference}).

The question whether to choose a \emph{cooperative} or \emph{non-cooperative} approach concerns the scheduling goal. Section~\ref{design:taxonomy:goal} derives the serverless scheduling goal to be a multi-objective problem with a common system objective and individual objectives of serverless functions, which suggests a \emph{cooperative} solution. Casavant and Kuhl further classify approaches by the method used (\emph{heuristic} or \emph{approximate}/\emph{optimal} \emph{enumerative}, \emph{graph theoretic}, \emph{mathematical programming} or \emph{queuing theoretic}). Applicable methods are further discussed in section~\ref{design:taxonomy:methods}.


\citet*{taxonomy:jobscheduling} provide a classification system with features of workload $\mathcal{W}$, resources $\mathcal{R}$ and scheduling requirements $\mathcal{Q}$ to define the scheduling problem $\mathcal{SP = (W,R,Q)}$.
The following classifies the serverless scheduling problem accordingly. Table~\ref{tab:classification} summarises the classification and gives option numbers from \cite{taxonomy:jobscheduling} for comparison.

Serverless \emph{workload} is typically multi-user and the multitude of functions can be considered different jobs $\mathcal{W}_1$.
Every event is considered a single, independent task $\mathcal{W}_2$, although events may trigger sequences of events with sequential data dependence.
The task execution time may actually vary from to local resource availability ($\mathcal{W}_3$ moldable), although some commercial platforms model seemingly rigid requirements to keep a fair service level.
Events arrive in intervals ($\mathcal{W}_4$ open) and with different demand, but follow the same serverless programming model and have the same execution structure ($\mathcal{W}_5$ same model/same structure).
Existing serverless platforms have different approaches to the quality of service and resource isolation $\mathcal{W}_6$/$\mathcal{W}_7$.
As a service, SLAs are inevitably a part of the problem but the platform scheduling only operates on a fair best-effort basis, i.e. customers can neither negotiate SLA nor customise real-time requirements of their serverless functions.



\begin{table}
\small
\begin{tabular}{p{\dimexpr 0.3\linewidth-2\tabcolsep}|p{\dimexpr 0.7\linewidth-2\tabcolsep}}
Feature & Serverless classification \\
\hline
\hline
$\mathcal{W}_1$ Source & (3) multi-user/multi-job \\
\hline
$\mathcal{W}_2$ Job structure & (1) single-task \\
\hline
$\mathcal{W}_3$ Job flexibility & (2) moldable \\
\hline
$\mathcal{W}_4$ Job arrival & (1) open arrival process (events come in at any time) \\
\hline
$\mathcal{W}_5$ Workload composition & (3) same programming model, same structure but varying demand\footnote{varying job length possible, but within same order of magnitude} \\
\hline
$\mathcal{W}_6$ Quality of service & (1) best-effort \\
\hline
$\mathcal{W}_7$ Real-time & (1) no real time \\
\hline
\hline
$\mathcal{R}_1$ Heterogeneity & (2) heterogenous \\ 
\hline
$\mathcal{R}_2$ Scaling & (4) machine shutdown \\ 
\hline
$\mathcal{R}_3$ Sharing & (3) dedicated containers / (5) OS sharing \\
\hline
$\mathcal{R}_4$ Geogr. coverage & (1) local \\
\hline
$\mathcal{R}_5$ Federation & (2) single domain \\
\hline
\hline
$\mathcal{Q}_1$ Goal & (1) optimisation criteria \\
\hline
$\mathcal{Q}_2$ Level & (2) task-level \\ 
\hline
$\mathcal{Q}_3$ Data locality & (4) node affinity \\
\hline
$\mathcal{Q}_4$ Failure model & (2) failure-aware \\
\hline
$\mathcal{Q}_5$ Adaptability & (2) adaptable \\
\end{tabular}
\caption{Serverless scheduling problem classification}
\label{tab:classification}
\end{table}


The resources for serverless platforms are typically Cloud infrastructures, i.e. heterogenous $\mathcal{R}_1$ to some degree and scalable $\mathcal{R}_2$.
The sharing model $\mathcal{R}_3$ of event executions is typically a dedicated container when based on \gls{CaaS} or OS-sharing for architectures with process isolation. 
The taxonomy should better distinguish from dedicated (spatially isolated) shares, temporal resource-isolation or processor sharing to distinguish different levels of performance isolation. 
Strong isolation helps accounting, but leaves the execution inflexible. 
The technology need not be container virtualisation but may as well be OS sharing when local admission control is used to prevent overload. 
In general, the sharing of allocated worker resources should be considered here, that is typically the isolation-level associated with containers, i.e. a dynamic workload scheduling (e.g. weighted fair resource sharing).
Serverless resources can be considered local $\mathcal{R}_4$ as propagation delay is typically not considered. The worker pool under control of event dispatching forms a self-contained scheduling domain $\mathcal{R}_5$. Note, that the provided problem specification is based on reviewed architectures. The design of serverless in other resource environments (e.g. edge computing) may classify differently.

Lastly, the requirements $\mathcal{Q}$ to serverless event scheduling have multiple optimisation criteria $\mathcal{Q}_1$ (the taxonomy does not distinguish any).
The scheduling level for serverless events is that of single tasks only $\mathcal{Q}_2$, i.e. unless workflows need to be considered in scheduling.
As every execution requires at least the code (and runtime) to be present as well as stored or cached input data, serverless scheduling heavily depends on data locality.
Todays systems employ distributed in-memory databases and JIT-compiled code resides in memory, so \emph{node affinity} makes a difference $\mathcal{Q}_3$.
Failure-awareness $\mathcal{Q}_4$, i.e. the ability to reassign jobs in case of failure is state-of-the-art in elastic Cloud systems.
As demand varies over time for serverless functions, so does the need to split or merge workloads upon resource scaling which requires adaptable scheduling $\mathcal{Q}_5$.


Although the taxonomy is weak w.r.t. workload performance requirements and resource sharing concepts, this classification of the serverless scheduling problems helps orientation to classify this work within the large research body.
\citet*{taxonomy:jobscheduling} has identified 10 groups for the top 109 scheduling problems, but none of these classifies the same as the serverless scheduling problem.
The majority (two thirds, 63\%, 68 out of 109) considers multiple tasks or workflow scheduling, which is a generalisation that can be made to find comparable approaches.
All groups are reported to use some combinatorial search solution but only a few of those are said to consider the delay involved with finding a solution.
Almost all solutions generate rigid schedules and use resource isolation to enforce this, i.e. they do not consider the execution moldable as in OS-sharing environments (i.e. processor sharing), which can be attributed to the size of jobs/tasks usually being large compared to serverless functions.
What makes the serverless event scheduling unique is the combination of an open arrival process with multiple users per function, similarly structured tasks with moldable (dynamic workload) execution times in an OS-sharing model that require affinity scheduling (to reuse instances, runtime dependencies and context).
It does not exactly fit any of the identified groups.
\textit{Presumably, the serverless scheduling problem is either unexplored and/or unpopular}.
However, applicable concepts can be found looking for approaches with similar design goals.

\subsection{Design Goal}
\label{design:taxonomy:goal}

The main goal of this thesis is \enquote{to improve resource efficiency by considering data locality and dependencies} (cmp. interim report\cite{interim}).
Formally, efficiency is the ratio of output over input, i.e. the resource efficiency of a serverless system is the ratio of the serverless events processed (output) over the resources spent (input).
With a total workload of events to process \glssymbol{product} and cost of allocated workers \glssymbol{cost}, the serverless scheduling efficiency \glssymbol{efficiency} should be defined as the events completed over the resource time covered:

\begin{equation}
\label{eq:efficiency}
\mathcal{E} = \frac{\mathcal{P}}{\mathcal{C}}
\end{equation}

Due to the open arrival process of the workload (\glssymbol{product}), 
the efficiency can only be increased by adapting the resource time coverage to handle the workload, 
i.e. the time a resource is used and/or the number of resources concurrently employed.
Resource planning would prepare for the maximum of concurrently required resources over the entire time frame $T$. Ideally, resources would be acquired and let go on demand, such that the resource time coverage reduces 
to a variable resource cost $\int{\mathcal{C}(t)} \leq \lceil\mathcal{C}\rceil \times T$.
In shared infrastructures, this means returning resources to the pool for use by other tenants\footnote{In energy-aware cost models, returning resources may include stepping down dynamic voltage frequency scaling (DVFS) or switching off servers}.
While it may be quick to release resources, a setup time is required to bring resources up.

\textbf{The dichotomy.} \citet*{taxonomy:scheduling} note a \enquote{resource/consumer dichotomy in performance analysis}, i.e. performance from a resource point of view would be concerned with throughput whereas performance from the consumer point of view would be concerned with minimising the response time.
In fact, a common metric in data centers (and current commercial serverless platforms) is \emph{resource utilisation} as a measure of productivity.
It is the time a resource has been actively used (generating revenue) over the total time that it has been employed (cost).
While resource utilisation is a meaningful business measure for a resource provider, it neglects application performance.
E.g. a work-conserving scheduler may achieve full utilisation, but increasing resource utilisation does not necessarily increase efficiency.
Simply because rigor utilisation of free resource capacity can cause additional data transfers, preemption and context switching, cache eviction, etc. and have a negative impact on the average response times.
Application task scheduling often uses the \emph{response time} as a performance metric\cite{grosu:2005,grosu:2008,widjaja:2013}.
Increasing the number of allocated resources reduces the makespan and increases the system's throughput but at the expense of scaling the system up, which incurs both setup time overhead and more idle periods that negatively impact the efficiency of the system.

With the serverless deployment model, the provider is entrusted to solve for both, the resource utilisation and the response time objective.
The conflict constitutes a multi-objective problem that is formalised in section~\ref{design:moop} to find solutions in the tradeoff between minimising both makespan and cost with the trajectory that reducing cost at some point sacrifices response time and vice versa.
Efficiency unifies both objectives, i.e. maximisation of efficiency marks a Pareto-optimal solution of the multi-objective optimisation problem.

\subsection{Method Classification}
\label{design:taxonomy:methods}

\citet*{taxonomy:jobscheduling} do not classify scheduling solutions for their method, but merely classify the goal be a set of multiple optimisation criteria.
\citet*{taxonomy:scheduling} classify the approaches as either optimal, sub-optimal approximate or suboptimal heuristic solutions, whereby only optimal and approximate solutions are said to employ enumerative methods (tree search, branch and bound), graph theory (coloring), mathematical programming (e.g. linear/mixed integer programming) or queuing theory (stochastic processes).

Metaheuristics, i.e. guided random search techniques, such as simulated annealing, ant colony optimisation, evolutionary algorithms, particle swarm optimisation, tabu search or variable neighborhood search are not explicitly named but can be considered hybrids of the classes.
According to \citet*{zhan:2015}, evolutionary algorithms have become popular with traditional Cloud scheduling.
Their survey distinguishes resource scheduling for the application, virtualisation and infrastructure layers, but 
the survey's application scheduling algorithms have only been applied to offline task and workflow scheduling.

With serverless, the event scheduling problem inadvertedly becomes an online multi-objective optimisation problem that includes the global system welfare of minimising resource consumptions and individual application response-time welfare of the different classes of functions provided by the tenants.
According to a recent survey on modeling and optimising multi-objective systems by \citet*{moop:survey}, the majority of key techniques to multi-objective problems for systems with identical individual payoff functions are game-theoretic.
Games are broadly classified as cooperative (global welfare) and non-cooperative (individual payoffs).
\citet*{tamer:1982} write, that the theory of differential games has been developed in parallel to optimal control theory and that cooperative games without side payments can be reduced to optimal control problems by the design of a single objective whereas non-cooperative games, on the other hand, do not consider such coalition of players.

The existing open-source serverless event dispatchers reviewed in section~\ref{systems:comparison} implement only heuristics.
The review of the survey\cite{moop:survey} suggests the design of a game to address the multi-objective problem.
On the basis of the literature search, the design an adaptive event dispatching for serverless requires the formalisation of the multi-objective optimisation problem, a review of heuristics for online event scheduling that address the locality objectives and exploration of online game-theoretic approaches.

\section{Multi-Objective Problem Formalisation}
\label{design:moop}

%
By the definition of the serverless scheduling problem, the scheduling goal comprises two objectives, 
i.e. to reduce resource time coverage (reduce cost) as a global system objective and to minimise response times (increase productivity) as objectives of the individual functions. 
This section formalises the multi-objective optimisation problem w.r.t. the isolation levels discussed in section~\ref{systems:operation}.

\citet{moop:survey} survey modeling techniques for systems with
multi-objective optimisation goals and adopt the definition from
\citet*{Coello:2000} for a multi-objective optimisation problem as a vector
function of non-commensurable objective criteria:
\begin{align*}
\bar{f}_k(\bar{x}_n) &= [o_1(\bar{x}_n), o_2(\bar{x}_n), \ldots, o_k(\bar{x}_n)]^\intercal \\
s.t.\qquad g_i(\bar{x}_n) &\geq 0, \quad i = 1, 2, \dots, m \\
     h_i(\bar{x}_n) &= 0, \quad i = 1, 2, \ldots, p
\end{align*}
The \textbf{serverless scheduling problem} can be defined as an offline version of the online job shop scheduling problem in which events 
\glssymbol{e:set} of classes $k$ arrive at times \glssymbol{e:time}.
A scheduling solution $X$ needs to be found that assigns events.
As discussed in section~\ref{systems:operation}, a platform may launch a worker (e.g. \gls{VM}) per instance or allocate worker capacity in bulk.
Section~\ref{design:moop:instances} formalises the scheduling problem to assign events to instances and 
section~\ref{design:moop:worker} extends the problem to also comprise the mapping of instances onto workers, 
followed by a review of the sojourn time definition of an event in section~\ref{design:moop:sojourn}.

\subsection{Event Scheduling on Instances}
\label{design:moop:instances}

When the serverless platform allocates a worker per instance, event dispatching would only need to assign events to instances.
The cost of such a \textit{worker-instance} comprises its setup time, the time a function is executing and the idle times between functions.
In other words, the resource time covered by a worker-instance starts with the event that triggers its creation and lasts until the completion of its last assigned event (after which it is ideally shut down).

The applicability of this model depends on how workers are accounted for.
For example, Microsoft Azure containers are charged each for creation and total
runtime, but Google Kubernetes Engine charges for the VMs used to scale the
customer's rented container engine.
Amazon Elastic Container Service offers both charging models.
In private Cloud sectors, such as Telecommunication Cloud infrastructures
based on OpenStack, infrastructure management would block resource shares
exclusively for the entire allocation time, making it unavailable for other tasks.

Let \glssymbol{c:set} be the set of worker-instances of type $k$.
Let the variable \glssymbol{e:decision} decide whether the event \glssymbol{e:set} is processed by instance \glssymbol{c:set}.
Equation \eqref{eq:sojourn} shows the sojourn time \glssymbol{sojourn} of an event, that is comprised of the queuing time \glssymbol{queue}, the setup time \glssymbol{setup} and the execution time \glssymbol{exec}.
\begin{align} 
S(e^k_i) &= (W(e^k_i)+I(e^k_i)+B(e^k_i)) \label{eq:sojourn}
\end{align}

The waiting and setup time may actually occur to the same event. The setup time depends on whether there exists a prior event that has already caused creation of the instance. Queuing may cause the arriving event to wait, but 
setup of the event isolation context (e.g. process start) may still be required, even if it is lower than creating a new container.

The resource time of a worker-instance \glssymbol{c:set} for a scheduling
solution $X$ starts with the first event assigned to the instance and lasts
until the last event assigned to the instance finishes processing.
The total resource time objective function is given by equation
\eqref{eq:instance:cost}. The response time objective function is simply taken
as the makespan, i.e. the sum of all sojourn times in the serverless system as
shown in equation \eqref{eq:instance:responsetime}.
\begin{align}
\mathcal{C}_1 &= \sum_{k\in K}\sum_{c\in C^k} \max_{t^k_i\in T^k}{ \{ t^k_i+S(e^k_i) \mid x^k_{i,c}=1 \} }-\min_{t^k_i\in T^k}{ \{ t^k_i \mid x^k_{i,c}=1 \} }
\label{eq:instance:cost} \\
\mathcal{C}_2 &= \sum_{k\in K}\sum_{e^k_i\in E^k} S(e^k_i)
\label{eq:instance:responsetime}
\end{align}
\begin{subequations}
The resulting multi-objective optimisation problem~\eqref{eq:instance:problem}
tries to find a scheduling solution $X$ that minimises both objectives, \eqref{eq:instance:cost} and \eqref{eq:instance:responsetime}, 
under the constraints that 
each event is only scheduled once~\eqref{eq:instance:problem:scheduledonce} 
and that no event starts processing (finishes waiting) on an instance while a previous event is occupying it~\eqref{eq:instance:problem:concurrency}.
\begin{alignat}{3}
\label{eq:instance:problem}
&\!\min_{X}        &\qquad& [\mathcal{C}_1, \mathcal{C}_2]^\intercal & \\
\label{eq:instance:problem:scheduledonce}
&\text{s.t.} &      & \sum_{c\in C^k}{x^k_{i,c}} = 1           & \forall{k},\forall{i\in E^k}\\
\label{eq:instance:problem:concurrency}
&                  &      & x^k_{i,c}(t^k_i + S(e^k_i)) \leq x^k_{j,c}(t^k_j + W(e^k_j)), \; & \forall{k},\forall{c},\forall{i,j\in E^k},t^k_i\leq t^k_j
\end{alignat}
\end{subequations}

\subsection{Event Scheduling on Workers}
\label{design:moop:worker}


By allocating workers and managing instances separately, the platform can reap multiplex gains. 
Instances only consume \gls{CPU} time during setup and execution time, i.e. they can be paused so that bringing them up requires less time than starting a new instance. 
While paused, they do not consume CPU capacity but only block memory.
The worker's memory capacity limits the total number of instances. 
With sufficient memory, a worker can contain more instances than it concurrently executes, which helps reducing setup times. 
When load shifts strongly between function types, a worker may still need to evict instances.

For example, the OpenWhisk controller schedules events to invokers, i.e. workers, and requires the worker pool to be dimensioned independently from event scheduling. The invoker manages instances separately. Container instances are always paused at the end of an invocation to not consume any more CPU time. But, the current implementation uses the same limit for the number of pooled instances and the number of concurrently executing instances and oversubscribes the worker by a factor of two. Also, reconfiguration of an OpenWhisk worker pool is difficult with the currently implemented scheduling heuristic, because it uses hash-based load balancing as will be discussed in section~\ref{design:heuristics}.

Event scheduling on workers should consider to assign events to instances such that the platform can adaptively scale the worker pool
\footnote{The problem to smooth online VM allocation in distributed Cloud infrastructures to minimise reconfiguration cost has been studied for example by \citet*{jiao:2017} and reduced to the problem of deciding for how long a VM should be kept during a decline of demand.}.
The offline version of the online job shop scheduling problem considers assignment of all instances $C$ onto a set of workers $W$,
such that the number of concurrently running instances on a worker does not exceed its \gls{CPU} capacity (e.g. 16 cores) to preserve execution \glspl{SLA} and that the total number of pooled instances (both active and inactive instances) does not exceed its memory capacity.
%


Let \glssymbol{c:decision} decide if instance $c$ belongs to worker $w$, i.e. the set $C_w = \{c\mid \forall k \in K, \forall c \in C^k, y_{c,w} = 1\}$ denotes all instances of worker $w$.
Let \glssymbol{w:total} be the set of pooled instances in worker $w$ at time $t$, i.e. there exist events assigned to an instance $c$ such that the instance processes before and after $t$.
\begin{align}
\label{eq:worker:pooled} 
C_w(t) &= \{c|\forall{k\in K,c\in C^k}, \quad y_{c,w}=1 \wedge \exists{e^k_i,e^k_j\in E^k},\quad t^k_i \leq t \leq t^k_j+S(e^k_i)\}
\end{align}
%
%
Let \glssymbol{w:active} be the set of concurrently processing instances in worker $w$ at time $t$, i.e. those instances that have an event processing at time $t$ (after waiting and before departing).
%
\begin{align} \label{eq:worker:concurrent} 
N_w(t) &= \{c \mid \forall c \in C_w(t), \exists e^k_i \in E_k, x^k_{i,c}=1 \wedge t^k_i+W(e^k_i) \leq t < t^k_i+S(e^k_i)\}
\end{align}
%
%
A worker \glssymbol{w:set} incurs cost for its entire lifetime, i.e. from its creation to the completion of the last event.
Let \glssymbol{w:setup} denote the setup time of a worker.
Let \glssymbol{sojourn} be the sojourn time of an event analog to \eqref{eq:sojourn} that comprises the event queuing \glssymbol{queue}, setup \glssymbol{setup} and execution time \glssymbol{exec}.
Then the resource time cost is given by equation~\eqref{eq:worker:cost}, which differs from the cost of a worker-instance~\eqref{eq:instance:cost} by a separate worker creation time that is no longer part of the first event's setup time.
The response time is given by equation~\eqref{eq:worker:responsetime} analog to~\eqref{eq:instance:responsetime}. Note, that although 
the response time definitions are congruent, their sojourn time comprises different waiting and execution times as will be discussed in section~\ref{design:moop:sojourn}.
\begin{align}
\label{eq:worker:cost}
\mathcal{C}_1 &= \sum_{w\in W} \max{ \{ t \mid t \in T, |N_w(t)| > 0 \} }-\min{ \{ t \mid t \in T, |N_w(t)| > 0 \} } + t_W
\\
\label{eq:worker:responsetime}
\mathcal{C}_2 &= \sum_{k\in K}\sum_{e^k_i\in E^k} S(e^k_i)
\end{align}
%
\begin{subequations}
Let \glssymbol{w:maxtotal} be the maximum allowed number of instances on a worker (memory bound).
Let \glssymbol{w:maxactive} be the maximum allowed number of concurrently running instances on a worker (compute bound).
The provider objective is to solve the optimisation problem in \eqref{eq:worker:problem}, such that every event is scheduled exactly once \eqref{eq:worker:problem:scheduledonce}, that every instance is mapped to a worker exactly once \eqref{eq:worker:problem:mappedonce} and that the compute capacity limit of each worker is not exceeded \eqref{eq:worker:problem:concurrency} as well as its pool capacity limit \eqref{eq:worker:problem:pool}.
\begin{alignat}{3}
\label{eq:worker:problem} 
&\!\min_{X,Y}      &\quad& [\mathcal{C}_1, \mathcal{C}_2]^\intercal & 
\\
\label{eq:worker:problem:scheduledonce}
&\text{s.t.} &      & \sum_{c\in C^k}{x^k_{i,c}} = 1 \qquad \; & \forall{k},\forall{i\in E^k}
\\
\label{eq:worker:problem:mappedonce}
&                  &      & \sum_{w\in W}{y^k_{c,w}} = 1 \qquad \; & \forall{k},\forall{c\in C^k}
\\
\label{eq:worker:problem:concurrency}
&                  &      & |N_w(t)| \leq z_N & \forall{w\in W},\forall{t \in T}
\\
\label{eq:worker:problem:pool}
&                  &      & |C_w(t)| \leq z_C & \forall{w\in W},\forall{t \in T}.
\end{alignat}
\end{subequations}

\subsection{Sojourn Times}
\label{design:moop:sojourn}

The above problem formalisations only use sojourn time for event response time.
The \emph{queuing time} \glssymbol{queue}, \emph{setup time} \glssymbol{setup} and \emph{execution time} \glssymbol{exec} are not further specified in equation~\eqref{eq:sojourn}. They depend largely on the event isolation mechanism and can have complex dependencies discussed in this section.

In fact, the end-to-end response time of an event comprises several latencies, e.g. scheduling, transfer, queuing, setup and execution delays.
The \textbf{scheduling delay} depends on the complexity of the scheduling algorithm 
and information acquisition (full/partial system state), and is not considered in the formulation of the objectives.
Scheduling delay has a comparatively small variation and is inevitable, so it only shifts event arrival times $t^k_i$ by an almost constant delay $t_{\text{sched}}$ and is omitted in the optimisation.
The time between an event arriving at the controller and its arrival at an instance or worker is entirely neglected, as it is typically small compared to setup and/or execution times.

\textbf{Queuing delay} \glssymbol{queue}.
On-demand instance setup is an essential feature of serverless and a platform is expected to scale instances infinitely unless bound by quota.
No specific behaviour has been established for serverless when capacity limits are reached.
Presumed that queuing is desired, a platform may serve requests in FCFS order, use classful queuing or implement priority queuing.

OpenWhisk, for example, queues events while OpenFaaS allows for processor sharing.
Public platforms yet provide different throttling behaviours when the limit is exceeded.
Amazon Lambda has recently introduced a limit to concurrent executions per functions and Microsoft Azure, IBM Cloud Functions and Google Cloud Functions provide a quota of concurrent executions per user or namespace.
When the limit is reached, Amazon Lambda rejects requests unless they are submitted asynchronously through a queuing service, in which case expected queuing latency is not further specified.

\textbf{Setup time} \glssymbol{setup} of an instance depends on its type and 
worker state. When an event is dispatched to an instance, the context 
to host the instance may need initialisation (cold start), it may miss only 
the function code (pre-warmed) or have already run the function (warm).
The first cold start of its type may require the worker to load custom 
function code (and library dependencies), which may require network transfer 
or disk access, while subsequent instance creation benefits from cache contents.
Other techniques may provide even faster copies of an instance (e.g. forking) or its
container context. 
Not only instance creation accounts for setup delay. OpenWhisk, for example, pauses containers between invocations, so events have at least a small unpausing setup delay, e.g.
a set of Python actions were measured to have an average of \SI{8.6}{\milli\second} wake up time in the range [\SI{0}{\milli\second},\SI{10}{\milli\second}] and a set of arbitrary docker actions was found to require \SI{7.6}{\milli\second} on average to be resumed ([\SI{0}{\milli\second},\SI{50}{\milli\second}]).
Additionally, OpenWhisk launches a new runtime process for every event in the container instance.
The total setup time may vary significantly depending 
on the steps required as will be shown by experience with OpenWhisk in section~\ref{design:heuristics:openwhisk}.

\textbf{Execution time} \glssymbol{exec} starts with the handover of the event
message to the function instance. Depending on its content, named data are
accessed from the platform's data store, e.g. remote or host-local replica,
residing on disk or in memory.
Although the function runtime is typically memory-heavy compared to the data
transfers in question, data access and concurrent modification can cause
\emph{access latency} that varies based on the location. Content population of
cold caches incurs additional execution overhead.
\citet*{duan:2014} consider the execution either processing-bound or transfer-bound,
i.e. processing and data access are assumed to run concurrently so that either
one defines the bound of the execution time. Actual data access is difficult to
estimate from the event parameters. Typically, a queue-theoretic model would 
rather assume some generic distribution of the execution time.

This thesis makes an attempt to isolate an access latency penalty. Assume that
the best execution time can be achieved if all events of a class $k$ are
processed by the same worker, i.e. sharing local cache with no consensus or
replication required. Equal distribution across all workers incurs the
highest synchronisation effort. Assume that the fraction of events processed at
worker $w$ is $\phi_w$, i.e. $\sum_w{\phi_w} = 1$. A function that has roots for any
$\phi_w=1$ and that peaks on equal distribution is $\sum_{w}{\phi_w(1-\phi_w)}$.
The product of the fraction processed at a worker and the fractions not
processed at the worker becomes zero if one worker gets all events $\exists w\in W, \phi_w = 1$, 
which also implies all other workers get no fraction, so the sum of products respectively becomes zero. 
Let $\beta_k$ be a parameter to this function that
describes the degree of synchonisation between function executions of type $k$.
Then the function execution time \glssymbol{exec} can be defined as sum of the 
expected processing time $p(e_{k,i})$ (cmp. \cite{duan:2014} $p_{ki}$) and 
the synchronisation penalty.

\begin{align}
B(e_{k,i}) = p(e_{k,i}) + \beta_k\cdot\phi_{k,w}(1-\phi_{k,w}) \\
\text{where} \, \phi_{k,w} = \frac{1}{|E^k|}\sum_{i\in E^k}{x^k_{i,w}}
\end{align}

The discusison shows, that modeling of the sojourn time composition is strongly 
related to the operation of the platform. Analysis of the scheduling problem often 
abstracts from its details \cite{grosu:2005,grosu:2008,widjaja:2013,duan:2014}, but 
should consider major delay components.

\subsection{Optimality}

While a provider would want to handle the workload spending less resources, the application wants to reduce response times.
The \textbf{Pareto frontier} of the optimisation problem constitutes the border at which, for every achievable response time objective value, 
no better cost can be achieved and vice versa.

In the established \gls{FaaS} business model, the provider is reimbursed for execution time, 
i.e. the revenue is the sum of all execution times $\sum_{k}\sum_{e^k_i \in E^k}\sum_{i}{B(e_{k,i})}$ and carries the cost of overheads.
Efficiency \glssymbol{efficiency}, i.e. maximising the quotient of product \glssymbol{product} over cost \glssymbol{cost}, counterintuitively increases when the function execution time increases.
From a customer viewpoint, the platform would not be considered more efficient if it requires more time to complete events, because that incurs more cost to the customer.
Luckily in the business model, \glspl{SLA} on the compute capacity of an instance prevent the provider from increasing the service time by oversubscribing workers, but the location penalty to the execution time for synchronisation and data accesses remains.

Serverless platforms in general need to scalarise the non-commensurable objectives of response time and resource time (cost). 
A potential solution is to consider energy efficiency, i.e. finding the minimum resource time which yields a certain response time or to set a constraint on the response time and try to minimise resource cost.
As serverless scheduling actually has an open arrival process, the problem requires online algorithms, such as heuristics, polynomial time approximation schemes, optimal control theoretic solutions or game theoretic approaches. 
Note, that the response time objective function can be described as identical, individual payoff functions for each function class $k$ of events, while the platform's cost objective is a global payoff function.
The survey by \citet*{moop:survey} identifies that the majority of techniques used for this class are game-theoretic approaches.
The following sections explore existing heuristics and game theoretic approaches that may be adopted in the design of a scheduling solution.

\section{Heuristics}
\label{design:heuristics}

Scheduling heuristics are designed with intuitive judgment and common sense, but does not guarantee optimal solutions.
It is a strategy based on experience with similar problems but is not free from cognitive bias.
However, basing the decision finding process on information at hand may provide satisfactory solutions in short time, 
hence heuristics can be superior to complex algorithms that need to gather a lot of information. 
For example, \gls{LRU} ordering of items is a sufficient heuristic in many cases to approximate least frequently used ordering (LFU) \cite{checconi:livemigration}.
Existing open-source serverless platforms (OpenWhisk, OpenFaaS, etc.) make use of web load balancing heuristics to design event dispatching.

\subsection{Event Classification and Aggregation}

Many heuristics exist to statistically load balance workload, e.g. round-robin or
random choice \enquote{with the philosophy that being fair to the hardware
resources of the system is good for the users of that
system}\cite{taxonomy:scheduling}, while they're actually avoiding probability 
of congestion during traffic bursts. In fact, many specific methods 
exist that try to reaggregate similar requests for context localisation, e.g. 
using hashing or session persistence. The reason for aggregation is to cluster 
the workload by its data dependencies to reuse cache contents, also known as 
locality-awareness.

To become locality-aware, the scheduler needs to classify the event, i.e.
dependencies need to be guessed.
Obviously, the serverless \emph{function} name identifies a tree of code
dependencies (function code, libraries, runtime, etc.) and the majority of
dependencies should be known from the time the user registers the function.
A \emph{session} subsumes actions (e.g. requests/transactions) of a common scope
or agenda.
The notion is commonly used in web browsing to bundle requests for a specific
intent under a session context, e.g. web applications establish session
identifiers used in messages to identify context stored at either side of the
session.
Eventually, a serverless event contains function \emph{parameters} - either by pass-by-value or pass-by-reference.
References carried in the message may be assumed to cause the data to be accessed, e.g. content URLs or database keys refer to stored data - either within or outside the boundaries of the platform.
In the following, these contextualising heuristics (function, session, data) are discussed and existing methods reviewed.

\textbf{Function aggregation.}
A serverless function's code and its runtime dependencies is a memory-heavy
component of the execution as the runtime (e.g. JIT compiler) often requires a
significant time to load code.
All existing serverless platforms try to cluster events of the same function
type on the same workers to reuse initialised function code.
OpenFaaS uses worker-instances that can be addressed each. The gateway resolves
a function's instances via DNS and simply chooses randomly from the set.
OpenWhisk, addresses each each worker (invoker) and simply hashes the function
type to map to a preferred worker in the pool - a simple yet effective heuristic to
localise functions execution.

\noindent\textbf{Session Context.} Web frameworks bundle requests from the same
client onto the same server to exploit caching and to avoid context transfers.
Application load balancers (reverse proxies) allow to hash arbitrary request
elements, track active connections, measure server response time, etc. and
provide various sophisticated static load balancing heuristics that can be
tailored to the application.
At IP level, ECMP~\cite{RFC2992} routing is used to have multiple hosts
share an IP address, whereby the router hashes flow identifiers (IP 5-tuple) to
load balance flows onto hosts persistently.
Web application load balancers use both techniques to aggregate requests of the
same session context.
There is currently no known serverless scheduler that honours sessions, maybe
because events are considered independent and stateless.
A platform could exploit event parameters to derive session context.




\noindent\textbf{Content caching.}
Content dissemination networks (CDN) direct requests for data items to the same
cache nodes to avoid cache reconfiguration, i.e. to improve efficiency of
content access.
For example, consistent hashing~\cite{Karger:1997} of the content identifier
ensures the load is statistically balanced across the hash space and preserves
mappings during scaling. The hash space is partitioned to bins and every bin 
assigned to a cache node. When scaling the number of nodes, only a few bins 
and the data items therein need to be reassigned.
Currently, no serverless platform is known to route events by content
identifiers in the event parameters.

\noindent\textbf{Conclusion.} Despite the use of round-robin or
pseudo-randomised server selection to balance request dispatching, many
approaches exist to classify and aggregate events to establish affinity or
localisation.
While hashing is susceptible to scaling of the server pool, consistent hashing
tries to minimise this effect.
Still, approaches are designed for a large variety of content identifiers.
Application data seem to versatile to be classified without support of the
developer.
In order not to break the resource transparency that serverless offers, a method
would be preferable that generically classifies events, e.g. using k-clustering
auto-classification methods, but this is out of the scope of this thesis.
As long as the runtime dependencies contribute the largest overhead, the
function type qualifies as the primary event classifier.

\subsection{Apache OpenWhisk Controller Scheduling}
\label{design:heuristics:openwhisk}

Apache OpenWhisk load balancing classifies as a global, distributed, cooperative,
suboptimal heuristic by the taxonomy\cite{taxonomy:scheduling}.
Multiple controllers share the responsibility of assigning
events to hosts (invokers). The OpenWhisk distributed controller
implementation \cite{openwhisk:loadbalancer} shares common host state
information, including the number of concurrently active invocations on each
host.\footnote{From 14/02/2018, an alternative load balancer implementation is
available that uses sharding to segment worker capacity and randomly assigns requests.}
A function name's hash $h$ is used to identify
a preferred host location in the worker pool. If the preferred worker is
saturated (the number of activations exceeds a threshold $\alpha$), the
controller pseudo-randomly iterates through to the next worker in the pool.
The hash $h$ is used again to select a \emph{step size} which is a generator to
the cluster size to ensure that eventually all hosts are considered in
pseudo-random order. It places the event on the first host that has a load level
below a given threshold (first-fit) and increases the threshold if necessary to find a suitable host, as the following
pseudo-code illustrates.


\begin{lstlisting}
select_host(h=$\textrm{hash}$, I=[]$\textrm{sites}$) {
  lvl=$\alpha$=16, G=[]
  for(i$= 0,\ldots,|I|$) { $\textrm{// create generators}$
    if($\forall g \in \text{G}, gcd(i,g) == 1$)
      G=G $\cup$ i
  }
  g=G[$\text{h} \pmod{|\text{G}|}$] $\textrm{// select generator}$
  while(lvl $\leq 3\alpha$){ $\textrm{// try up to 3 times busy threshold}$
    for(k$= 0,\ldots,|I|$) {
      x = I[$\text{h} + \text{k}*\text{g} \pmod{|\text{I}|}$]
      if($\textrm{events at}(\text{x})$ < lvl)
        return x
    }
    lvl += $\alpha$
  }
  return $\textrm{random site}$
}
\end{lstlisting}

If no host is below the final busy level threshold ($3\alpha$), a random site would be picked.
This compelling first-fit, hash-based load balancing heuristic solves
statistically the locality, distribution and overflow of function allocations
across hosts.

\begin{table}[H]
\centering
\begin{tabular}{|l|c|c|}
\hline
\textbf{Operation} & \textbf{Python action} & \textbf{Docker action} \\
\hline
container creation & [\SI{448}{\milli\second},\SI{1370}{\milli\second}] & [\SI{504}{\milli\second},\SI{3208}{\milli\second}] \\
\hline
function init & [\SI{307}{\milli\second},\SI{612}{\milli\second}] & [\SI{206}{\milli\second},\SI{308}{\milli\second}] \\
\hline
cold cache init  & $\approx$\SI{1475}{\milli\second} & - \\
\hline
container resume & [\SI{0}{\milli\second},\SI{10}{\milli\second}] & [\SI{0}{\milli\second},\SI{52}{\milli\second}] \\
\hline
warm start & [\SI{1}{\milli\second},\SI{88}{\milli\second}] & [\SI{1}{\milli\second},\SI{103}{\milli\second}] \\
\hline
\end{tabular}
\caption{OpenWhisk operation steps time ranges}\label{tab:openwhisk_timing}
\end{table}

Table~\ref{tab:openwhisk_timing} shows experience values for Python and Docker actions measured on an OpenWhisk installation.
Timings are collected from the log files of invocations.
OpenWhisk uses different containers for every runtime (e.g. Java, Python, NodeJS, etc).
A Python action uses a unique container image, the creation of a container can take more than a second.
After creation, the invoker loads the Python user function code from a database into the container. While this takes typically takes less than a second, 
it may take another \SI{1.5}{\second} the first time it pulls the code from the database.
Once the runtime loads a function code, the container can only be used for this function or it can be torn down. 
After the invocation, OpenWhisk pauses the container processes and keeps the initialised container in memory for a fixed timeout (default 5 min) in case it can be reused.
To unpause a container is usually fast but may also take \SI{10}{\milli\second}.
Timings are different for custom container images. A user may provide a Docker container image. Upon container creation, OpenWhisk always checks the container image 
repository for the latest version, which prolongues the creation to up to 3 seconds. It runs the container init function that is provided by the user for sanity checks.
Unlike with Python source code, the worker does not cache any data for the initialisation of a custom container image.
Resuming such a custom image shows can take longer in some cases, but this may also depend on the load of the worker, i.e. how fast the OS can resume the custom user processes inside the container instance.
In addition to these setup times, every event invocation launches a new runtime process in the container instance that needs to load the function's code (or precompiled byte code). The total time seen to start 
event processing in case of an existing container instance ranges from \SI{1}{\milli\second} to \SI{88}{\milli\second} for Python and \SI{103}{\milli\second} for custom docker instances, 
but the average of this series of invocations is about \SI{10}{\milli\second}.

\textbf{Summary.} By hashing the function name to identify a \emph{home invoker}, OpenWhisk aggregates by the function for instance reuse and cache hit rate improvements. 
The first fit heuristic with a function-specific step size for scaling also increases the likelihood to reuse warm containers when the scheduler needs to scale out.
Hands-on experience with an OpenWhisk installation shows the large extent of setup time variation with the isolation technologies used.
As indicated in the serverless systems overview (chapter~\ref{systems}), isolation technologies with shorter setup times are making their debut.
But for now, setup delays of instances need to be factored in the design of serverless event dispatching.

\subsection{Online Bin Packing}
\label{design:heuristics:binpacking}

The serverless scheduling problem can be regarded an online bin packing
problem that assigns each event to a worker (bin) with sufficient residual compute capacity.
According to \citet*{Seiden:2002}, the simple heuristics first fit, best fit and
next fit were originally researched by Ullman\cite{ullman:1971} and Johnson\cite{Johnson:1974}.
To date, these heuristics serve for comparison in online scheduling, 
because of their low complexity to make a scheduling decision.

Each worker has a computing capacity \gls{w:maxactive} of concurrently running instances.
The online bin packing problem is to assign an event with capacity $1$ to a
worker $w$, such that the number of concurrently running tasks does not exceed
the worker's capacity (cmp. constraint \eqref{eq:worker:problem:concurrency}).

Assuming the scheduler is notified of the finishing of a message, 
it can implement the \emph{first fit} heuristic such that it progresses the list of workers
from a single starting point and maps the event to the first worker that has sufficient residual capacity.
The \emph{next fit} heuristic can be implemented analogously, but would use the last worker that it has scheduled an event to as starting point.
The \emph{best fit} heuristic would assign the event to the worker with the least residual capacity.

These heuristics have been originally researched in the context of memory page
allocation and did not consider overbooking of resources.
To facilitate queuing, by the example of OpenWhisk, the capacity limit
\glssymbol{w:maxactive} needs to be raised by an integer factor until a worker can been
found, whose sum of active instances and queued events is less than the new limit.
Let $L_w(t)$ be the length of the queue and \glssymbol{w:active} be the number
of active instances at worker $w$. To schedule an event, a minimial factor $a
\in \mathcal{N}^+$ needs to be found such that $\exists w \in W, L_w(t)+N_w(t) <
a*z_N$. Note, that if $a\in \mathcal{R}$, the heuristics would all find the
earliest start time.

Expectations on performance of these simple heuristics is low, as they do not consider setup time or execution time estimation and do not 
distinguish classes of events. 
Without aggregation, the first-fit and best-fit heuristic could cause many instance evictions when event classes vary often.
Next fit has the same problem but would move on when arrivals surpass the capacity of a worker. 
After completing one round, it faces the same risk of evicting instances unnecessarily.
However, these heuristics provide fast decisions, 
which is crucial to the scheduling delay that an event experiences as part of its response time. 
If coupled with aggregation policies, they have great potential to make an efficient scheduling heuristic.

\section{Game-Theoretic Methods}
\label{design:game}



Game theory provides means to solve online optimisation problems by designing an
iterative algorithm for the game players to converge towards control-optimal solutions. This
section reviews two such game designs. \citet*{duan:2014} design a cooperative
sequential game for task scheduling (\gls{BoT} workflows) and
\citet*{grosu:2005} design a noncooperative online request load balancing game.
By brief analysis, only the latter is found applicable to serverless by mapping customer
functions to players that compete for minimum response times in serverless event
scheduling. 

\subsection{Multi-Objective Game-Theoretic Workflow Scheduling}
\label{design:game:duan}

\citet*{duan:2014} design a global, static, optimal, mathematical programming solution to task scheduling (cooperative game).
The sequential, cooperative game tries to reduce both response time and cost.
Supposedly, the response time can be lowered by employing high performance resources, which tend to also be more expensive.

They design a game of $K$ (\gls{BoT}) players and $M$ sites.
The game finds a task distribution (placements) $\left(\delta_{ki}\right)_{K\times M}$.
For every distribution $\Delta$, the players derive a new allocation strategy $\Theta^{S(l)}(\Delta^{S(l-1)})$.
They calculate potential allocations $\left(\theta_{ki}\right)_{K\times M}$ 
using different weights for processing ($pw_{ki}$), cost ($cw_{ki}$), bandwidth ($bw_{ki}$) and storage ($sw_{ki}$).
Objective functions for every such resource type ($f,c,h,g$ respectively) 
 are used to decide on the allocation strategy with the best gain, which is in turn used to derive a new task distribution $\Delta^l(\Theta^l)$.
\textit{Formulae of allocation weights, objectives and task distribution are left out for brevity and can be found in \cite{duan:2014}.}

\textbf{Discussion.}
The authors evaluate their game multi-objective (GMO) approach and report
significantly shorter scheduling times than various task selection strategies
that place the selected task on the machine with earliest completion time.
Having implemented and tested the algorithm, there are a few things to note.
The algorithm uses linear relaxation and often arrives at a partial solution indicating
only a preference where to place tasks, so it needs to be ensured that all tasks
are placed eventually when it doesn't converge to a definite solution.
The weights are calculated using the processing, bandwidth and storage
requirements of the application and the site cost, namely, the expected time of
a task $k$ to compute on site $i$ ($p_{ki}$), the cost from the use of a site
(with cost $\varphi_i$), the network utilisation from the data a task needs to
transfer to site $i$ ($d_{ki}$) and the storage utilisation at site $i$ from its
storage requirements ($sr_{k}$). When the weights do not differ across tasks and
sites, the allocation strategies provide no sufficient gain and the algorithm stops early
with the initial distribution $\Delta^0$.
Cost ($\varphi_i$) naturally differs only across sites and the storage
requirement of a \gls{BoT} ($sr_{k}$) differs only across \glspl{BoT}. A variable that
differs for both \glspl{BoT} and sites is necessary to play this game, so that the
expected time to compute ($p_{ki}$) and/or data ($d_{ki}$) required to transfer
differs enough for the allocation strategy to provide a performance gain,
otherwise the solution stays with the initial distribution.

\begin{figure}
\centering
\begin{subfigure}{.5\textwidth}
  \includegraphics[width=\linewidth]{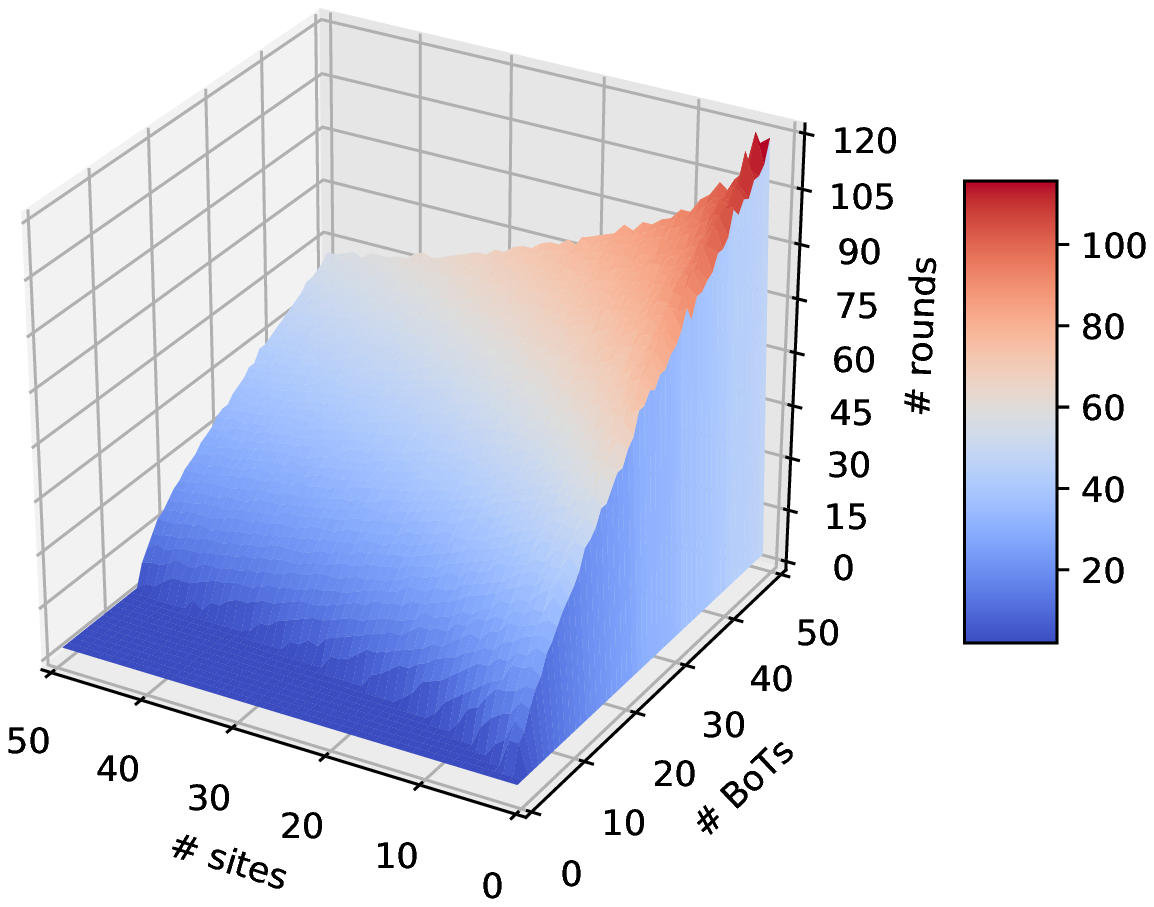}
  \caption{maximum number of rounds}
  \label{fig:duan:rounds}
\end{subfigure}%
\begin{subfigure}{.5\textwidth}
  \includegraphics[width=\linewidth]{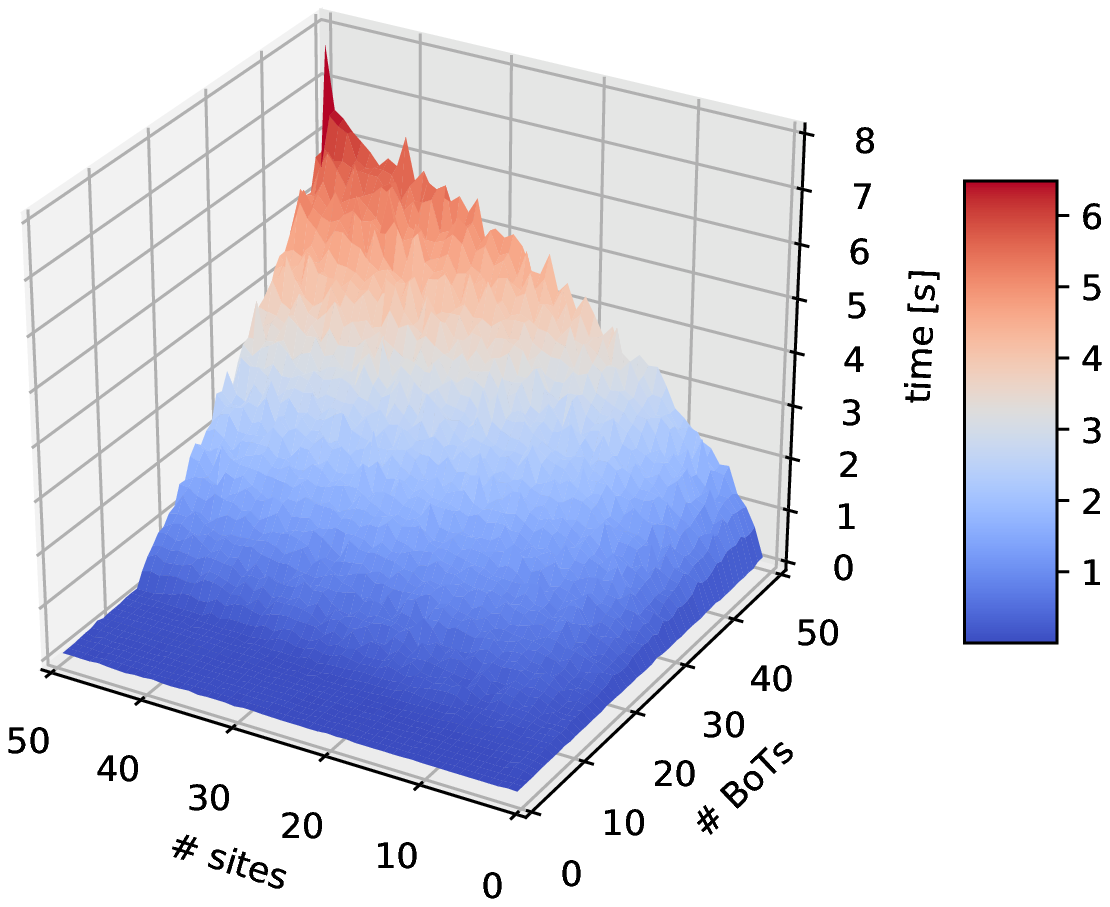}
  \caption{maximum time to compute}
  \label{fig:duan:times}
\end{subfigure}
\caption{Game-theoretic task scheduling evaluation}
\label{fig:duan}
\end{figure}

\textbf{Evaluation.} The algorithm has been implemented, verified to produce the example result in \cite{duan:2014} and evaluated generating random games.
In this evaluation, half of the machines have 8 and 16 cores each with proportional cost.
Bags (e.g. serverless functions) of 8 tasks have an expected computation time of \SI{200}{\milli\second} 
plus a random location-dependent delay in the range [\SI{0}{\milli\second},\SI{50}{\milli\second}].
With a threshold of $\epsilon = 10^{-3}$, 1k games were generated per configuration of 
number of \glspl{BoT} and number of sites.
\citet*{duan:2014} report fast computation of $10^5$ tasks (actually only 100 \glspl{BoT} with $10^3$ tasks each) on $10^3$ processors (actually only 10 sites with 100 cores each) that requires only 476 iterations.
The designed, randomised games require even less rounds, event in the worst case ($\approx 100$) and with more sites, there is less competition for resources and less rounds are required.
However, the time to compute can be much higher as the complexity of the algorithm is $\mathcal{O}(l \cdot K \cdot M)$.
Figure~\ref{fig:duan} shows the results of different game configurations.
The worst case out of a thousand generated games of each configuration is plotted in Figure~\ref{fig:duan:rounds}.
The time to compute the rounds for small problems is similar to the numbers reported by Duan et al.
However, the times grow with the number of sites $M$ and the number of \glspl{BoT} $K$ as depicted in Figure~\ref{fig:duan:times} to
almost 6 seconds to schedule 50 \glspl{BoT} on 40 sites with only 60 rounds. 
The average of the games played in this configuration lies at $\approx$\SI{4.5}{\second}.

\citet*{duan:2014} mention their algorithm could be adapted for flow scheduling, but
initial tests raise doubts on the applicability to online event scheduling. 
The multiple objectives cover those of the serverless 
optimisation problem, i.e. both request completion time and cost of employing a resource are considered.
However, the use of the expected completion time matrix requires independent tasks. In \gls{BoT} scheduling, 
each task is considered to require an individual amount of processor time and data transfer to complete.
Serverless event aggregation exploits reuse of context.
This would require the scheduler to re-estimate processing time in every placement strategy, 
because the processing time of events depends on whether they run on the same instance, 
the setup time of instances depend on whether they are mapped to the same worker.

\subsection{Noncooperative Load Balancing in Distributed Systems}
\label{design:game:noncoop}

\citet*{grosu:2005} have designed a non-cooperative game
approach to balance multi-user flow allocations for optimal response time in a
distributed server environment. Each player (e.g. load balancer) collects
information on the service time and allocations at every host to calculate an
optimal split of its own perceived arrival rate in response to the other
players' allocations. The noncooperative approach yields Nash Bargaining Solutions for a set
of heterogenous M/M/1-type servers (a cooperative version is
described in \cite{grosu:2008} that creates Pareto-optimal solutions).

Every player $j=1,\ldots,m$ has a request stream with arrival rate $\Phi$ and
seeks to balance fractions $s_{ji}$ to sites $i=1,\ldots,n$.
In each round, a player $j$ uses a machine's service rate ($\mu_i$) and the
other ($k$) players' allocations $\sum^m_{k=1,k\neq j}\Phi_k$ to calculate
residual capacity $\mu^j_i$ to optimally rebalance its own arrival rate
$\Phi_j$. The players rebalance their arrival rates in turns until the 
overall change in (estimated) response times is minimised ($\epsilon$-constrained).
The result constitutes a Nash Bargaining Solution in which neither player 
can change an allocation without worsening response times.

\noindent\textbf{Evaluation.} The single-objective game has a low complexity.
The algorithm used to calculate an optimal response (BEST-REPLY) needs to sort
the residual capacities of $n$ sites and hence has a complexity of
$\mathcal{O}(n\log n)$.
The total game requires $m$ players to play $l$ rounds, which results in a total
complexity of $\mathcal{O}(l\cdot m\cdot n\log n)$.
The single-objective function and strategy calculation has less complexity than
the multi-objective game designed by \citet*{duan:2014}.
Under ideal system behaviour, i.e. if the service rates are known and if flow
allocation yields ideal system response, the algorithm's convergence time
depends on the heterogeneity of resources. For example, if all machines have the
same service rate, the players should balance allocations equally and the
algorithm converges after the first round regardless the differences in player's
rates. If machines have different service rates, the convergence of the
algorithm depends on the overall system utilisation. With high utilisation, the
competition for resources prolongues the game.

The noncooperative game approach can be applied to serverless computing by
making each function a player to compete in the allocation for minimum response
time. The scheduler would evaluate periodically the arrival rates and service
rates of events and play the game to find new allocations. The recomputation
loop of rates is independent from event dispatching but should be timely to
react to changes in arrival rates.

To briefly evaluate the algorithm for fitness, static games with randomised,
steady arrival rates were played and the highest number of rounds out of 1000
games reported for different game configurations.

\begin{figure}
\centering
\begin{subfigure}{.5\textwidth}
  \includegraphics[width=\linewidth]{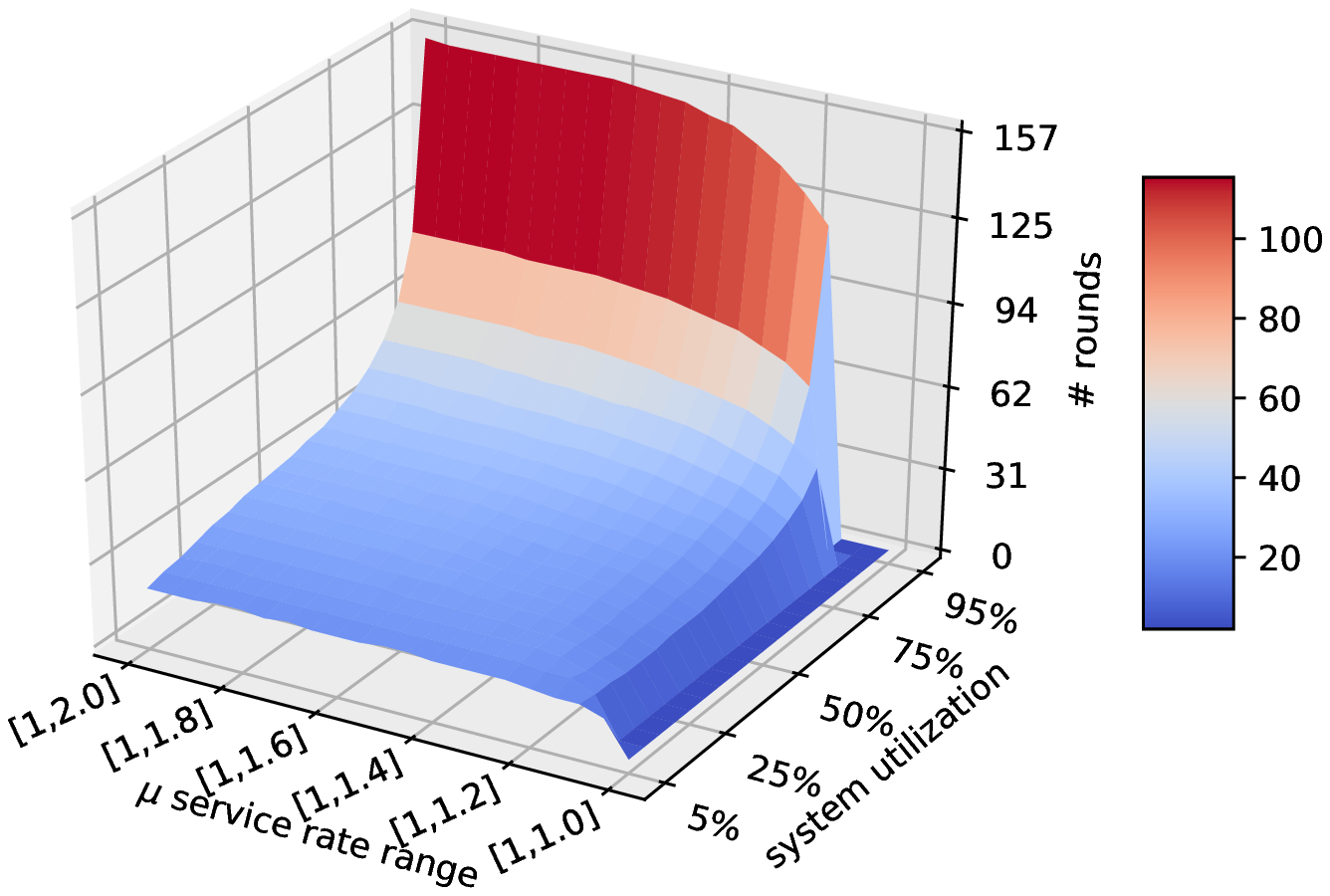}
  \caption{$10\times 10$}
  \label{fig:noncoop_test}
\end{subfigure}%
\begin{subfigure}{.5\textwidth}
  \includegraphics[width=\linewidth]{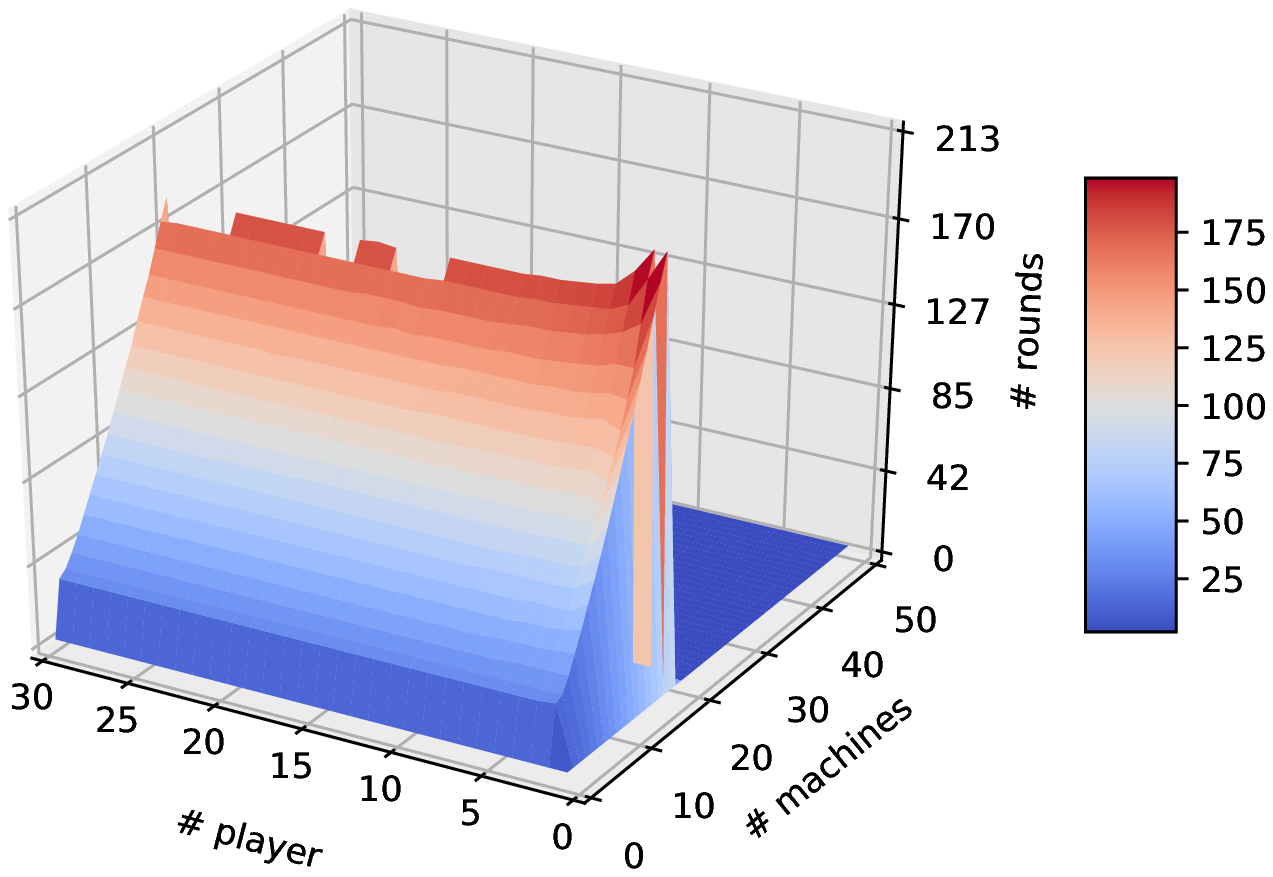}
  \caption{Scaling games}
  \label{fig:noncoop_test2}
\end{subfigure}
\caption{Noncooperative load balancing evaluation}
\label{fig:noncoop}
\end{figure}

\noindent\textbf{10 $\times$ 10 games.} The first evaluation series displayed in
Figure~\ref{fig:noncoop_test} shows the number of rounds required for static
games of 10 players on a pool of 10 machines to converge with a threshold
$\epsilon = 10^{-4}$ (same as used in \cite{grosu:2005}) under varying system
utilisation and service rate deviation.
Players' static arrival rates are
randomly generated picking a weight from the interval $[1.0,1.5]$ and
normalising them to the desired total system utilisation (cmp. \cite{grosu:2005}
uses fractions in $[0.01,0.3]$).
Different service rate ranges of machines were tested with
linear distribution in intervals from $\mu_i \in [1.0,1.0]$ (same service rate)
to $\mu \in [1.0,2.0]$ (cmp. \cite{grosu:2005} uses $[10,100]$).
Utilisations were tested from 5\% to 95\%.
The evaluation shows, that without deviation of machine's service rates, the
players simply equally distribute their arrival rates and are done within 2
rounds (second round to notice no further change).

\noindent\textbf{Scaling games.} The second evaluation displayed in
Figure~\ref{fig:noncoop_test2} tests different system sizes with players
$j=1,\ldots,30$ and machines $i=1,\ldots,50$. Player's steady rates were
randomly choses in the rate interval $[1.0,2.0]$ and normalised to a fixed system utilisation of 90\%.
The evaluation shows a remarkable drop of rounds with more than 16 players.
Closer investigation has revealed, that with large player numbers, the
individual player contributions to the system utilisation are small w.r.t. the
total system capacity. The achievable gains in response to the other players'
allocation is less than $\epsilon = 10^{-4}$ and the game finishes early.

\noindent\textbf{Discussion.} The brief evaluation makes several assumptions,
such as an ideal system response to the reallocation of rates and perfect
knowledge of the perceived service rates. Most importantly, the evaluation uses
static games to estimate fitness for purposse. In an online system such as the
simulation by the authors \citet*{grosu:2005}, the game is played continuously
to adapt players' allocations and perceived service rates that vary under
sampling.
Although the approach has only a single objective, it seems fit to have different
classes of serverless events adapt their individual event dispatching across the
worker pool to achieve the minimal response time objective.
\chapter[Noncooperative Online Allocation Heuristic]{Noncooperative Online Allocation \\ \mbox{Heuristic}}
\label{noah}

The review of scheduling literature in section~\ref{design} for serverless
event dispatching has motivated the design of a new solution that is based on 
the following insights of the analysis.

First, classifying the serverless scheduling problem with a job scheduling
taxonomy puts serverless in a unique position among popular job scheduling
research~\cite{taxonomy:jobscheduling}. The task structure and the open arrival
process put it at the boundary between job scheduling and request load
balancing. The global system welfare and identical objective functions of
individuals suggest the use of game-theoretic principles.

Second, the formalisation of the offline multi-objective optimisation problem
shows, that setup and waiting time of the event are crucial components in both the
response time objective of the customer \emph{and} the cost objective of the provider.
Colocation of instances may not only reduce the setup time but
may also reduce data synchronisation overhead during execution. 
Both objectives are at the hands of the provider and need
to be pursued by a serverless platform to earn customers' trust.

Third, review of common heuristics for request aggregation and review of 
OpenWhisk shows, that it may be desirable to 
allow for flexible event classification. For now, the weight of the 
function instance (its runtime and library dependencies) make the function 
type the preferred classifier to aggregate events.
The OpenWhisk load balancer is a field-tested 
example of such a heuristic and shows strong event aggregation properties.

Fourth, considering the multi-objective problem as a game with global welfare and
identical individual objective functions helps to identify related
game-theoretic approaches. The review of a multi-objective cooperative job scheduling game
\cite{duan:2014} indicates long convergence time for large systems. The design
of noncooperative load balancing \cite{grosu:2005} has compelling features, such
as the exploitation of queue-theoretic optimal control theory to decouple the 
event dispatching from rate allocation.

This has inspired the design of a noncooperative online allocation heuristic
(NOAH), which uses allocation placement and a minimum completion time heuristic to
design a configurable solution to the serverless scheduling problem.

The heuristic is based on the following intuition:
\begin{itemize}
  \item The number of workers covered should be contained to avoid costly underutilisation of reserved infrastructure capacity. To achieve this, the required capacity could be estimated applying queuing theory. 
  \item Instances should be gathered to improve colocation. The more instances of the same class are running on the same worker, the less synchronisation overhead.
  \item To keep setup times at a minimum, instance context should be reused whenever possible. In fact, it may be beneficial to wait in aspiration of an instance finishing instead of launching a new instance.
\end{itemize}

\begin{figure}[t]
\centering
\includegraphics[width=.7\textwidth]{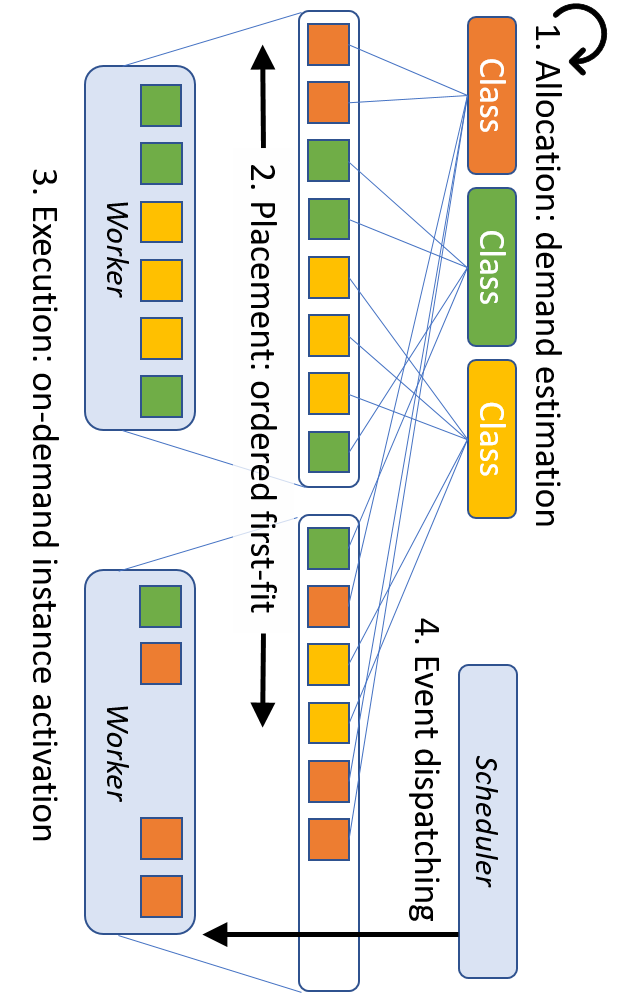}
\caption{NOAH design overview}
\label{fig:noah_design}
\end{figure}

The approach is designed as noncooperative, distributed scheduling of event classes
(players) with individual objectives to reduce response time and a global system
objective to minimise resource usage. The heuristic is implemented as part of
the platform, hence player's strategies are trusted to obey the system
objective.
Figure~\ref{fig:noah_design} summarises the design of the noncooperative online
allocation heuristic, which can be understood as 4 separate mechanisms:
\begin{enumerate}
  \item Each class of events continuously estimates its required number of instances.
  \item The classes play a game to place allocations virtually, in which they can
try to colocate, but are bound to the provided workers.
  \item Workers autonomously decide between waiting or
launching a new instance, based on the demand.
  \item The scheduler dispatches events according to virtual allocations.
\end{enumerate}

\section{Allocation Estimation}

Queuing theory is often applied for rate allocation~\cite{grosu:2005,grosu:2008,widjaja:2013,Tantawi:1985}.
Ideally, a serverless platform would scale infinitely like an $M/M/\infty$ model, but setup times incur resource cost overhead.
Among queue models, the $M/M/c$ model provides a server parameter $c$ to scale the modeled system,
i.e. the model can be applied to calculate the number of instances required per event class to contain queuing time within the class of events.
The $M/M/c/setup$ model with variants regarding delayed shutdown or sleep of a resource has been analysed by \citet*{Gandhi:2013}.
These model variants apply well to the container setup and unpause delays in OpenWhisk.
However, instances could be evicted by other instances and do not consume resources while idling, so the model can at least provide an approximation. 
Besides, the estimate is only used virtually and not to make exclusive resource reservations.

Each event class $k$ (player) calculates the mean arrival rate $\hat\lambda_k$ and mean service rate $\hat\mu_k$ from the exact interarrival times and execution times of processed events, 
e.g. $\hat\mu_k = \frac{n}{\sum_{i=1}^{n}B(e^k_i)}$. Note, that only the execution time without waiting or setup time is used. 
These execution times need to be provided by the worker. 
The scheduler uses obtained means to calculate an optimal number of instances $c_k$ applying a multiple server queuing model. 
In case of model variants that allow setup times, the respective metrics need to be collected as well.

For an initial design\footnote{Regarding other models, e.g. \citet*{Franx:2001} has shown that response times of a fixed service time
queue ($M/D/c$) model can be numerically computed, though with substantial complexity.}, the simple $M/M/c$ model is used that is much easier to solve numerically. 
Assuming a Poisson arrival process with rate $\lambda$ and exponentially distributed service times with mean service rate $\mu$, 
the allocation should at least satisfy the stability condition of the model: $\frac{\lambda}{c\mu} < 1 \implies c >
\frac{\mu}{\lambda}$. To contain the average waiting time of events, the Erlang C formula can be used to find $c_k$ such
that the expected waiting time of queuing remains below threshold $\alpha_k$ \eqref{eq:mmc_wait}.
Note, that the estimation does not depend on any other class of events.

\begin{equation}
\label{eq:mmc_wait}
\frac{C(c_k,\frac{\hat\lambda_k}{\hat\mu_k})}{c_k\hat\mu_k - \hat\lambda_k} < \alpha_k
\end{equation}



\section{Virtual Allocation Placement}

Having identified the required number of allocations $c_k$ for class $k$ events,
the system needs to map $c_k$ virtual instances to workers $W$
with a maximum number of allowed allocations per site $z_C$.

Each function constitutes a player in a noncooperative game for the location of the allocations.
Players try to aggregate instances of their class but may not do so at the
expense of the system cost. This constraint does not allow for many variation in the 
allocation strategy. However, dynamic scaling of event classes leads 
to fragmentation in the system allocation that can be exploited.
A player can prefer locations with large allocations over other locations where it has no allocations.

When it scales out, allocation management tries to colocate allocations to keep a low setup time overhead.
When a function scales in, the host with the least allocations is reduced.
This keeps the bulk of events colocated and decreases expected waiting time
of an arriving event. The implementation of the scheduler estimates required allocations and places
them in the worker pool using a first fit heuristic to keep the number of
employed workers minimal.

The system would have an incentive to block workers when it wants to
scale down the worker allocation. Such defragmentation strategies in which the system claims free spots on
underutilised workers are possible, but reconfiguration strategies are currently not
implemented. Functions that require timely synchronisation of a shared data context would have a stronger incentive to aggregate
allocations. Coalitions to swap allocations (tit-for-tat) between classes could be tested in future versions.

\section{Worker Queuing and Instance Creation}


The resulting allocations are only virtually mapped, i.e. no actual
resource reservations are made at the worker and no instances are spawned by mapping an
allocation.
Instead, each worker manages its instance pool autonomously to handle arriving events,
deciding whether to enqueue a request or spawn a new instance.

The reason for using this heuristic is simply the multiplex gain expected from having events share the entire worker.
As observed in the design of \gls{BoT} scheduling \cite{duan:2014},
allocations can be treated virtually, i.e. no actual resource share allocations
have to be made.
Virtual allocations merely serve to identify locations to which events may be
dispatched to. A blocking resource reservation would mean more harm
than benefit as it increases the idle overhead whereas sharing the worker resources
 allows for statistical multiplexing.

\noindent\textbf{Queuing Discipline.} The worker implements a simple queuing heuristic. The
execution time of events \glssymbol{exec} is measured to estimate an average $\hat\mu^k$,
that is computed as the total average of event execution time $\frac{1}{n}\sum_{i=1}^{n}B(e^k_i)$.

A worker implements a queue for every event class and tries to schedule events on two occasions.
Firstly, when an event of type $k$ arrives, it enqueues the event and tries to schedule an event of the same type.
Secondy, when an event of type $k$ finishes, the worker tries to schedule an event of the same type. If the queue
is empty, it tries to schedule an event from the longest queue\footnote{serving the longest queue on a loaded host when another queue is empty may cause desired reconfiguration of the instance pool, i.e. a switch of types}. 

When an attempt is made to schedule an event of type $k$, it attempts the first event in the queue to maintain FCFS order within a type.
If there is an idle instance of that type, it dequeues the event and launches it.
If there is no idle instance available, it checks if the worker has already reached the limit of concurrently executing instances.
If so, there is no point in launching another instance, so the attempt stops and leaves the event queued.
However, if there is a free spot to launch an instance, the worker estimates first, if the queue would be drained before the new instance has been started, i.e. 
if the expected starting time of all events in the queue is within the setup time of a new instance. 
If the worker expects the queue to drain before the new instance can become available, it rather leaves the event queued.

\section{Event Dispatching}

Having laid out the allocation heuristic and the independent worker operation,
the scheduler needs to decide which worker to send an event to. Primarily, the
scheduler seeks to schedule an event to a worker with a free instance. A queuing 
scoreboard as designed by \citet*{McGrath} could be used to find free instances 
or workers could selectively report free instances to controllers.

Following the OpenWhisk example of distributed
worker state information, the current implementation tries to dispatch events 
to workers with free instances. If there is no idling instance reported, the 
scheduler dispatches the next event to the worker that has the
smallest ratio of active allocations over total (virtual) allocations, which can
be considered a weighted feedback load balancing heuristic.
This design requires the scheduler to closely monitor the worker state 
to achieve a good event balance.


%

\chapter{Serverless simulation}
\label{simulation}

To evaluate the newly designed approach of chapter~\ref{noah} against the identified methods
of noncooperative load balancing (section~\ref{design:game:noncoop}), 
OpenWhisk hash-based first-fit load balancing heuristic (section~\ref{design:heuristics:openwhisk}) and
online bin packing heuristics (section~\ref{design:heuristics:binpacking}), either an implementation or simulation of the system is required.
On the one hand, designing the simulation has the risk of underspecifying the model.
On the other hand, it is common in algorithmic design to evaluate by simulation because it is feasible to test larger systems.
This chapter starts by laying out the requirements for a serverless simulator (section~\ref{simulation:requirements}) and evaluates the 
CloudSim framework for applicability as a serverless simulation platform (section~\ref{simulation:cloudsim}). It is found that 
no current fork of CloudSim fits the requirements of serverless,
so a simulator is designed on the SimPy library and verified (sections~\ref{simulation:design},\ref{simulation:verification}).


\section{Requirements}
\label{simulation:requirements}

The simulation should evaluate a serverless platform's efficiency under different scheduler implementations.
To evaluate the effect of event scheduling on serverless function excection, 
the simulation may implement abstractions of event processing and data access.
The \textbf{abstraction level} should be just low enough to capture the effects discussed 
in section~\ref{design:moop:sojourn} on sojourn times, such as code initialisation and data synchronisation among others.

\textbf{Processing model requirements.}
To measure the effects of scheduling decisions on the response times of serverless functions, the abstracted simulation model has to preserve 
the following characteristics of an invocation:
\begin{itemize}
  \item A function instance has dependencies to initialise the runtime and libraries prior to event processing
  \item Event processing may include data access (read/write), which may cause synchronisation overhead with replica located at other resources
  \item The event processing comprises varying CPU workloads depending on the event parameters and accessed data
\end{itemize}
This level of abstraction is comparatively close to actual processing. E.g., \citet*{duan:2014} chose to abstract from data only being loaded onto a machine, but do not consider synchronisation. \citet*{grosu:2005} have abstracted processing entirely by considering each machine an M/M/1 type server. 
The multi-objective problem formulation provided in section~\ref{design:moop} however consideres both variable setup \glssymbol{setup} and execution time \glssymbol{exec} as a function of the particular event, i.e. its class, arrival time, and the instance it is scheduled to. This level of detail is required to support different application types, which may range from entirely stateless and independent invocations to concurrent modifications of large data application contexts.

\textbf{Data model requirements.}
To capture library, runtime and data dependencies of customer functions, the simulator needs to 
manage item sizes, dependencies and resource capacity. E.g. the first instance of a function on a worker
may require to load missing dependencies, whereas the second instance can reuse local copies of data for initialisation.
Processing of an event may create, read, update or delete named data items. Such context data 
may be reused by other events, so synchronisation of remote replica should be supported by the simulator to 
capture the effect of data locality.

\textbf{Resource metrics} should be collected to evaluate the resource
utilisation and response time objectives of the optimisation problem
(section~\ref{design:moop:worker}). Some schedulers also require to capture the
lineage of event execution, i.e. interarrival times at the scheduler and
queuing, setup and execution times. To compare resource efficiency, the
simulation needs to capture the time a resource is utilised and the time it is
employed by the system. Eventually, the simulator needs to facilitate different
\textbf{scheduler} and \textbf{worker} models, e.g. with allocations,
concurrency limits, instance pool limits, per-worker or per-class queues and
custom queuing disciplines.

\section{CloudSim Evaluation}
\label{simulation:cloudsim}

CloudSim\cite{cloudsim} is a discrete event simulation toolkit developed for data
center resource allocation and utilisation simulation.
Various forks exist that extend the toolkit by different aspects of Cloud computing.
The following section evaluates their applicability to serverless event processing.

\textbf{Core engine.} The toolkit simulates data center resource utilisation
under custom workload modeled as Cloudlets. Virtual entities (Cloudlets, VMs)
are deployed and provsioned with pluggable placement and scheduling strategies.
Initially designed to model virtual machine placements, every
deployment has a \emph{broker} receive the data center's resource status, create VMs,
submit Cloudlets and wait for completion. The host's resource admission 
scheduling decides on the progress of placed workloads. CloudSim implements three such scheduling principles.


\begin{itemize}
  \item The space-shared scheduler exclusively allocates MIPS from processing
  elements (CPU cores), i.e. at VM level, allocation either succeeds or fails and workload allocation
  (Cloudlets) to a running VM is queued until the required amount of MIPS is
  available.
  \item The time-shared scheduler implements processor sharing, i.e. VM and  
  workload allocation share the overall machine capacity, so a VM receives
  MIPS capacity proportionally to the number it demands and the same applies to
  time-shared workload scheduling inside VMs.
  \item Dynamic workload scheduling respects workloads with varying resource
  utilisation and features over-subscription of resources, i.e. a VM can
  allocate more resources than available but is given its time-share according
  to utilisation but no more than its maximum allocation.
\end{itemize}

Dynamic workload scheduling is the most realistic to capture varying
CPU utilisation by running processes. Dynamic hosts even consider a 10\% service
degradation for VMs under live migration. Unfortunately, estimated file transfer time
for required files is factored as additional CPU time into scheduling, so it
varies in the time-shared and the dynamic workload model and it does not consider 
to incur workload on the transfer's source host.

\textbf{CloudSim extensions.} A framework of projects has evolved around the
toolkit that simulates various other aspects of Cloud computing. CloudSim 
natively supports only brokers and data centers information exchange, 
i.e. the data center collectively subsumes
allocation, scheduling and workload execution. Its focus on Cloud
resource allocation makes modeling of fine granular effects difficult, such
as tasks, data exchange or caching.

WorkflowSim addresses workflow scheduling in a spin-off project by \citep{Chen:2012} based
on version 3.0.3 without container support. Workflows are DAG graphs
that are planned, deployed and executed without further interaction.

After CloudSim had introduced switched networking in version 3.0.3, CloudSimSDN
(2015) had been developed with support for virtual links.
FogSim\cite{Gupta:2016} is based on CloudSimSDN to simulate a
sensor-controller-actor stream processing model in which distributed
applications consist of multiple VMs (modules) and exchange tuples on a tree
host structure. Tuples are either forwarded uplink or on all downlinks of
the tree structure.


\textbf{Container extension.}
In its latest release (4.0), the CloudSim toolkit has introduced container
virtualisation \cite{piraghaj:2017}, yet the code quality has been decreasing
as open tasks and piling issues remain unaddressed.
While the VM implementation had allowed concurrent execution of Cloudlets in a
single VM (time-sharing), the container implementation only creates a
container for every Cloudlet at startup, and can't handle dynamic submission
of Cloudlets to container instances (to simulate processing of e.g. events or requests).


\textbf{Conclusion.}
For the CloudSim toolkit to become applicable, it needs to
complete the container technology implementation. Yet, Cloudlet processing would still not 
consider data access at runtime. It would also be beneficial if 
resource provisioning (scheduling, allocation policies) could be developed independently 
from the resource type (Host, VM, container).

Network delay is considered in CloudSim when sender and receiver of a simulation
event are not the same, but actual network delay can vary depending on whether
the entities share a host (loopback device) or reside on different hosts. The
networking resolution in CloudSim is designed for wide area networks.

Data access is considered at creation time and is abstracted from by modelling an estimated 
transfer time as a local processing workload. Actually, data 
synchronisation between tasks requires a shared data layer model that allows access 
and replication which affects both ends of a transfer.

\begin{table}
\begin{tabular}{l|c|c|c}
 & CloudSim 4.0 & WorkflowSim & FogSim \\
\hline
Job model & Cloudlets & Task workflows & Stream processing \\
\hline
Container & VM-like & - & - \\
\hline
Placement & \shortstack{online \\ offline} & \shortstack{offline} &
\shortstack{offline} \\
\hline
Communication & \shortstack{Broker \\ Datacenter} & \shortstack{WorkflowPlanner
\\ ClusteringEngine \\ WorkflowEngine \\ WorkflowDatacenter} &
\shortstack{FogBroker \\ FogDevice \\ Sensor \\ ApplicationModule \\ Actuator}
\\
\hline
\end{tabular}
\caption{CloudSim extension overview}
\label{table:cloudsim}
\end{table}

Table~\ref{table:cloudsim} summarises the features of CloudSim and the two reviewed extensions.
The framework can well simulate data center resource allocation requests of various types (for workflows or streams), but
to test event dispatching and load balancing algorithms, the event communication
level would need to be lowered from data center brokers to hosts
exchanging events to capture the side effects of colocation. iFogSim achieves
the stream level only by modelling each host as a datacenter, but yet lacks to model a
request load balancer and resource scaling. Neither WorkflowSim not iFogSim allow online scheduling decisions for workloads.
CloudSim has many benefits, e.g. in modelling the concurrency of VMs and 
oversubscription in Cloud infrastructures. Its strength to simulate large 
infrastructures are a compelling argument to base serverless platform simulation on CloudSim.
However, its gaps in data dependency modelling and the coarse granularity of both
allocation and processing would require to modify its core design elements.

\section{Simulator Design and Implementation}
\label{simulation:design}

The gap between CloudSim infrastructure simulation and serverless event
processing has spurred the design of an independent simulator that focuses on
the model outlined by the serverless terminology, operation description and
reference architecture provided in chapter~\ref{systems}. The serverless
simulator was developed using SimPy (v3.0.1) and is based on the experience with OpenWhisk. 
It implements three layers. At the
core, it implements work-conserving, processor-sharing CPU scheduling and a caching memory
architecture. The platform operation layer implements a data layer, code and instances 
to launch processes, to execute serverless events and to
synchronise data transfer and data access.

\subsection{Processing Model}
\label{simulation:design:processing}

The base processing model is implemented consisting of workers that each have CPUs and memory to host execution.
Container contexts can be provided to isolate executions.

\textbf{Worker CPU.}
Based on the serverless operation model, a \textbf{worker} is a capacitated computing resource and can be assumed to use work-conserving scheduling. The typical OS model uses a processor sharing discipline, i.e. concurrent workloads share the capacity of virtual multicore machines (e.g. $n$ virtual cores). To split function invocations into data access and execution operations, the worker processing model should allow to schedule single-threaded \textbf{executions} with a required \gls{CPU} time.
The effect of \gls{CPU} models on the execution is abstracted from, i.e. requested \gls{CPU} time is the same on each worker.
Single-threaded execution means, that a workload proceeds at the speed of a virtual \gls{CPU} unless there are more concurrent workloads ($m$) than  virtual cores ($n$), in which case they proceed each with $\frac{m}{n}\textit{th}$ of the \gls{CPU} speed.
Context switching and the interleaving of processes is abstracted from, i.e. workloads proceed concurrently.

\textbf{Worker memory.}
Each worker implements a \textbf{memory} that provides allocations. An allocation can be read from concurrently, but it can be written only by a single \textbf{execution}. \textbf{Reads} and \textbf{writes} are executions that occupy \gls{CPU} time equivalent to the amount of data read or written at the read/write speed of the memory. The memory allocation implements priority queueing of read and write executions, i.e. a write preempts all current reads. Multiple writes are queued in \gls{FCFS} order and read requests are queued until after the writes have been processed.
The memory implements \gls{LRU} eviction, i.e. unlike OS memory management, allocations are not blocked when the memory is full. Instead, the least recently used allocations are evicted to fit the new allocation and processes holding references are interrupted. This model eases instance eviction as well as implementation of caching data layers.

\textbf{Container context.}
The model implements container contexts as aliases of workers, i.e. executions are attributed to the parent \glspl{CPU} and allocations are held in the parent memory. However, container names are used to log execution time, event processing, data locations, etc.

\subsection{Serverless Platform Operation}
\label{simulation:design:platform}

Based on the processing model, several platform elements have been implemented. A \textbf{data layer} is provided that abstracts from data subsystems (container image repository, serverless function repository, key-value stores, event messaging, etc.).
\textbf{Instances} are an abstraction from processes that require to load runtime libraries, library dependencies and code files.

\textbf{Data layer.}
An in-memory \textbf{data} layer is implemented that allows named data items (container images, function code, messages, context data) to be replicated between workers and containers. Simply, if a piece of \textbf{data} is required at a worker (or container) where it has not been accessed before, the item is replicated. The replication requires a memory \textbf{read} at the source, a \textbf{write} at the target and a \textbf{transfer} delay depending on the locations. If a replica exists on the same worker, it is preferred over other locations to avoid network transfer. Different transfer speeds apply for disk access, network transfer or memory access. The simulation abstracts from the transfer protocol, i.e. replication is considered complete after the last of the three actions (read at source, write at target and transfer delay) has completed. Write access to a replica updates remote locations immediately and concurrently, so it may preempt reads in remote locations. \textbf{Messages} are also data items that incur network transfer delay when dispatched to a host.
The real architecture uses Kafka for reliable, distributed messaging. However, as discussed in section~\ref{design:moop:sojourn} on sojourn times, delays for scheduling are not considered.

\textbf{OpenWhisk code and instances.}
A piece of \textbf{code} is a data item with \textbf{dependencies} to other data items. When it is accessed, it ensures replication of all of its dependencies. Furthermore, when an execution enters the code, it subscribes with all dependencies, so that eviction of a dependency causes failure of the execution. A code may also provide a function that is executed when an instance of the code is created.
An \textbf{instance} is the abstraction of a serverless function instance. Any code with an entry point function can be instantiated.
Upon initialisation, it reads the entire code and dependencies once (at memory speed) to mimic process initialisation. This is a comparatively coarse-grained abstraction.
System methods such as forking or shared libraries could reduce the process context initialisation, while actual runtime or code initialisation requires a lot more execution that the simple write of its memory contents.
For example, a container instance would not spend time to read the container image entirely but requires additional OS context and container daemon operations. Thus, the time for OpenWhisk on Docker to create a container instance takes about \SI{504}{\milli\second}, while reading the image contents (290MB at 12.800MB/s) would take only \SI{22.7}{\milli\second}. To account for the difference, the instance's function adds an additional execution with a workload of \SI{482}{\milli\second}.
Also, the initialisation of a base container image with the user-provided function runtime requires at least \SI{300}{\milli\second} in OpenWhisk due to the code being injected through a \gls{REST}ful container network interface. As the additional delay is mostly attributed to I/O, the delay is accounted for by a timer (no workload) parallel to code replication (small workload).
So, the resulting simulation mimics the OpenWhisk model by adding execution delays where necessary.

\textbf{A neutral deterministic workload function.}
To measure the effects of replication, initialisation, oversubscription, etc., a simple function with deterministic execution time is implemented. Although the function has a deterministic workload of \SI{200}{\milli\second}, the execution time \glssymbol{exec} can vary upon oversubscription of the \gls{CPU} capacity, the different setup times \glssymbol{setup} that optionally comprise container instantiation and initialisation affect the overall response time. The closest model to this behaviour would be a deterministic service time processor sharing with setups and sleeps $M/D/c/setup/sleep-PS$, i.e. the expected timing is neither a completely deterministic service time (e.g. $M/D/c-FCFS$) nor exponentially distributed processor sharing (e.g. $M/M/c-PS$). However, this function allows to visualise the combined effects of queuing and overload on the response time by providing a fixed, ideal execution time, that marks the lower bound of achievable execution times.

\subsection{Scheduling Implementation}

On top of the serverless platform operation that mimics OpenWhisk, several scheduling approaches have been implemented.

\textbf{OpenWhisk}(\ref{design:heuristics:openwhisk}) uses a controller-invoker infrastructure. An invoker processes events in FCFS order and implements a single 
queue. When it dequeues, an event may cause eviction of a container and cause the start of a new one. The controller implements the hash-based first-fit heuristic that progresses pseudo-randomly from the preferred host that the function name hashes to.

\textbf{First fit, next fit and best fit} (\ref{design:heuristics:binpacking}) reimplement the controller and check the load (active instances and queue length) against the capacity limit (factor of the number of cores) to make scheduling decisions.

\textbf{Noncooperative load balancing} (\ref{design:game:noncoop}) also reuses the OpenWhisk invoker and implements an alternative controller. The controller implements a player for each function type that measures the interarrival times and their service times at each invoker (worker). Unlike the sampling method proposed in \cite{grosu:2005}, arrival rates and service rates are measured exactly from all processed events. 
It plays the game every \SI{100}{\milli\second} to optimally redistribute arrival rates to workers.

\textbf{NOAH} (\ref{noah}) implements its own invoker and controller.
The new invoker logs the service time and setup time of events. The average setup time is required to decide between queuing an event or launching a new instance. The service time is required at the controller to estimate the required number of instance allocations.
Further, the invoker reports free instances whenever it has drained a queue and no more events to schedule.
This allows the controller to immediately schedule an arriving event to a free instance. 
Otherwise, the controller balances events according to allocations that it estimates from mean interarrival and service times of the function type.

\section{Verification}
\label{simulation:verification}

To verify the implementation of the serverless simulator, two test suites have been implemented.\footnote{a third test suite also compares the default OpenWhisk model with a real OpenWhisk installation, but is left out for the sake of brevity.}

\textbf{Execution.}
The first set of tests verifies the worker's work-conserving, processor-sharing execution scheduling using the $M/M/1$ queue model for a single-core test and $M/M/c$ model for multi-core tests.
When executions are spawned at a Poisson arrival rate $\lambda$ with exponentially distributed CPU time demand rate $\mu$ (avg. service time $\frac{1}{\mu}$) in the \gls{PS} execution model on a single worker with a \textit{single} \gls{CPU},
 the $M/M/1$ model applies, i.e. the worker has a utilisation of $\rho = \frac{\lambda}{\mu}$ and its mean response time is $\frac{1}{\mu-\lambda}$.
A series of 1k tests with 10k executions each, $\lambda = 8$ and $\mu = 10$ using independent random number generators for the arrival and service rates yields a set of 1k observations of the mean response time.
The 95\% confidence interval of the set using the Student's t distribution ($999$ degrees of freedom) is [\SI{497}{\milli\second}, \SI{502}{\milli\second}]. The expected response time is \SI{500}{\milli\second}.

Analogously to the single-core tests, a worker with $4$ cores is tested with $\lambda = 32$ and $\mu = 10$ that is expected an average response time of $C(c,\frac{\lambda}{\mu})+\frac{1}{\mu}$ according to the $M/M/c$ model.
The 95\% confidence interval of the observed set of means using the Student's t distribution with $n-1$ degrees of freedom is [\SI{173}{\milli\second}, \SI{175}{\milli\second}]. The expected response time is \SI{174.6}{\milli\second}.


\textbf{Allocation and data management.}
To verify that data allocations, reads and writes operate as described, logical tests have been implemented that verify the features.
Read and write operations have been tested to allow concurrent reads, read preemption by writes, write queuing and read-after-write enqueueing.
To test cache eviction behaviour under memory limits, allocation eviction of the least recently used allocation is tested when memory is full and the notification of processes (e.g. executing code) subscribed with evicted allocations (library dependencies) has been verified to abort the execution.
Data replication has been tested for different latency timings, i.e. worker-to-worker, worker-to-container (on the same and different worker) and container-to-container replication.
More importantly, effects on replication and synchronisation duration have been tested where either the read at the replication source or the write at the replication target is being delayed.
And of course, replication and synchronisation failures upon either preemption or eviction at the source or target allocation have been verified.
Detailed test specifications are omitted for brevity.

These unit tests are crucial to ensure a working simulator implementation, because bugs in long running simulations could lead to wrong conclusions on the scheduler performance.




\chapter{Evaluation}
\label{evaluation}

Finally, the designed and verified serverless simulator is used to evaluate the different scheduling solutions.
For the evaluation, a dynamic workload scenario is designed (section~\ref{evaluation:scenario}) that ramps up the workload to a specified maximum.
As NOAH allocation estimation is configurable with different waiting time thresholds $\alpha$, the range will be compared using
thresholds \SI{10}{\milli\second}, \SI{1}{\milli\second}, \SI{100}{\micro\second} and \SI{10}{\micro\second} in section~\ref{evaluation:noah}.
The configuration $\alpha$=\SI{100}{\micro\second} is then compared with results for classic online bin packing heuristics in section~\ref{evaluation:heur} and 
eventually is compared against the results of the OpenWhisk field-tested hash-based first-fit load balancer and noncooperative load balancing in section~\ref{evaluation:comp}.

\section{Concurrently Scaling Workload}
\label{evaluation:scenario}

The evaluation can choose from multiple configuration parameters, e.g.
platform resources, customer-provided functions and event workloads.
The important metric to evaluate is platform efficiency, i.e. the resource cost
and the response time objectives.

It needs to be noted that the OpenWhisk load balancer is designed for a fixed
worker pool size. A reconfiguration of the pool at runtime would cause rehashing
of function names to workers and would cause additional container setup times
for the rehashed functions following the reconfiguration, which would cause an
unfair disadvantage. Hence, the worker pool size is fixed for all experiments.
Instead, workloads are scaled to explore the behaviour and operational limits of
the schedulers with the given resource set, which actually makes it easier to
compare qualities of the approaches.

A pool of 10 workers is simulated with typical server configurations of $16$ cores and $48$ GB memory.
Data replication simply considers ideal throughput with 12.8GB/s memory speed (\gls{DDR3}), 711MB/s disk speed (\gls{SSD}) and $1135$MB/s network
transfer throughput (\gls{10GbE}) to model replication latencies.
The effects of scheduling decisions on response times need to be evident, so the neutral
deterministic workload function discussed in section~\ref{simulation:design:platform} 
is used with an ideal execution time of \SI{200}{\milli\second}. 
Ten functions with independent random Poisson-distributed
 arrival processes are simulated.
Given the total $160$ cores, the maximum, ideal upper-bound service rate of the simulated system is $800$ events per second.

\begin{figure}
\centering
\includegraphics[width=0.6\textwidth]{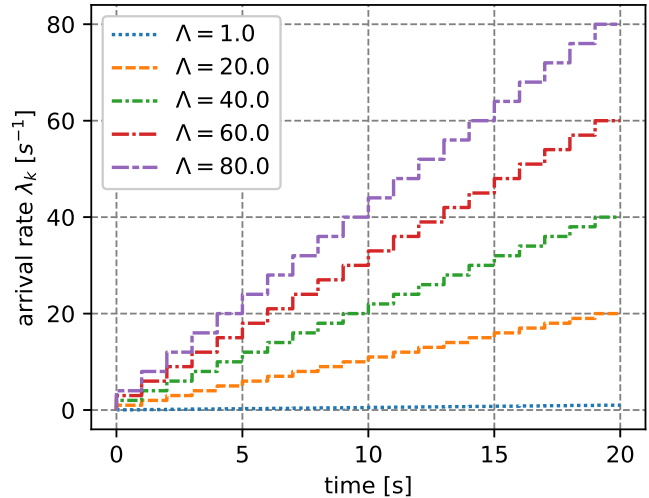}
\caption{Simulation workload increase} 
\label{fig:explain}
\end{figure}

With this configuration, experiments are run that each ramp up the workload to a certain maximum.
The purpose is to evaluate the scheduling under dynamic scaling conditions.
In each experiment, the interarrival rate $\lambda_k$ of the $10$ independent functions grows linearly every second from zero to a maximum rate $\Lambda s^{-1}$ over a period of $20$ seconds and then drops.
Each experiment tests a different maximum arrival rate \glssymbol{maxarrival}.
Figure~\ref{fig:explain} exemplifies the increasing interarrival rate $\lambda_k(t) = \frac{\lceil t \rceil}{20} * \Lambda$, ($0\leq t\leq 20$) for experiments \glssymbol{maxarrival}=$\{1,20,40,60,80\}$.
The experiment is continued until all events have been processed.
With $10$ independent arrival processes, the absolute maximum arrival rate \glssymbol{maxarrival} per process at $t=20s$ is $80$ events per second, i.e. the upper bound theoretical system limit of $800$ events per second. 

This concurrent scaling scenario is designed to show various aspects.
The number of workers and number of functions are chosen to allow OpenWhisk to balance load using the pseudo-random order first-fit with pseudo-random starting points on all workers under high load.
All approaches are expected to struggle when \glssymbol{maxarrival} reaches the maximum theoretical limit.
Workloads are generated before the experiment, i.e. the evaluation uses identical workloads for every scheduler.
Greedy approaches will try to utilise the entire pool from the start, while adaptive approaches leave workers untouched under low workload.
Of course, none of the approaches will manage to achieve the theoretical limit because of container unpausing or container creation and code initialisation delays.
Hence, the scenario drops arrivals after the maximum rate (sawtooth) and gives the time required to complete queued events.

\section{Simulation Results}

This section presents experiment results.
The two main objectives are response times and resource cost.

The \emph{response time} is measured as the \textbf{average response time} of events in
an experiment. Since all events have an identical ideal function execution time of
\SI{200}{\milli\second}, the average response time indicates the effects of
scheduling decisions on the function response time objective. 

The \emph{cost} is evaluated using multiple metrics.
As OpenWhisk does not natively support worker scaling, the pool size has a fixed
number of 10 workers.
To show the effectiveness of adaptive approaches, the number of \textbf{workers covered} is plotted, i.e. only those used eventually for event processing.
First-fit, best-fit and NOAH approaches are likely to spare hosts if there are
not too many concurrent events.
Greedy approaches (next fit, OpenWhisk) are likely to employ all hosts provided. 
In addition, the number of \textbf{total instances} spawned is compared as this gives a hint to how often the scheduling solution requires 
the workers to start a new instance.
To also measure how the solution makes use of allocated memory, \textbf{instance utilisation} is measured.
Because instances are usually kept for a timeout of 5 minutes after their last processing, instance utilisation measures the time an instance actively processes events over the period from its creation until it completes its last message processing only, discarding any idle time after.
It has been ensured that none of the experiments runs into memory overload and that all events complete successfully.

In the following, results on the four metrics \textbf{average response time}, \textbf{workers covered}, \textbf{total instances} and \textbf{instance utilisation} are discussed in comparison of different NOAH thresholds (\ref{evaluation:noah}), NOAH vs. online bin packing heuristics (\ref{evaluation:heur}) and NOAH vs OpenWhisk and noncooperative load balancing (\ref{evaluation:comp}).

\subsection{NOAH Configurations}
\label{evaluation:noah}

\begin{figure}[H]
\centering
\includegraphics[width=0.8\textwidth]{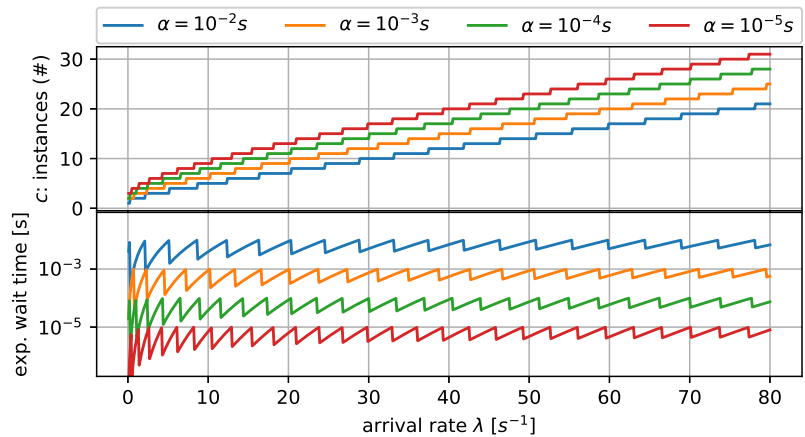}
\caption{NOAH allocation estimation with} 
\label{fig:noah:scaling}
\end{figure}

This section presents the results of different configurations of the NOAH allocation estimation 
($\alpha$=$10^{-2}s$, $\alpha$=$10^{-3}s$, $\alpha$=$10^{-4}s$, $\alpha$=$10^{-5}s$). 
As described in chapter~\ref{noah}, the estimation uses the measured interarrival time and execution time of a function execution and
an $M/M/c$ model to contain expected waiting time under the threshold $\alpha$. 
Figure~\ref{fig:noah:scaling} shows the allocations that would be made with an ideal execution time of \SI{200}{\milli\second} (rate $\mu = 5$) and exact interarrival times (rate $\lambda \in ]0,80[$).

\begin{figure}[t]
\centering
  \begin{subfigure}{.495\textwidth}
  \centering
  \includegraphics[width=\linewidth]{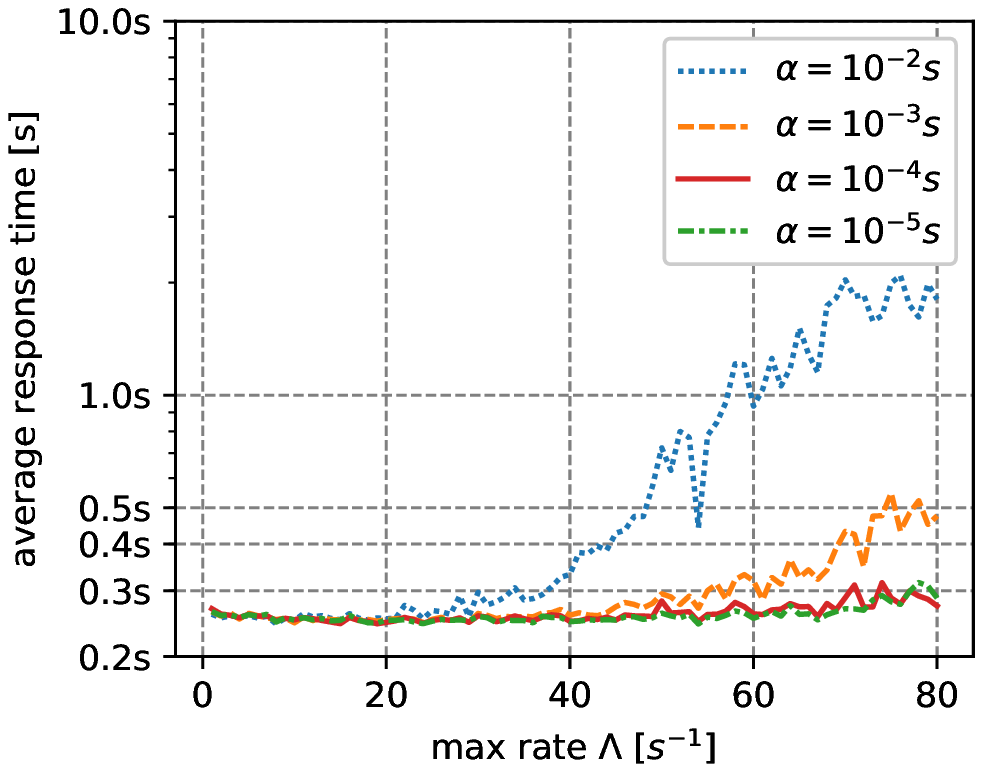}
  \caption{Average Response Time}
  \label{fig:noah:response_time}
  \end{subfigure}
  \begin{subfigure}{.495\textwidth}
  \centering
  \includegraphics[width=\linewidth]{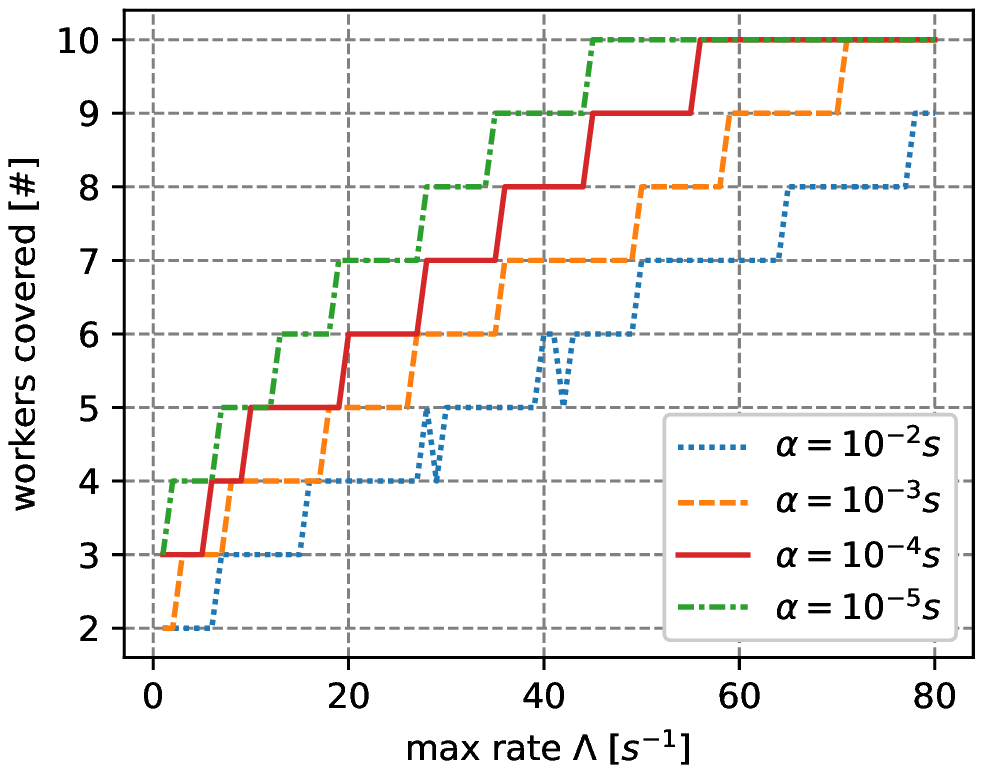}
  \caption{Workers Covered}
  \label{fig:noah:workers_covered}
  \end{subfigure}
  \begin{subfigure}{.495\textwidth}
  \centering
  \includegraphics[width=\linewidth]{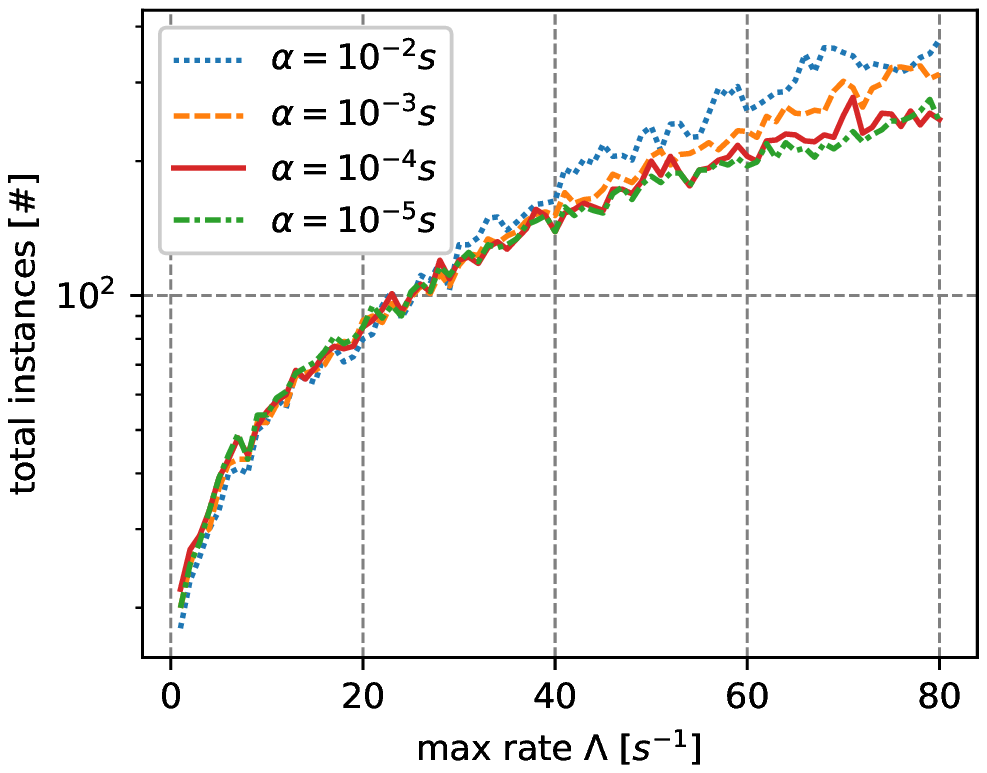}
  \caption{Total Instances}
  \label{fig:noah:instances}
  \end{subfigure}
  \begin{subfigure}{.495\textwidth}
  \centering
  \includegraphics[width=\linewidth]{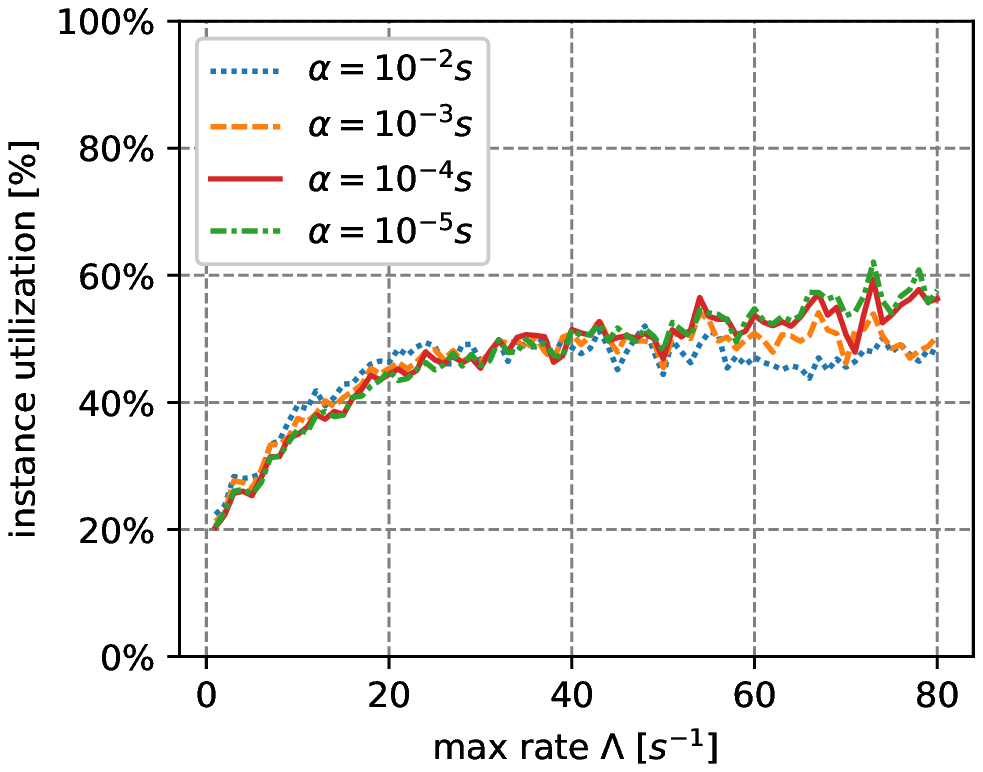}
  \caption{Instance utilisation}
  \label{fig:noah:instance_utilisation}
  \end{subfigure}
  \caption{Comparison of NOAH configurations}
  \label{fig:noah}
\end{figure}

The results are given in Figure~\ref{fig:noah}.
The average response time (\ref{fig:noah:response_time}) climbs for $\alpha$=$10ms$ up to two seconds. 
The allocation estimation does not consider setup times in the model. 
Under high load, the initial setup time is more likely to be carried over till the last events in the experiment. 
Simply put, the relaxed threshold ($\alpha$=$10^{-2}$) underestimates the load.
This can also be seen in the number of workers covered (\ref{fig:noah:workers_covered}) as the configuration tries to cover the maximum system load $\Lambda$=$80$ with only $9$ out of $10$ workers.
A much better threshold of $\alpha$=$10^{-4}$ keeps the average response time visibly under \SI{300}{\milli\second} while not saturating the pool before $\Lambda$ reaches $\approx 55$.
The total number of instances (\ref{fig:noah:instances}) scales similarly for all approaches, because the worker implementations decides autonomously whether it is better to wait for an instance to finish rather than starting a new instance. 
However, thresholds $\alpha=\{10^{-2},10^{-3}\}$ start launching more actual instances.
This can be attributed to underestimation of the load, which causes more instance evictions and in turn more instance creations as workers  oscillate between different function types.
The case of evicting an instance to start another is called instance \emph{churn}.
Additional churn also accounts for the increased response times seen in Figure~\ref{fig:noah:response_time}.
Instance utilisation is better with tighter thresholds $\alpha$=$\{10^{-4},10^{-5}\}$. 
With less instance churn, these configurations suffer less from setup time overhead that would contribute negatively on utilisation.
The maximum utilisation reached is not very high, because a worker allows more total instances than actively processing instances to 
reduce the number of instance starts \glssymbol{w:maxtotal} $>$ \glssymbol{w:maxactive}.
The scores of schedulers in the utilisation metric need to be regarded relative to one another as the metric mixes both the setup time overhead as well as memory allocation overhead to the time actually used to process events.

\newpage

\subsection{NOAH vs. Online Bin Packing Heuristics}
\label{evaluation:heur}

\begin{figure}[H]
\centering
  \begin{subfigure}{.495\textwidth}
  \centering
  \includegraphics[width=\linewidth]{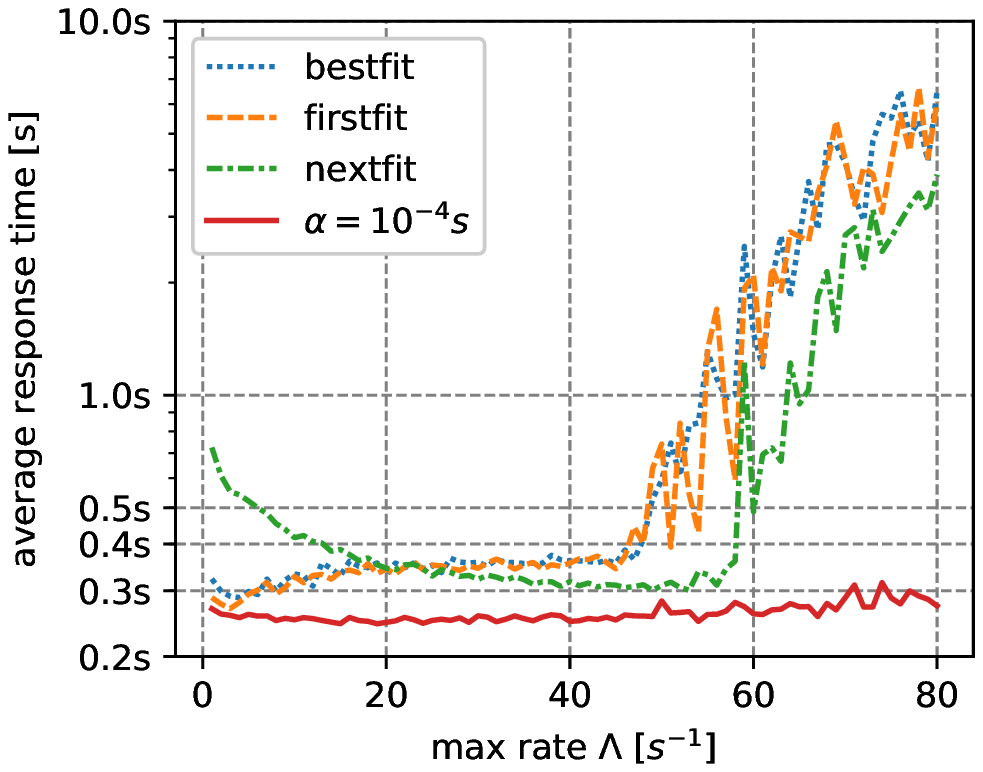}
  \caption{Average Response Time}
  \label{fig:heur:response_time}
  \end{subfigure}
  \begin{subfigure}{.495\textwidth}
  \centering
  \includegraphics[width=\linewidth]{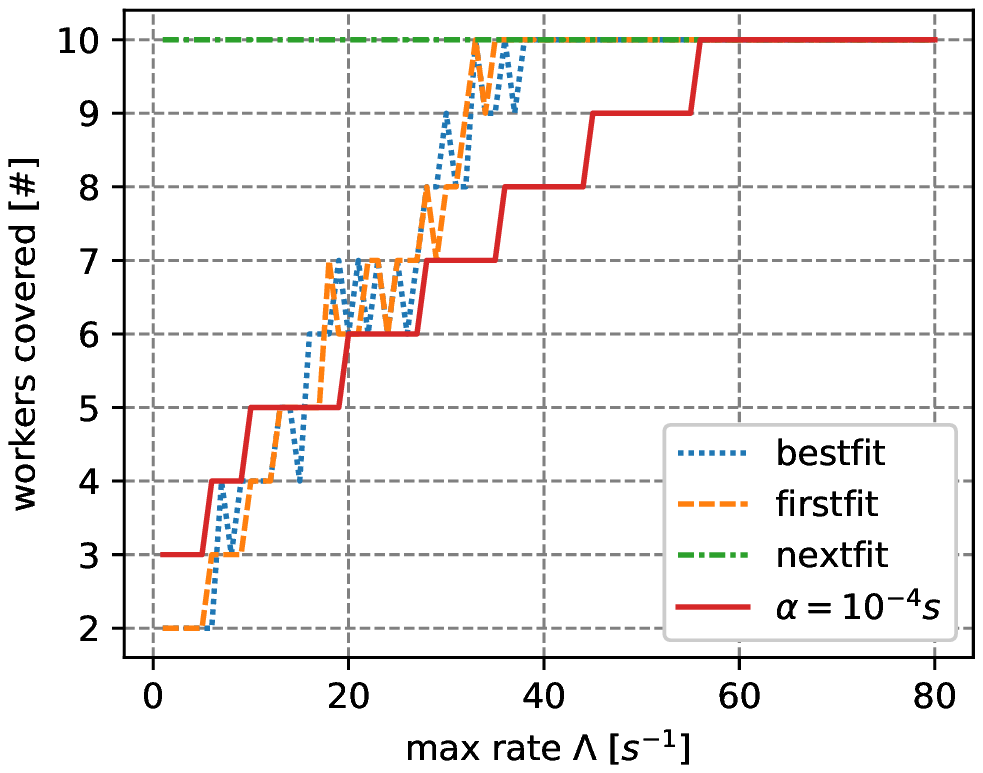}
  \caption{Workers Covered}
  \label{fig:heur:workers_covered}
  \end{subfigure}
  \begin{subfigure}{.495\textwidth}
  \centering
  \includegraphics[width=\linewidth]{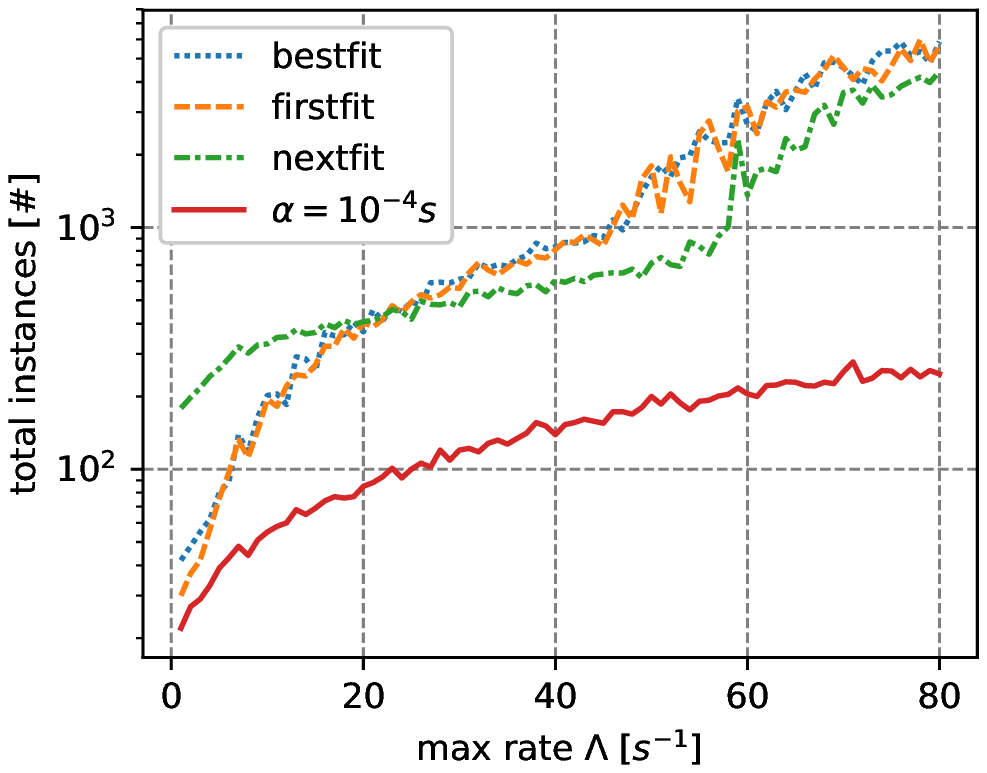}
  \caption{Total Instances}
  \label{fig:heur:instances}
  \end{subfigure}
  \begin{subfigure}{.495\textwidth}
  \centering
  \includegraphics[width=\linewidth]{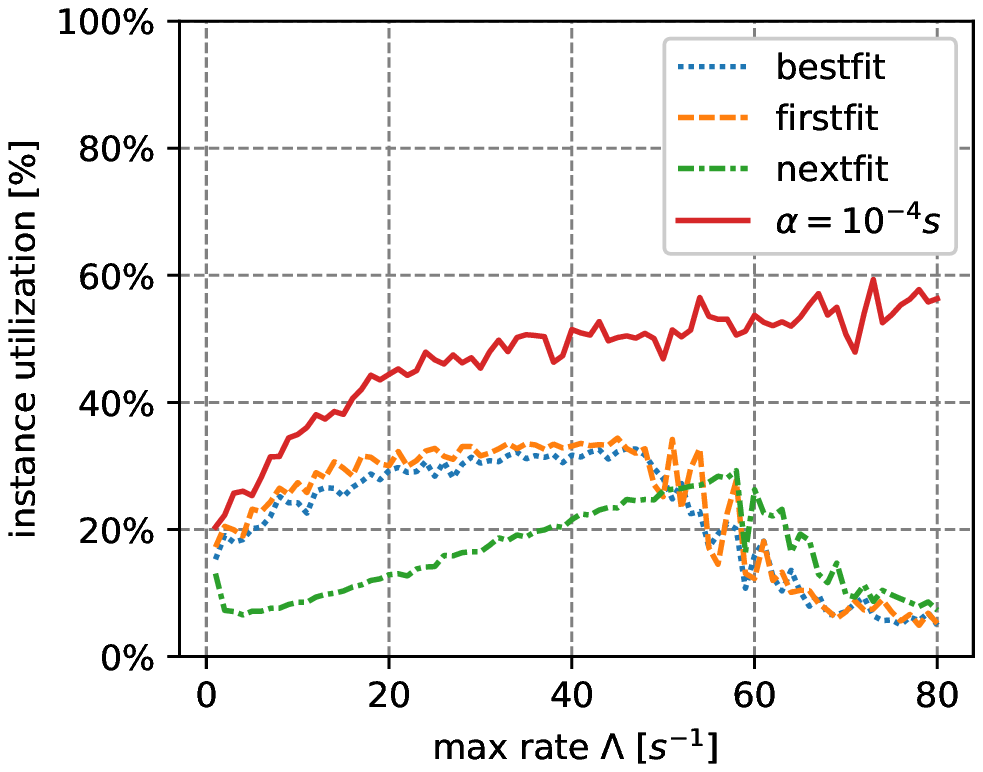}
  \caption{Instance utilisation}
  \label{fig:heur:instance_utilisation}
  \end{subfigure}
  \caption{Comparison of NOAH and online bin packing heuristics} 
  \label{fig:heur}
\end{figure}

To compare with standard online bin packing heuristics, the results of the NOAH configuration $\alpha=10^{-4}$ is shown against
best fit (\gls{BF}), first fit (\gls{FF}) and next fit (\gls{NF}) in Figure~\ref{fig:heur}.
The average response time of \gls{FF} and \gls{BF} starts low as it is more likely to reuse a started instance than \gls{NF} with low loads, whereas \gls{NF} progresses through the entire pool every time it sees more than 16 concurrent events.
Likewise, \gls{NF} always employs the entire worker pool, while \gls{BF} and \gls{FF} keep it at a minimum. 
Estimations of NOAH start with 3 workers, i.e. more than with \gls{BF} and \gls{FF}, but it also sees less churn and achieves better response times.
Neither of the classic heuristics are classful and suffer from high churn, so the total number of instances is much higher and utilisation suffers from more setup time overhead accordingly.

\subsection{NOAH vs. OpenWhisk vs. Noncooperative Load Balancing}
\label{evaluation:comp}

\begin{figure}[H]
\centering
  \begin{subfigure}{.495\textwidth}
  \centering
  \includegraphics[width=\linewidth]{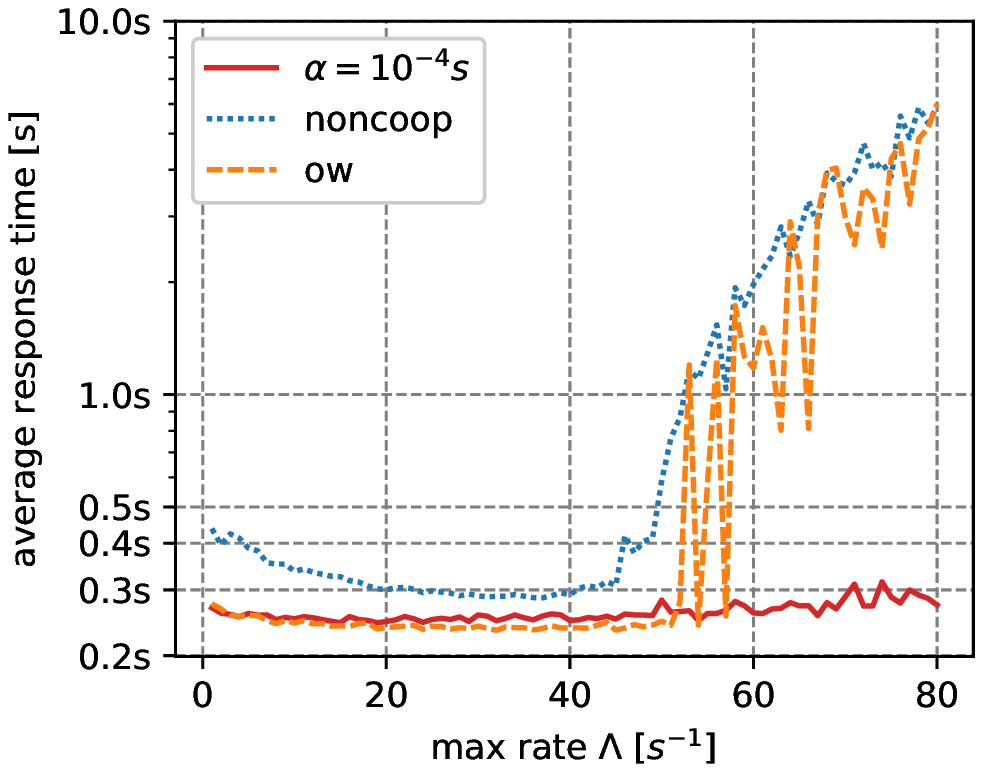}
  \caption{Average Response Time}
  \label{fig:comp:response_time}
  \end{subfigure}
  \begin{subfigure}{.495\textwidth}
  \centering
  \includegraphics[width=\linewidth]{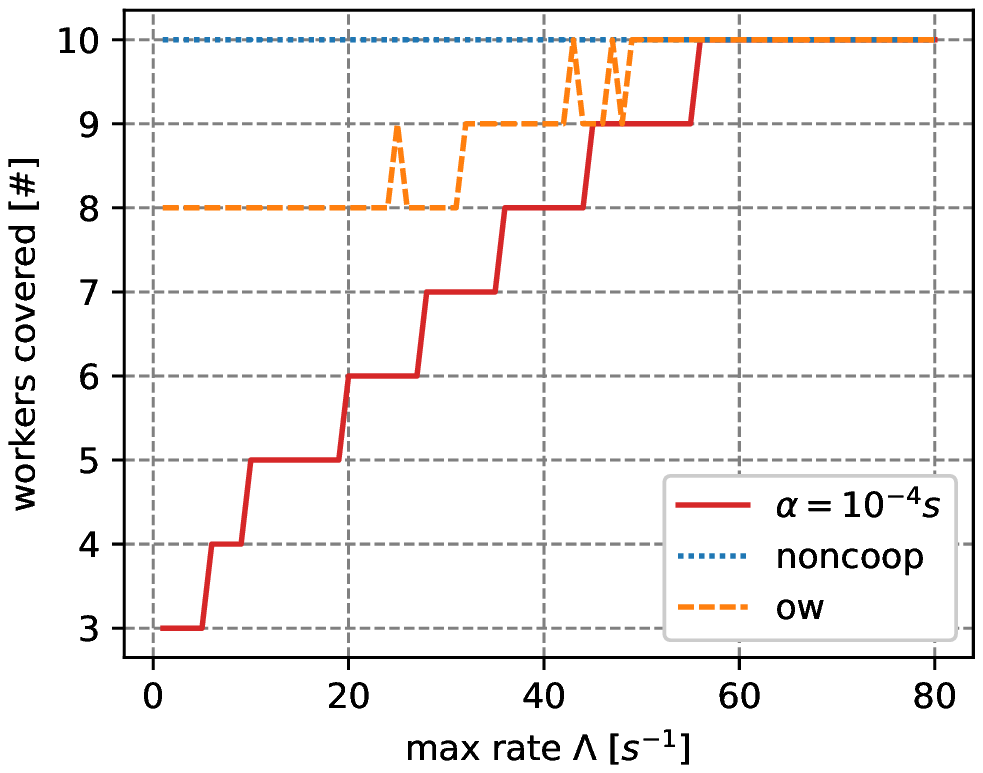}
  \caption{Workers Covered}
  \label{fig:comp:workers_covered}
  \end{subfigure}
  \begin{subfigure}{.495\textwidth}
  \centering
  \includegraphics[width=\linewidth]{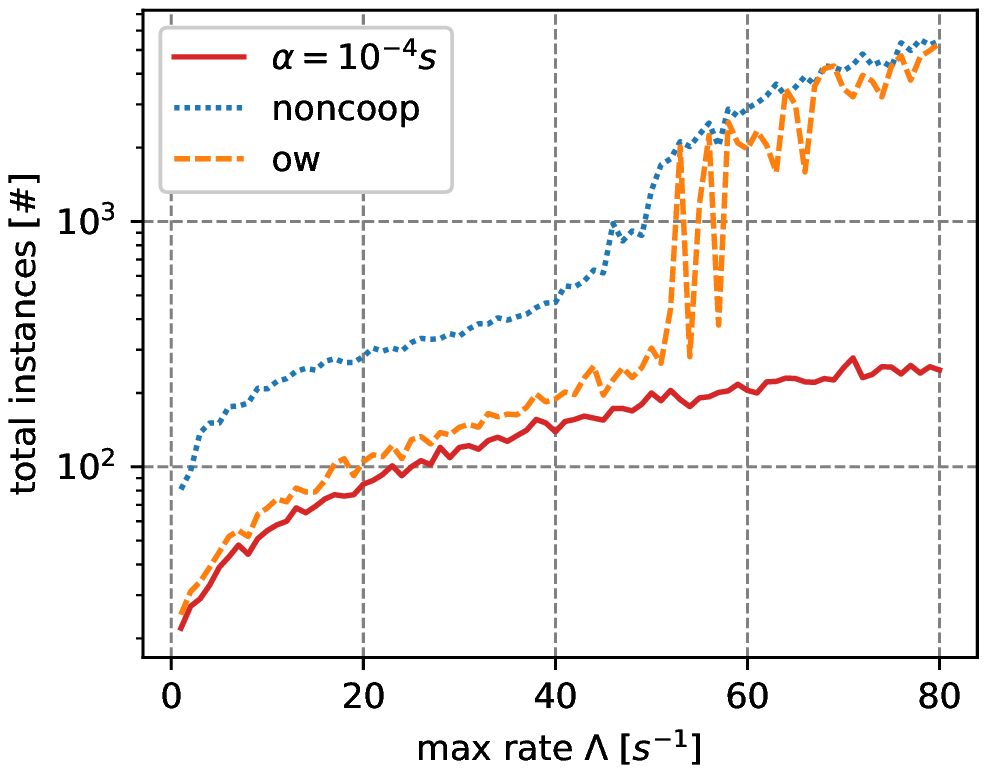}
  \caption{Total Instances}
  \label{fig:comp:instances}
  \end{subfigure}
  \begin{subfigure}{.495\textwidth}
  \centering
  \includegraphics[width=\linewidth]{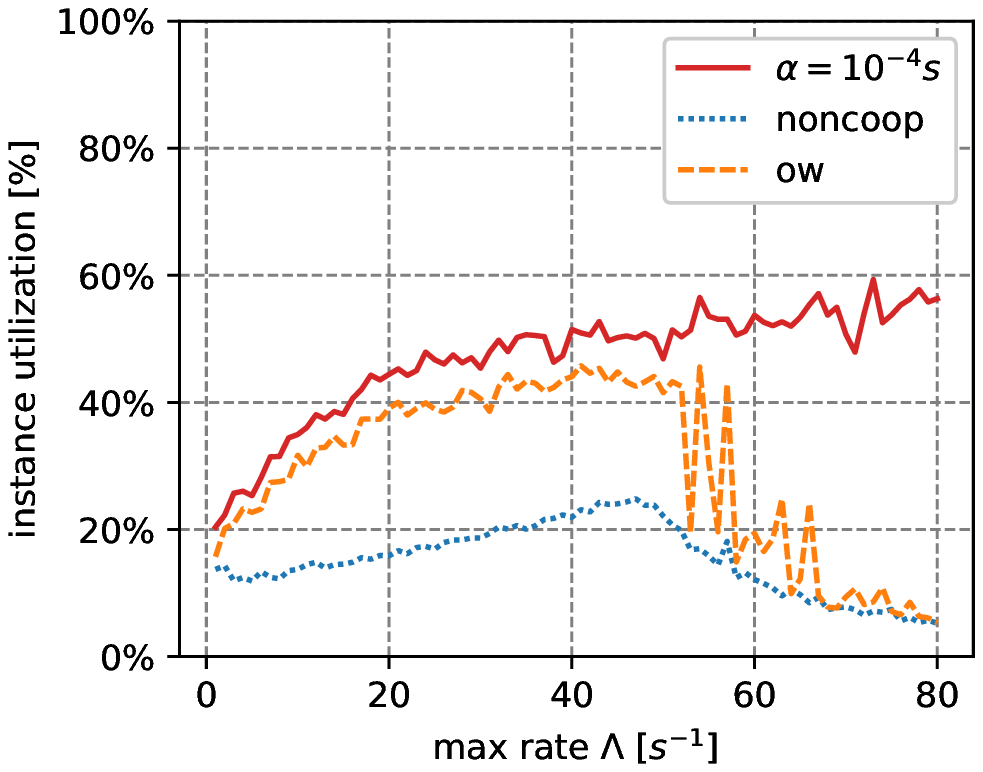}
  \caption{Instance utilisation}
  \label{fig:comp:instance_utilisation}
  \end{subfigure}
  \caption{Comparison of NOAH, OpenWhisk and Noncooperative load balancing} 
  \label{fig:comp}
\end{figure}

Figure~\ref{fig:comp} shows the average response times of different NOAH configurations in comparison to OpenWhisk (\emph{ow}) and noncooperative load balancing (\emph{noncoop}). 
Regarding response times (Figure~\ref{fig:comp:response_time}), the results are mixed.
OpenWhisk can maintain the best response times under moderate load.
Message transfer and container unpausing seem to be the only delays to the execution time.
NOAH is slightly above, accepting the occasional queuing delay.
Noncooperative load balancing can not score good response times at all, as it tries to balance the 10 classes evenly across all hosts.
Unknowingly, the lack of event class aggregation causes unnecessary setup delays.
Under very low load, the additional delay can be attributed to starting instances on all hosts, whereas under moderate load ($\approx 50$), increasing instance churn seems to add to the delay.
Starting from $\Lambda$$\approx$$50$, which is only about two thirds of the theoretical bound, both noncooperative load balancing and OpenWhisk have uncontrollably high average response times.
Setup times climb excessively and no longer differ from the performance of a classic first-fit heuristics when OpenWhisk random progression starts to search for free spots (cmp. Figure~\ref{fig:heur:response_time}).
Although slightly increasing, NOAH maintains average response times, because its virtual allocations constrain the scheduler to reuse locations and accept queuing instead of searching for a free spot that would also cause creation of an instance.
As expected, noncooperative load balancing covers the entire pool of workers (Figure~\ref{fig:comp:workers_covered}).
OpenWhisk hashing to hosts seems to cause two collisions as it uses only $8$ out of $10$ workers under low load and expands to the entire pool to schedule overflow loads.
NOAH shows a slightly concave, almost linear proportion of demand and resource cost, which stems from the allocation scaling depicted in Figure~\ref{fig:noah:scaling}.
The workload aggregation on few workers allows to release workers entirely under low load and shows good adaptive behaviour.
The total number of instances created as shown in Figure~\ref{fig:comp:instances} seconds the interpretation that increasing response times under high load are caused by increasing instance churn rates.
The oscillating increase in the number of instances created for OpenWhisk matches its unstable response time increase. The more instances created, the higher the average response time.
Note, that both the response time graph and the graph with the number of created instances use a log scale, so the real increase is exponential.
Also note, that OpenWhisk creates more instances under moderate load than NOAH where it also achieves better response times.
NOAH, on the other hand, constrains the allocation to less workers than OpenWhisk and accepts occational queuing delays.
This relation nicely shows the tradeoff between cost and performance.
Lastly, Figure~\ref{fig:comp:instance_utilisation} shows utilisation performance.
The results are not surprising.
Because of the high numbers of instances created by noncooperative load balancing and the inclined setup times, it can not achieve good utilisation.
OpenWhisk likewise has a slightly lower utilisation to realise slightly better response times and degrades terribly when it starts to uncontrollably place events across the worker pool in search for a first fit.

%
%

\textbf{Conclusively}, the evaluation shows great results for NOAH, especially when it comes to high system utilisation.
The scenario tests only a short timeframe (a single ``sawtooth'' load increase)\footnote{When the maximum load $\Lambda$ is kept for a longer period, it only becomes more evident under which load levels the approaches end up keep piling up work. \SI{200}{\second} periods were also tested but are not discussed for brevity.}.
However, short load bursts of a few seconds are not entirely uncommon. 
None of the other approaches among classic heuristics, noncooperative load balancing (which lacks aggregation) or OpenWhisk hash-based balancing
(that lacks allocations) can contain the setup times in high workload conditions. NOAH's use of virtual allocations 
stabilises the platform behaviour.


\chapter{Conclusion}
\label{discussion}

The evaluation shows great results for the newly developed Noncooperative Online Allocation Heuristic, 
particularly when it comes to load bursts.
This means, NOAH can be used to run much higher utilisations of the platform resources and allows a smoother scaling of the worker pool.
Competitive heuristics require the platform to scale the worker pool at two-thirds of the theoretical load limit, which means they 
need to overprovision by 50\% of the workload.
Surprisingly, this matches both anectdotal operation limits of telecommunication provider equipment\footnote{Operation of servers is considered safe until $\approx$70\%}
as well as the data center utilisations reported by \citet*{koomey:2015} and their updated report \cite{koomey:2017}, which still does not show full utilisation of allocated resources despite getting rid of ``comatose'' servers.
Raising the safe operation limits of resources for higher utilisations is exactly what has motivated the research for a better event scheduling solution (cmp. section~\ref{intro:motivation}).

The first objective to \textbf{exploit locality} is addressed by NOAH with the aggregation of event classes using virtual allocations.
The function runtime and its library dependencies are large (290MB for the OpenWhisk container image, 140MB for a Python runtime) and instance creation (measured \SI{504}{\milli\second} for the container and \SI{307}{\milli\second} for function initialisation)
contributes major delays as compared to data transfer (e.g. up to 900MB can be transferred in \SI{800}{\milli\second} at 1135MB/s in a \gls{10GbE} network).
There are two important developments to this objective.
First, alternative approaches to context isolation and instance provisioning achieve much lower setup times (cmp. \cite{lightvm,enddominance})
 than OpenWhisk action containers,
so the serverless platform would rather need to consider data location and aggregate for lower context synchronisation overhead.
Second, recent research on resource disaggregation tries to lower the transfer speed barrier between servers using combinations of \gls{RDMA} and \gls{NVMe} to relax the requirements on data locality constraints (cmp. \cite{disaggregation},\cite{crail}).
Disaggregated data access means, the serverless platform would need to aggregate for function types again.
The current approach leaves room for improvement, e.g. to provide a generic event classification that can be configured depending on the application type.

The second objective to \textbf{adapt to varying demand} is achieved by having the algorithm scale the virtual allocations and therefore aggregate demand to the estimated resource set.
The scaling behaviour shows clear potential to keep the worker pool size at the required (estimated) minimum.

The third objective to \textbf{find a tradeoff between response times and resource allocation} is discussed throughout the thesis. 
It is reflected in the resource/application dichotomy noted by \citet*{taxonomy:scheduling}. 
It is inherent to the goal discussion for improving resource efficiency. 
And it is the basis to the formulated multi-objective optimisation problem (section~\ref{design:moop}). 
The evaluation of NOAH configurations (section~\ref{evaluation:noah}) shows the obvious tradeoff between response times and resource cost (worker coverage) using different configuration parameters $\alpha$.
In the chosen scenario, $\alpha$=$10^{-4}$ shows a visibly good efficiency tradeoff under all workload levels. The choice depends on the expected execution time of functions and the response time requirement of the application.

\textbf{Future directions.} The research summarised in chapter~\ref{design} has provided a thorough classification, problem definition and overview of methods.
However, more research may be worth exploring to refine the approach presented in this thesis.
First, research of the online job shop scheduling problem has been dominated by \glspl{PTAS} since the emergence of the primal-dual approach and now has various advancements in exploring multi-dimensional bin packing, design of competitive online algorithms, etc.
For example, \citet*{widjaja:2013} formulate an online load reallocation scheme for distributed data centers using duality and derive decentralised iterative algorithms. 
Similarities to optimal control problems or games with a single, global objective function mentioned by \citet*{tamer:1982} exist.
Second, game theory has inspired the developed heuristic (noncooperative allocation), but leaves open questions to whether coalition formation, tit-for-tat allocation exchanges or bidding for free spots can improve the approach, especially for aggregation of events.

\textbf{Contribution summary.}
This thesis makes several contributions.
As serverless is still being defined, this thesis contributes a design reference for serverless systems based on the state-of-the-art of Cloud ecosystems and contributes a clear definition of the serverless scheduling problem, considering the stakeholder objectives and operation aspects of a serverless system.
The thesis presents a small subset of the plentiful applied and theoretical research that exists on scheduling to help explore possible solutions to the serverless scheduling problem.
Also, a serverless simulator is contributed that implements the approaches and is likely to be reused for future research.

The main contribution is the new serverless scheduling solution NOAH, whose considerate design has proven 
to address the objectives and ultimately the aim of this thesis. The compelling results have been summarised 
in a short paper and submitted for peer review at an internationally well recognised workshop~\cite{hotcloud18}.

Serverless online event scheduling remains an interesting problem;
not only because of the complexity of the problem and the variety of methods that try to solve it, 
but also because it seems that existing implementations might still be based on simple web request load balancing heuristics despite the
fact that event execution can bear a significant setup time and data synchronisation overheads.
As contemporary work\cite{cncf:serverlesswhitepaper} positions serverless as a new paradigm to application deployment in different ecosystems, 
it is likely that we will see more research on this topic in the near future.

%
%


\setstretch{1}
\bibliographystyle{apalike}
\bibliography{ms}

\begin{thebibliography}{}

\bibitem[Alex~Reznik, 2017]{etsi:mec}
Alex~Reznik, Rohit~Arora, M. C. L. C. W. F. R. F. F. G. S. K. A. L. D. S. C. T.
  Z.~Z. (2017).
\newblock Developing software for multi-access edge computing.
\newblock Technical report, European Telecommunication Standards Institute.

\bibitem[{Apache OpenWhisk project}, 2018]{openwhisk:loadbalancer}
{Apache OpenWhisk project} (2018).
\newblock {OpenWhisk Container Pool Balancer}.
\newblock
  \url{https://github.com/apache/incubator-openwhisk/blob/aa85fc9b9bf32eb78309df5c9390a195375cfb7d/core/controller/src/main/scala/whisk/core/loadBalancer/LoadBalancerService.scala}.

\bibitem[Ba{\c{s}}ar and Olsder, 1982]{tamer:1982}
Ba{\c{s}}ar, T. and Olsder, G. (1982).
\newblock {\em Dynamic Noncooperative Game Theory}.
\newblock Mathematics in science and engineering. Academic Press.

\bibitem[Calheiros et~al., 2009]{cloudsim}
Calheiros, R.~N., Ranjan, R., Rose, C. A. F.~D., and Buyya, R. (2009).
\newblock Cloudsim: {A} novel framework for modeling and simulation of cloud
  computing infrastructures and services.
\newblock {\em CoRR}, abs/0903.2525.

\bibitem[Casavant and Kuhl, 1988]{taxonomy:scheduling}
Casavant, T.~L. and Kuhl, J.~G. (1988).
\newblock A taxonomy of scheduling in general-purpose distributed computing
  systems.
\newblock {\em IEEE Transactions on Software Engineering}, 14(2):141--154.

\bibitem[Checconi et~al., 2010]{checconi:livemigration}
Checconi, F., Cucinotta, T., and Stein, M. (2010).
\newblock Real-time issues in live migration of virtual machines.
\newblock In {\em Proceedings of the 2009 International Conference on Parallel
  Processing}, Euro-Par'09, pages 454--466, Berlin, Heidelberg.
  Springer-Verlag.

\bibitem[Chen and Deelman, 2012]{Chen:2012}
Chen, W. and Deelman, E. (2012).
\newblock {WorkflowSim}: {A} toolkit for simulating scientific workflows in
  distributed environments.
\newblock In {\em 8th {IEEE} International Conference on E-Science, e-Science
  2012, Chicago, IL, USA, October 8-12, 2012}, pages 1--8.

\bibitem[Cho et~al., 2017]{moop:survey}
Cho, J.~H., Wang, Y., Chen, I.~R., Chan, K.~S., and Swami, A. (2017).
\newblock A survey on modeling and optimizing multi-objective systems.
\newblock {\em IEEE Communications Surveys Tutorials}, 19(3):1867--1901.

\bibitem[{CNCF Serverless Working Group}, 2018]{cncf:serverlesswhitepaper}
{CNCF Serverless Working Group} (2018).
\newblock {CNCF WG-Serverless Whitepaper v1.0}.

\bibitem[Coello, 2000]{Coello:2000}
Coello, C. A.~C. (2000).
\newblock An updated survey of ga-based multiobjective optimization techniques.
\newblock {\em {ACM} Comput. Surv.}, 32(2):109--143.

\bibitem[Commission, 2017]{H2020}
Commission, E. (2017).
\newblock {Horizon 2020 Work Programme 2018-2020 - Information and
  Communication Technologies}.
\newblock
  \url{http://ec.europa.eu/research/participants/data/ref/h2020/wp/2018-2020/main/h2020-wp1820-leit-ict_en.pdf}.

\bibitem[Duan et~al., 2014]{duan:2014}
Duan, R., Prodan, R., and Li, X. (2014).
\newblock Multi-objective game theoretic schedulingof bag-of-tasks workflows on
  hybrid clouds.
\newblock {\em IEEE Transactions on Cloud Computing}, 2(1):29--42.

\bibitem[Ellis, 2017]{openfaas}
Ellis, A. (2017).
\newblock \url{https://blog.alexellis.io/functions-as-a-service/}.

\bibitem[Fox et~al., 2017]{wosc}
Fox, G.~C., Ishakian, V., Muthusamy, V., and Slominski, A. (2017).
\newblock Status of serverless computing and {Function-as-a-Service} ({FaaS})
  in industry and research.
\newblock Technical report, First International Workshop on Serverless
  Computing (WoSC) 2017.

\bibitem[Franx, 2001]{Franx:2001}
Franx, G.~J. (2001).
\newblock A simple solution for the m/d/c waiting time distribution.
\newblock {\em Oper. Res. Lett.}, 29(5):221--229.

\bibitem[Gandhi et~al., 2013]{Gandhi:2013}
Gandhi, A., Doroudi, S., Harchol-Balter, M., and Scheller-Wolf, A. (2013).
\newblock Exact analysis of the m/m/k/setup class of markov chains via
  recursive renewal reward.
\newblock In {\em Proceedings of the ACM SIGMETRICS/International Conference on
  Measurement and Modeling of Computer Systems}, SIGMETRICS '13, pages
  153--166, New York, NY, USA. ACM.

\bibitem[Grosu and Chronopoulos, 2005]{grosu:2005}
Grosu, D. and Chronopoulos, A.~T. (2005).
\newblock Noncooperative load balancing in distributed systems.
\newblock {\em J. Parallel Distrib. Comput.}, 65(9):1022--1034.

\bibitem[Grosu et~al., 2008]{grosu:2008}
Grosu, D., Chronopoulos, A.~T., and Leung, M.~Y. (2008).
\newblock Cooperative load balancing in distributed systems.
\newblock {\em Concurr. Comput. : Pract. Exper.}, 20(16):1953--1976.

\bibitem[Gupta et~al., 2016]{Gupta:2016}
Gupta, H., Dastjerdi, A.~V., Ghosh, S.~K., and Buyya, R. (2016).
\newblock ifogsim: {A} toolkit for modeling and simulation of resource
  management techniques in internet of things, edge and fog computing
  environments.
\newblock {\em CoRR}, abs/1606.02007.

\bibitem[Han et~al., 2013]{disaggregation}
Han, S., Egi, N., Panda, A., Ratnasamy, S., Shi, G., and Shenker, S. (2013).
\newblock Network support for resource disaggregation in next-generation
  datacenters.
\newblock In {\em Proceedings of the Twelfth ACM Workshop on Hot Topics in
  Networks}, HotNets-XII, pages 10:1--10:7, New York, NY, USA. ACM.

\bibitem[Hopps, 2000]{RFC2992}
Hopps, C. (2000).
\newblock Analysis of an equal-cost multi-path algorithm.
\newblock RFC 2992, RFC Editor.

\bibitem[{HyperHQ}, 2015]{hypercontainers}
{HyperHQ} (2015).
\newblock {Hyper - Make VM run like Container}.
\newblock \url{https://hypercontainer.io}.

\bibitem[Iorga et~al., 2017]{nist:fog}
Iorga, M., Feldman, L., Barton, R., Martin, M.~J., Goren, N., and Mahmoudi, C.
  (2017).
\newblock {SP 800-191. The NIST Definition of Fog Computing}.
\newblock Technical report, National Institute of Standards \& Technology,
  Gaithersburg, MD, United States.

\bibitem[Jiao et~al., 2017]{jiao:2017}
Jiao, L., Tulino, A.~M., Llorca, J., Jin, Y., and Sala, A. (2017).
\newblock Smoothed online resource allocation in multi-tier distributed cloud
  networks.
\newblock {\em {IEEE/ACM} Trans. Netw.}

\bibitem[Johnson, 1974]{Johnson:1974}
Johnson, D.~S. (1974).
\newblock Fast algorithms for bin packing.
\newblock {\em J. Comput. Syst. Sci.}, 8(3):272--314.

\bibitem[Karger et~al., 1997]{Karger:1997}
Karger, D., Lehman, E., Leighton, T., Panigrahy, R., Levine, M., and Lewin, D.
  (1997).
\newblock Consistent hashing and random trees: Distributed caching protocols
  for relieving hot spots on the world wide web.
\newblock In {\em Proceedings of the Twenty-ninth Annual ACM Symposium on
  Theory of Computing}, STOC '97, pages 654--663, New York, NY, USA. ACM.

\bibitem[Koller and Williams, 2017]{enddominance}
Koller, R. and Williams, D. (2017).
\newblock Will serverless end the dominance of linux in the cloud?
\newblock In {\em Proceedings of the 16th Workshop on Hot Topics in Operating
  Systems}, HotOS '17, pages 169--173, New York, NY, USA. ACM.

\bibitem[Koomey and Taylor, 2015]{koomey:2015}
Koomey, J. and Taylor, J. (2015).
\newblock {30\% of Servers are Sitting "Comatose"}.
\newblock
  \url{https://blog.anthesisgroup.com/30-of-servers-are-sitting-comatose}.

\bibitem[Koomey and Taylor, 2017]{koomey:2017}
Koomey, J. and Taylor, J. (2017).
\newblock {Zombie/Comatose Server Redux}.
\newblock \url{https://blog.anthesisgroup.com/zombie-servers-redux}.

\bibitem[Lopes and Menasc\'e, 2016]{taxonomy:jobscheduling}
Lopes, R.~V. and Menasc\'e, D. (2016).
\newblock A taxonomy of job scheduling on distributed computing systems.
\newblock {\em IEEE Transactions on Parallel and Distributed Systems},
  27(12):3412--3428.

\bibitem[Madhavapeddy et~al., 2013]{unikernel}
Madhavapeddy, A., Mortier, R., Rotsos, C., Scott, D., Singh, B., Gazagnaire,
  T., Smith, S., Hand, S., and Crowcroft, J. (2013).
\newblock Unikernels: Library operating systems for the cloud.
\newblock In {\em Proceedings of the Eighteenth International Conference on
  Architectural Support for Programming Languages and Operating Systems},
  ASPLOS '13, pages 461--472, New York, NY, USA. ACM.

\bibitem[Manco et~al., 2017]{lightvm}
Manco, F., Lupu, C., Schmidt, F., Mendes, J., Kuenzer, S., Sati, S., Yasukata,
  K., Raiciu, C., and Huici, F. (2017).
\newblock My {VM} is lighter (and safer) than your container.
\newblock In {\em Proceedings of the 26th Symposium on Operating Systems
  Principles}, SOSP '17, pages 218--233, New York, NY, USA. ACM.

\bibitem[McGrath and Brenner, 2017]{McGrath}
McGrath, G. and Brenner, P.~R. (2017).
\newblock Serverless computing: Design, implementation, and performance.
\newblock In {\em 2017 IEEE 37th International Conference on Distributed
  Computing Systems Workshops (ICDCSW)}, pages 405--410.

\bibitem[MSV, 2017]{behindopenwhisk}
MSV, J. (2017).
\newblock An architectural view of apache openwhisk.
\newblock
  \url{https://thenewstack.io/behind-scenes-apache-openwhisk-serverless-platform/}.

\bibitem[{OpenStack Foundation}, 2017]{katacontainers}
{OpenStack Foundation} (2017).
\newblock {Kata Containers - The speed of containers, the security of VMs}.
\newblock \url{https://katacontainers.io}.

\bibitem[Piraghaj et~al., 2017]{piraghaj:2017}
Piraghaj, S.~F., Dastjerdi, A.~V., Calheiros, R.~N., and Buyya, R. (2017).
\newblock Containercloudsim: An environment for modeling and simulation of
  containers in cloud data centers.
\newblock {\em Softw., Pract. Exper.}, 47(4):505--521.

\bibitem[Rahman et~al., 2011]{taxonomy:autonomicmanagement}
Rahman, M., Ranjan, R., Buyya, R., and Benatallah, B. (2011).
\newblock A taxonomy and survey on autonomic management of applications in grid
  computing environments.
\newblock {\em Concurrency and Computation: Practice and Experience},
  23(16):1990--2019.

\bibitem[Seiden, 2002]{Seiden:2002}
Seiden, S.~S. (2002).
\newblock On the online bin packing problem.
\newblock {\em J. ACM}, 49(5):640--671.

\bibitem[Stein, 2017]{interim}
Stein, M. (2017).
\newblock Interim report on adaptive event dispatching in serverless computing
  infrastructures - interim report.
\newblock \url{https://arxiv.org/abs/1901.02680}.

\bibitem[Stein, 2018]{hotcloud18}
Stein, M. (2018).
\newblock The serverless scheduling problem and {NOAH}.
\newblock \url{https://arxiv.org/abs/1809.06100}.

\bibitem[Stuedi et~al., 2017]{crail}
Stuedi, P., Trivedi, A., Pfefferle, J., Stoica, R., Metzler, B., Ioannou, N.,
  and Koltsidas, I. (2017).
\newblock Crail: {A} high-performance {I/O} architecture for distributed data
  processing.
\newblock {\em {IEEE} Data Eng. Bull.}, 40(1):38--49.

\bibitem[Tantawi and Towsley, 1985]{Tantawi:1985}
Tantawi, A.~N. and Towsley, D. (1985).
\newblock Optimal static load balancing in distributed computer systems.
\newblock {\em J. ACM}, 32(2):445--465.

\bibitem[Ullman, 1971]{ullman:1971}
Ullman, J. (1971).
\newblock {\em The Performance of a Memory Allocation Algorithm}.
\newblock Technical report. Dept. of Electrical Engineering, Computer Sciences
  Laboratory, Princeton University.

\bibitem[Widjaja et~al., 2013]{widjaja:2013}
Widjaja, I., Borst, S.~C., and Saniee, I. (2013).
\newblock Building an elastic cloud out of small datacenters.
\newblock In {\em 13th {IEEE/ACM} International Symposium on Cluster, Cloud,
  and Grid Computing, CCGrid 2013, Delft, Netherlands, May 13-16, 2013}.

\bibitem[Wiggins, 2011]{12factorapp}
Wiggins, A. (2011).
\newblock The twelve-factor app.
\newblock \url{https://12factor.net}.

\bibitem[Zhan et~al., 2015]{zhan:2015}
Zhan, Z.-H., Liu, X.-F., Gong, Y.-J., Zhang, J., Chung, H. S.-H., and Li, Y.
  (2015).
\newblock Cloud computing resource scheduling and a survey of its evolutionary
  approaches.
\newblock {\em ACM Comput. Surv.}, 47(4):63:1--63:33.

\end{thebibliography}

\end{document}